\documentclass[draft]{agujournal2019}
\usepackage{url} 
\usepackage{lineno}
\usepackage{amsmath}
\usepackage[inline]{trackchanges} 
\usepackage{soul}
\draftfalse

\journalname{Agricultural and Forest Meteorology}

\begin{document}

%
%


\title{Impact of Heterogeneity on Scalar Flux Variance Relations Across Diverse Ecosystems}

%
%




\authors{T. Waterman\affil{1}, I. Stiperski\affil{2}, L. Torres-Rojas\affil{3}, M. Calaf\affil{1}}

\affiliation{1}{Mechanical Engineering, University of Utah, Salt Lake City, Utah, United States of America}
\affiliation{2}{Atmospheric and Cryospheric Sciences, Universität Innsbruck, Innsbruck, Tyrol, Austria}
\affiliation{3}{Program in Atmospheric and Oceanic Sciences, Princeton University}




\correspondingauthor{Dr. Tyler Waterman}{tyswater@gmail.com}




%
%

%
%


\begin{abstract}
Monin-Obukhov Similarity Theory (MOST), the traditional surface layer theory used to understand the behavior, scaling and exchange of heat, water vapor and carbon dioxide between the land surface and atmosphere relies on a number of commonly broken assumptions. In particular, traditional theory breaks down under three different forms of heterogeneity highlighted in this work: spatial heterogeneity in the sources of the scalars, heterogeneity in the Reynolds stress tensor (turbulence anisotropy), and temporal heterogeneity (non-stationarity). The work explores the relationship between the idealized flux-variance relations and these three forms of heterogeneity across a diverse network of 47 flux towers representing a broad range of ecosystems including forests, agricultural land, grasslands, tundra, tropical and arid: the National Ecological Observation Network (NEON). Results use high resolution spatial data (1 meter resolution) to show a direct relationship between spatial heterogeneity and deviation from traditional scaling relations. Prior work indicates a close relationship between turbulence anisotropy and velocity scaling. Results from this work show that the turbulence anisotropy dependence for variance scaling effectively extends from the velocity to moisture, heat and carbon. The study also indicates an interplay between stationarity and anisotropy, with the non-dimensionalized scalar variance scaling more strongly with anisotropy under more non-stationary turbulence conditions. Updated flux-variance relations that leverage turbulence anisotropy for the scaling of temperature are introduced, as are novel anisotropy-generalized scalings for water vapor and carbon dioxide. The novel scalings show significant improvement over traditional relations. The work also explores in detail how the scaling relations, and their relationship with heterogeneity, vary across the diverse sites in the NEON network. Deviations from traditional theory in carbon dioxide scaling in particular are well correlated with the bioactivity of the site. Results have important implications for development of improved surface layer parameterizations in large scale atmospheric models and flux-variance based flux measurements.
\\
\textbf{Keywords: } surface layer scaling, heterogeneity, turbulence anisotropy, stationarity, carbon dioxide, water vapor
\end{abstract}


%
%

%


%
%
%
%

\section{Introduction}
Understanding the transport of carbon dioxide, water vapor, and heat from the land surface to the Atmospheric Boundary Layer (ABL) is critical for a broad range of applications, from agriculture and forest management to weather and climate prediction. The behavior of these three species in the atmospheric surface layer (ASL) has been used for decades to develop applied and theoretical models and parameterization schemes \cite{foken_50_2006,sfyri_scalar-flux_2018,shih_challenges_2025,ohtaki_similarity_1985}. Many of these parameterizations, including the ones used in nearly all Earth System Models (ESMs) for both Numerical Weather Prediction (NWP) and long-term climate simulations \cite{nakanishi_development_2009,larson_clubb-silhs_2022,cheng_improved_2015}, ultimately rely on Monin-Obukhov Similarity Theory (MOST) and therefore implicitly accept the core MOST assumptions: (1) The flow is stationary; statistics do not change at the timescales of analysis, (2) the flow is horizontally homogeneous; statistics do not change horizontally in space, and (3) there is no subsidence; no mean vertical flow \cite{monin_basic_1954}. Many key violations of these assumptions for the scalars can be broken down into different forms of heterogeneity, both spatial and temporal, affecting the scalars. Some forms of heterogeneity, in particular heterogeneity in the Reynolds stress tensor, or anisotropy, have been shown to improve MOST-based models for velocity variances when properly integrated into existing scalings \cite{stiperski_generalizing_2023,waterman_evaluating_2025} as well as the velocity and heat gradients and turbulent spectra \cite{mosso_fluxgradient_2024,charrondiere_spectral_2024}. In this work, we examine how heterogeneity, non-stationarity and turbulence anisotropy are related to deviations from traditional MOST-based surface layer scaling for the variances of heat, moisture and carbon across the wide range of terrains and ecosystems that should be well represented in our global models.

MOST-based models are widely used and implemented, particularly in ASL schemes. While more traditional boundary layer schemes employ first-order, 1.5-order or occasionally second order closure schemes \cite{cohen_review_2015,lock_new_2000}, many modern schemes leverage higher order closures that resolve the variances of temperature and moisture prognostically and require complete surface layer scalings \cite{waterman_examining_2022,larson_clubb-silhs_2022,nakanishi_development_2009,cheng_improved_2015}. MOST, specifically the flux-variance relations, is also leveraged to make measurements of ecosystem fluxes (latent heat and carbon fluxes) using less expensive instrumentation than the standard eddy covariance method requires \cite{fischer_merging_2023,shih_challenges_2025,buttar_fetch_2019,buttar_estimation_2022}. MOST-based methods are common due to the simple and powerful relationships they enable. MOST flux-variance relations state that the variance of any quantity $s$ in the surface layer, ($\sigma_s^2$), non-dimensionalized by the flux of $s$ ($\overline{w's'}$) and friction velocity ($u_*$), will be a universal function of only one parameter $\zeta$, or
\begin{equation}
    \Phi_{s}=f(\zeta)
    \label{eq:phi_eq}
\end{equation}
where $\Phi_{s}$ is defined as 
\begin{equation}
    \Phi_{s}=\frac{\sigma_s}{\overline{w's'}}u_*
    \label{eq:phi_def}
\end{equation}
with $\sigma_s$ as the square root of the scalar variance (standard deviation) and with stability parameter 
\begin{equation}
    \zeta=\frac{z}{L}=\frac{z_h-d}{-u_*^3\overline{\theta_v}/\kappa g \overline{w'\theta_v'}}
    \label{eq:zeta}
\end{equation}
 with height $z_h$ as height above the ground, $d$ as the zero plane displacement height, $\theta_v$ as the virtual potential temperature, $g$ as the acceleration due to gravity, $\kappa=0.4$ as the von Kármán constant, $\overline{w'\theta_v'}$ as the buoyancy flux, and friction velocity $ u_*=(\overline{u'w'}^2+\overline{v'w'}^2)^{1/4}$.  The form $f(\zeta)$ is obtained by curve fitting through observational results, with these fit curves applied in the aforementioned MOST-based schemes \cite{monin_basic_1954,wyngaard_budgets_1971}

\par While the flux-variance relations are an elegant solution for understanding surface exchange, there are significant and common deviations from MOST in observations. The surface layer scaling for temperature is relatively well studied, and its deviations from traditional theory fairly well understood. In the near neutral regime, as the buoyancy flux approaches zero, the temperature variance can remain finite due to small scale variability in surface temperatures which can be exacerbated by non-stationarity in the time series \cite{chor_flux-variance_2017,sfyri_scalar-flux_2018,waterman_examining_2022,kroon_crau_1995,wyngaard_budgets_1971}. Temperature also fails to scale under stable conditions\cite{mahrt_stratified_1999}. Intermittent turbulence, especially caused by sub-mesoscale motions and small secondary circulations but also internal mechanisms and Kelvin-Helmholtz instability \cite{linden_businger_2020}, violate the assumptions of stationarity and cause a breakdown of the flux-variance relations and surface layer scaling \cite{mahrt_stratified_1999,lee_influence_2009,moraes_nocturnal_2004}. Studies also show that complex topography can cause failures in the temperature scaling \cite{de_franceschi_analysis_2009,nadeau_similarity_2013,sfyri_scalar-flux_2018}, as can surface heterogeneity in thermal and vegetative properties, even in terrain typically considered homogeneous \cite{moraes_nocturnal_2004,kroon_crau_1995,sfyri_scalar-flux_2018,waterman_examining_2022}.

\par Many of the deviations from theory discussed for temperature also apply to the bioactive scalars of  water vapor and carbon dioxide, including differences in scaling in complex topography and spatial heterogeneity \cite{nadeau_similarity_2013,sfyri_scalar-flux_2018,detto_surface_2008,detto_scaling_2010}. Complete understanding of moisture and carbon scaling is essential for understanding biogeochemical cycles, as well as accurate parameterization of moisture exchange in models. While most surface layer turbulence experiments require only a 3D-sonic anemometer, to explore carbon and moisture scaling additional, at times expensive, infrared gas analyzers or similar devices are needed. As such, comparatively fewer field campaigns examine the surface layer scaling of water vapor and carbon dioxide. Studies on moisture scaling show that there is no strong consensus on the details of scaling constants and forms of the scaling relations \cite{sfyri_scalar-flux_2018,ramana_surface_2004}. Carbon scaling is also less studied, with most studies showing inconsistencies in scaling, large scatter in the observations around the scaling curves, a failure to scale when non-turbulent transport and small fluxes are prevalent, and a notable dissimilarity between moisture, temperature and carbon \cite{zahn_relaxed_2023,chor_flux-variance_2017,detto_scaling_2010,rannik_surface_1998,ohtaki_similarity_1985}. Many studies use site placement, data pruning, and elimination of non-stationary regimes, to fulfill MOST assumptions. However, for application in atmospheric models which must frequently represent non-ideal conditions, stronger understanding of the scaling in these regimes is needed. 

\par While MOST was not originally derived from the budgets for variances in the atmosphere, it can be examined and even derived from a surface budget perspective \cite{siqueira_estimating_2002} which can help illuminate how spatial, temporal, and Reynolds stress heterogeneity may impact the surface layer scaling of scalar variances. If we assume streamwise coordinate axes, the variance budget for some scalar $s$ can be written as \cite{stull_mean_1988}:
\begin{equation}
    \underbrace{\frac{\partial\overline{s'^2}}{\partial t}}_{\text{Storage}}+
    \underbrace{\overline{u}\frac{\partial\overline{s'^2}}{\partial x}+\overline{w}\frac{\partial\overline{s'^2}}{\partial z}}_{\text{Advection}}=
    \underbrace{-2\overline{s'u'}\frac{\partial s}{\partial x}-2\overline{w's'}\frac{\partial s}{\partial z}}_{\text{Production}}
    -\underbrace{2\epsilon_q}_{\text{Dissipation}}-\underbrace{\frac{\partial \overline{u's's'}}{\partial x}+\frac{\partial \overline{v's's'}}{\partial y}+\frac{\partial \overline{w's's'}}{\partial z}}_{\text{Turbulent Transport}}
    \label{eq:var_budget}
\end{equation}
For potential temperature variance, there is an additional radiative term which will not be considered here. MOST assumptions of horizontal homogeneity ($ds/dx=0$), no subsidence ($\overline{w}=0$) and stationarity ($ds/dt=0$) remove the advection, storage, horizontal production and horizontal turbulent transport terms. Using the argument from \citeA{detto_scaling_2010} we can also assume the vertical turbulent transport $\frac{\partial \overline{w's's'}}{\partial z}$ would be negligible, leading to a balance of production and dissipation that is assumed to hold for MOST conditions.  

\par For heterogeneous surface conditions, the horizontal production terms and the advection term in the variance budget equations become significant and MOST is not expected to hold. Spatial heterogeneity has long been shown to be highly relevant for surface layer dynamics \cite{finnigan_boundary-layer_2020,bou-zeid_persistent_2020}. Surface heterogeneity at the fine scales relevant for ASL scaling is difficult to examine quantitatively.  Since the spatial heterogeneity relevant for surface layer scaling is small scale ($O(0.1-10\ m)$), commonly available spatial products from satellite based remote sensing are too coarse resolution to adequately assess ASL scales for velocity or scalars. A number of studies have examined scalar variance scalings in heterogeneous terrain, often without quantification of the degree of heterogeneity. \cite{moraes_nocturnal_2004,de_franceschi_analysis_2009,nadeau_similarity_2013,grachev_airsealand_2018, martins_turbulence_2009}. Studies tend to focus on velocity or temperature, occasionally moisture, and rarely carbon dioxide. Recent studies have also used process of elimination \cite{sfyri_scalar-flux_2018} to rule out other sources of error in the scaling and identify heterogeneity as a major source of deviation from traditional theory for heat and moisture. Others have examined potential proxies of spatial heterogeneity, specifically solar altitude $\alpha_{solar}$, to connect spatial patterns to deviations from MOST for all three scalars; at low solar altitude, patchworks of shaded areas will appear, especially in forested environments, that can generate complex patterns in the sources of heat, moisture and carbon \cite{chor_flux-variance_2017,zahn_scalar_2016}. These studies commonly find that spatial heterogeneity generally results in higher $\Phi$ values than would be predicted by MOST, as additional surface variability is included in measured variances. Limited attempts have been made at a more direct relationship between spatial patterns and scaling, most notably \citeA{detto_surface_2008}, which directly related high resolution spatial patterns and turbulent timescales to deviations from MOST scaling.

\par The second form of heterogeneity we wish to examine in this study is turbulence anisotropy, a form of heterogeneity in the Reynolds stress tensor. Turbulence anisotropy is correlated with deviations in the surface layer scaling of a number of parameters, including the velocity variances \cite{stiperski_dependence_2018,stiperski_anisotropy_2021,stiperski_generalizing_2023,waterman_evaluating_2025}, temperature variance \cite{stiperski_generalizing_2023}, gradients of heat and momentum \cite{mosso_fluxgradient_2024} and the spectra \cite{charrondiere_spectral_2024}. In these studies, the Reynolds stress tensor $\overline{u_i'u_j'}$ is normalized and decomposed into isotropic and anisotropic components. In Einstein notation, this is written as $b_{ij}=\frac{\overline{u_i'u_j'}}{\overline{u_l'u_l'}}-\frac{1}{3}\delta_{ij}$ \cite{pope_turbulent_2000} where the $\overline{u_lu_l}$ is twice the turbulence kinetic energy and $\delta$ is the Kronecker delta. From this tensor, we extract three invariants. While the first is null by definition, the second and third invariants can be mapped into a non-linear representation following the original work of \citeA{lumley_return_1977} or the linear representation from \citeA{banerjee_presentation_2007}. This linear representation allows examination of the behavior of turbulence and scaling within the invariant map known as the barycentric map of the lumley triangle. In this approach from \citeA{banerjee_presentation_2007}, the invariant $y_b$ is defined as a function of the smallest eigenvalue of the anisotropy tensor, and represents the degree of anisotropy, such that
\begin{equation}
    y_b=\sqrt{3}/2(3\lambda_3+1).
    \label{yb}
\end{equation}
The second invariant $x_b$ is related to the form of anisotropy (one or two-component), and is written as
\begin{equation}
    x_b=\lambda_1-\lambda_2+\frac{1}{2}(3\lambda_3+1)
    \label{xb}
\end{equation}
where $\lambda_1$, $\lambda_2$, $\lambda_3$ are the eigenvalues of the normalized anisotropy tensor $b_{ij}$ sorted from largest to smallest. Previous work mostly focuses on the invariant $y_b$, with small values indicating very anisotropic turbulence in one or two components and larger values (up to a maximum of $y_b=\sqrt{3}/2$) indicating isotropic turbulence. When $x_b$ is small, this indicates primarily two component turbulence and when it is large, turbulence is primarily one component where a single eigenvalue dominates over the others. By leveraging $y_b$, \citeA{stiperski_generalizing_2023} and \citeA{waterman_evaluating_2025} were able to produce modified scalings with significant improvement over traditional MOST, where instead of a function $\Phi$ as in equation \eqref{eq:phi_eq}, new curves are fitted with 
\begin{equation}
    \Phi_{s}=f(\zeta,y_b)
    \label{eq:phi_eqyb}
\end{equation} 
where the constants in traditional MOST relations were shown to be functions of $y_b$. These modified curves saw improvement in the scaling of the velocity variances on the order of $40\%$. While modified curves have been produced for temperature variance, work has yet to relate turbulence anisotropy to water vapor or carbon dioxide scaling.

\par It is relatively clear how spatial heterogeneity or temporal heterogeneity would affect the variance budget in equation \eqref{eq:var_budget}. While for turbulence anisotropy it is less straightforward, anisotropy would certainly have an effect on the turbulent production terms, as the Reynolds stress components appear in the budget for the scalar fluxes \cite{stull_mean_1988}. For the velocity variance, anisotropy has a clear connection to the turbulent transport term, and is indeed often leveraged for the closure of this term, although it is not obvious that this would extend readily to the scalars. Rigorous theory is still lacking connecting anisotropy directly to the scalar variance budget, however as others and this manuscript reveals, there are certainly relevant correlations, theory and relationships that can be surmised \cite{stiperski_generalizing_2023,waterman_evaluating_2025}. 

\par The third form of heterogeneity we explore in this work is stationarity, or the temporal heterogeneity of turbulence statistics. Non-stationarity is directly connected to the storage term in equation \eqref{eq:var_budget}, which is assumed to equal zero for traditional MOST. Studies consistently show higher error and elevated $\Phi$ values under non-stationary conditions, due to both changing mean values over longer periods, as well as highly intermittent flows \cite{babic_fluxvariance_2016,mahrt_non-stationary_2020,detto_scaling_2010,falocchi_refinement_2018,waterman_examining_2022,lee_influence_2009,zahn_relaxed_2023}.There are a number of metrics for stationarity, which can have significant differences in their performance when used to filter data for non-stationary conditions \cite{babic_fluxvariance_2016}. It is very common to use one, or multiple, of these metrics to remove non-stationary data to achieve agreement with MOST assumptions. One such metric, which will be used in this study, is the test described in \citeA{foken_tools_1996}:
\begin{equation}
    \xi_s=\left| \frac{\overline{s's'}_{5\ min}-\overline{s's'}_{30\ min}}{\overline{s's'}_{30\ min}} \right| \times100
    \label{eq:xi}
\end{equation}
When the $\xi_s$ metric is below 30\%, it is generally considered sufficiently stationary to accept the stationarity assumption of MOST. 

\par To properly understand the impacts of these three forms of heterogeneity on surface layer scaling of the variances of heat, water vapor, and carbon, this work leverages the vast and broad National Ecological Network (NEON) of towers across ecosystems in North America \cite{metzger_neon_2019}. The wide range of ecosystems, and degrees and form of heterogeneity, across nearly six years at 47 sites allows for a fuller picture of ASL dynamics when compared to more limited field campaigns. In section \ref{sec:data}, we will further describe the NEON data and how it is used it for further analysis. In the results, the general surface layer scaling of the 47 site ensemble (\ref{sec:trad_scaling}) is examined together with our three forms of heterogeneity: spatial heterogeneity (\ref{sec:spatial_het}), anisotropy (\ref{sec:anisotropy}), and non-stationarity (\ref{sec:stationarity}). The final results section explores trends and features of the surface layer scaling across the different sites in NEON and relates these results to the three forms of heterogeneity explored (section \ref{sec:sites}). The study finishes with a brief discussion on the interplay between the three forms of heterogeneity, and their relationship with traditional scalings (section \ref{sec:discussion}).

\section{Data and Methods}\label{sec:data}
\subsection{Data: National Ecological Observation Network}
\begin{figure}
\centering
\includegraphics[width=5.5in]{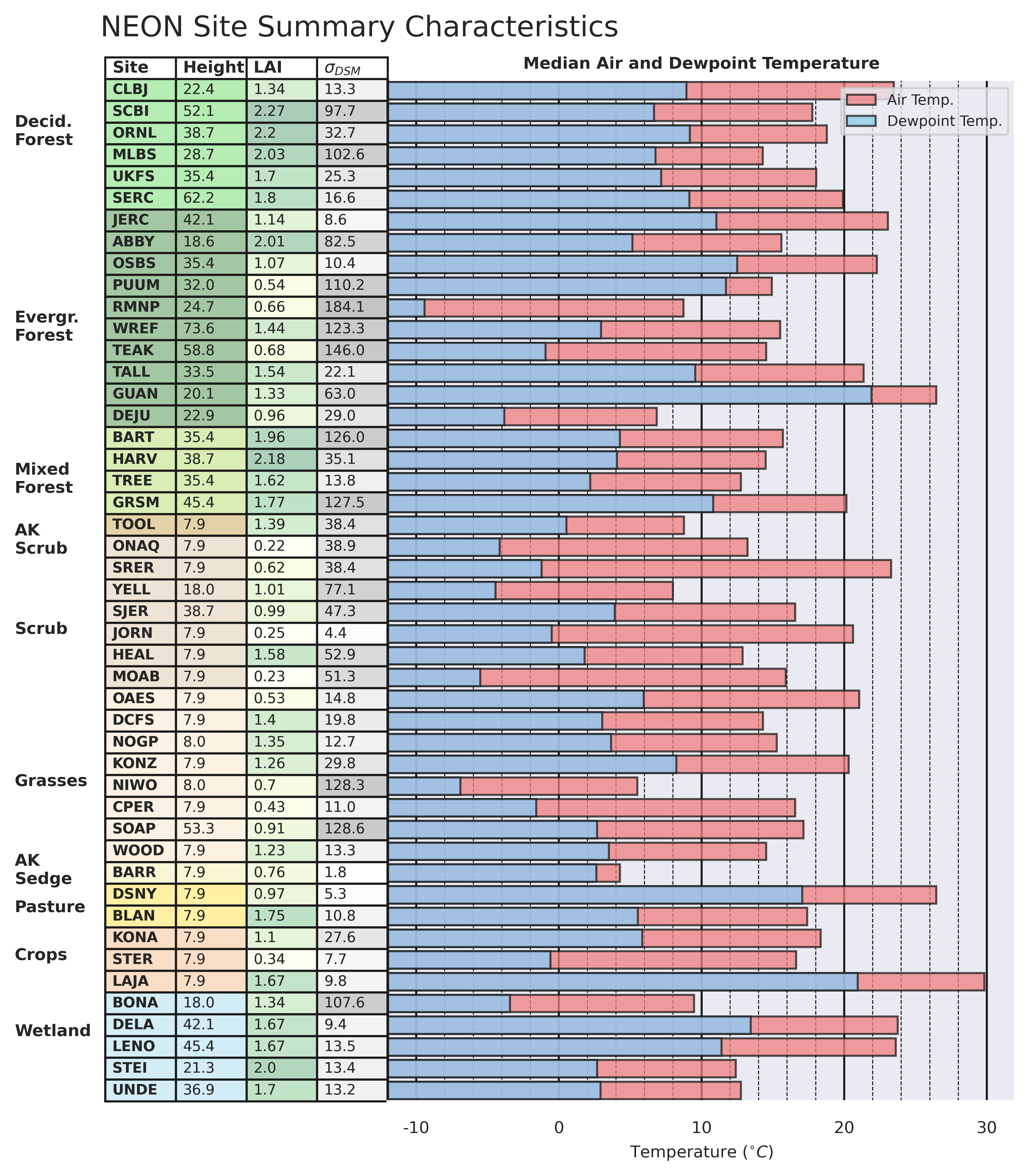}
\caption{Table and bar plot illustrating key environmental characteristics of the NEON sites. Sites are grouped, ordered and colored according to the dominant NLCD land cover type in a 2km box around the tower. From left to right, measurement height for the 3D sonic anemometer, mean growing season Leaf Area Index (LAI), and the standard deviation of the Digital Surface Model (DSM), a measure of complexity, are shown. The blue bar indicates the median daytime dewpoint temperature, and the red bar indicates median daytime air temperature. Three sites have NLCD land cover that may not be representative of the tower area: YELL, SJER and SOAP are all relatively sparse canopies, that produce a scrub/grass landcover type despite significant treecover.}
\label{fig:site_summary}
\end{figure}

The data for this study comes primarily from the National Ecological Observation Network (NEON), which is an NSF funded network with sites placed in representative ecosystems across the territory of the United States. The network includes locations that are not frequently instrumented in major networks such as five cold weather sites in Alaska (BARR, DEJU, HEAL, TOOL, and BONA) and three tropical island sites (PUUM - Hawaii, LAJA and GUAN - Puerto Rico). This diverse ecosystem representation supports cross comparison and validation of surface exchange theories in the non-conventional terrain where theories were not developed but are nonetheless applied in atmospheric models. Figure \ref{fig:site_summary} provides a summary of the NEON sites, including dominant landcover, height of the 3D sonic measurements designed to lie above the momentum roughness sublayer, growing season Leaf Area Index (LAI), the standard deviation of the Digital Surface Model (DSM), and the mean daytime dewpoint and air temperatures at each site. We use the $\sigma_{DSM}$ value as a simple representation of site topographic and canopy complexity. NEON provides thirty-minute fluxes, means and variances of moisture, carbon dioxide, temperature, and velocity and one-minute means and variances of those quantities. More information about the general site design and instrumentation can be found in \citeA{metzger_neon_2019}, and further details on the eddy covariance instrumentation in \citeA{metzger_neon_2022}.

\par This work leverages published and quality assured data from the NEON eddy-covariance product \cite{metzger_neon_2022} as well as the raw, 20Hz sonic anemometer and infrared gas analyzer (IRGA) data \cite{national_ecological_observatory_network_neon_bundled_2025}. NEON also publishes, in collaboration with the PhenoCam Network, NDVI and solar angle ($\alpha_{solar}$) from photos taken from automated cameras at 15 minute intervals and radiometry equipment near the top of the towers \cite{national_ecological_observatory_network_neon_phenology_2025}. Finally, the work also makes use of NEON airborne remote sensing data to provide information about the spatial structure, complexity and spectral properties of both the canopy and terrain at 1 meter resolution around each site. This includes NEON products for remotely sensed water indicies \cite{national_ecological_observatory_network_neon_canopy_2025,national_ecological_observatory_network_neon_canopy_2025-1}, albedo \cite{national_ecological_observatory_network_neon_albedo_2025}, and Leaf Area Index (LAI) \cite{national_ecological_observatory_network_neon_lai_2025,national_ecological_observatory_network_neon_lai_2025-1}. Outside of the NEON data product suite, 30 meter land cover is acquired from the National Land Cover Database (NLCD) using the 2021 data product for the Continental United States, 2016 NLCD data for Alaska, and 2011 data for Puerto Rico and Hawaii \cite{jin_overall_2019}.

\begin{table} \label{tab:equations}
    \caption{Table showing the traditional MOST equations used in this work for the scalars under unstable and stable conditions, as well as the anisotropy generalized modified versions of MOST in the second column. Models sourced from \citeA{sfyri_scalar-flux_2018},\citeA{stiperski_generalizing_2023}, and \citeA{ramana_surface_2004}.}
    \centering
    \begin{tabular*}{\linewidth}{@{\extracolsep{\fill}} c|cccc }
    \hline
    \multicolumn{5}{c}{\textbf{Unstable Conditions}} \\
    \hline
    & \multicolumn{2}{c}{\textbf{Traditional MOST}} & \multicolumn{2}{c}{\textbf{Modified MOST}} \\
    \hline
    $\Phi_{\theta}=$ & \multicolumn{2}{c}{\begin{tabular}{@{}c@{}} $.99(.067-\zeta)^{-1/3},(\zeta<-.05) $\\ $.015(-\zeta)^{-1}+1.76, (\zeta>-.05)$ \end{tabular}} & \multicolumn{2}{c}{\begin{tabular}{@{}c@{}} $1.07(.05+|\zeta|)^{\frac{-1}{3}}+$ \\ $[-1.1+a(y_b)|\zeta|^{\frac{-9}{10}}][1-tnh(10|\zeta|^{\frac{2}{3}})$ \end{tabular}} \\ 
    $\Phi_q=$ & \multicolumn{2}{c}{$2.74(1-8\zeta)^{-1/3}$ } & \multicolumn{2}{c}{$a(y_b)(1-c(y_b)\zeta)^{-1/3}$ } \\ 
    $\Phi_c=$ & \multicolumn{2}{c}{$4.1(1-8\zeta)^{-1/3}$} & \multicolumn{2}{c}{$a(y_b)(1-8\zeta)^{-1/3}$} \\
    \hline
    \multicolumn{5}{c}{\textbf{Stable Conditions}} \\
    \hline
    & \multicolumn{2}{c}{\textbf{Traditional MOST}} & \multicolumn{2}{c}{\textbf{Modified MOST}} \\
    \hline
    $\Phi_\theta=$ & \multicolumn{2}{c}{$0.00087(\zeta)^{1.4}+2.03$} & \multicolumn{2}{c}{$a(y_b)\zeta^{-1.4}+e(y_b)$} \\
    $\Phi_q=$ & \multicolumn{2}{c}{$2.74$} & \multicolumn{2}{c}{$a(y_b)+b(y_b)\log_{10}\zeta-0.04(\log_{10}20\zeta)^{3}$} \\ 
    $\Phi_c=$ & \multicolumn{2}{c}{$4.1$} & \multicolumn{2}{c}{$a(y_b)+b(y_b)\log_{10}\zeta-0.0326(\log_{10}\zeta)^{3}$} \\
     \hline
    \end{tabular*}
\end{table}

\subsection{Data processing} \label{sec:dataproc}
\subsubsection{Turbulence Data}
This work primarily leverages publicly available, post-processed products from NEON. The NEON turbulence processing workflow generally follows established protocols, and is replicated and repeated by the authors to compute fluxes for 5 minute averaging periods as well as the 5 minute and 30 minute Reynolds stresses needed to assess the anisotropy of turbulence. Details on the quality assurance and processing steps for this work, as well as the choice of 5 minute averaging periods for analysis in stable conditions, can be found in \citeA{waterman_evaluating_2025} and \citeA{metzger_neon_2022}, and generally follow established protocols for processing of turbulence data including planar-fit coordinate rotation and lag correction for carbon and moisture. In addition, we further eliminate averaging periods during rain events, or when relative humidity exceeds the sensor operating range ($RH>95\%$) for the scaling of carbon and moisture. With the exception of the analysis in section \ref{sec:stationarity} which looks at scaling across degrees of stationarity, all averaging periods with $\xi_s>30\%$, as defined in equation \eqref{eq:xi}, are removed. The same stationarity metric applied to the fluxes was also briefly examined, however we have chosen to exclude detailed analysis from the study. Previous studies into stationarity and flux-variance relations focus on variance stationarity \cite{zahn_relaxed_2023}. In general, we also found that variance stationarity was less common than flux stationarity, with only 2\% of the data reporting non-stationary fluxes when variance stationarity was met. Averaging periods are assessed independently at each site, and we do not require all towers have valid observations at the same time. After all quality checks, for stable conditions (5 minute averaging period) the mean number of averaging periods available at each site are 79,225 averaging periods for temperature, 45,615 for water vapor, and 81,688 for carbon dioxide. For unstable conditions (30 minute averaging) mean site availability is 14,121 averaging periods for temperature, 9,605 for water vapor, and 11,744 for carbon dioxide. All sites, under all conditions, have at least 3,000 valid averaging periods. 
\subsubsection{Airborne Remote Sensing and Lengthscale of Heterogeneity} \label{sec:airborne}
To assess spatial heterogeneity at the different sites in the network, we leverage products for Leaf Area Index (LAI), Normalized Difference Water Index (NDWI), and albedo from the NEON Imaging Spectrometer. NDWI is a remotely sensed water index, specifically designed to capture liquid water present in canopies, and therefore could relate closely with spatial patterns of water sources \cite{gao_ndwinormalized_1996}. Flights are conducted approximately once a year at most sites, at 90\% of peak greenness or greater, with datasets produced by NEON providing 1 meter resolution products in an approximately 10 by 10 km area around each tower. For analysis focused on spatial heterogeneity in section \ref{sec:spatial_het}, only turbulence data from the same month as the flights are used, and the remotely sensed data is cropped to an area with radius approximately twice the fetch of the tower as determined from the NEON statistics of the flux footprint.
\par To directly assess how spatial patterns interact with surface layer scaling, we compute lengthscales of spatial heterogeneity $\ell_{het}$ for each metric, and each flight, at each site. A common metric to understand spatial heterogeneity of 2D fields, especially of vegetation characteristics, is lacunarity \cite{salmaso_canopy_2025}. Lacunarity is a scale dependent measure that quantifies the deviation of a pattern from translational invariance \cite{plotnick_lacunarity_1996}. There are a number of methods to compute lacunarity; we leverage the gliding box method where lacunarity $\Lambda$ of a 2D gridded variable $x$ for some scale $r$ is defined as 
\begin{equation}
    \Lambda_x(r)=\sigma^2_x(r)/\mu_x(r)
    \label{eq:lacunarity}
\end{equation}
where, $\sigma^2_x(r)$ is the variance of the quantity $x$ that has been calculated at a scale $r$ and $\mu_x(r)$ is the mean of the quantity $x$ that has been calculated at a scale $r$. Conceptually, when $\Lambda_x(r)$ is computed across many scales $r$ ranging from the native grid resolution to the grid size, the resulting $r$-$\Lambda_x$ function can provide an indication of the variance or variability in $x$ contained within each scale $r$. This curve is then integrated over all scales $r$ to provide an integral lengthscale of heterogeneity, $\ell_{het}$. This integral lengthscale is evaluated across three remotely sensed variables chosen to be closely correlated to the sources for the three scalars: Albedo for potential temperature, NDWI for water vapor, and LAI for carbon dioxide. The integral heterogeneity length scale is similar to an integral lengthscale from an autocorrelation function as we still integrate the energy contained across scales. The values are expected to be similar, but not identical, to an integral lengthscale based on the 2D autocorrelation function, however the lacunarity based integral lengthscale is computationally simpler to compute and more resilient to the presence of no data values which are occasionally found in the remotely sensed data.

\subsection{Curve Fitting}
To develop new empirical scaling curves for the variances of temperature, moisture and carbon (function $f$ in equation \ref{eq:phi_eqyb}) that consider and leverage the anisotropy of turbulence, we fit equations from the literature to the NEON data. For determining a fit curve, additional quality steps are taken on the data in addition to those described in section \ref{sec:dataproc}. Averaging periods with very small magnitudes for the scalar fluxes ($F_x=\overline{w'\theta'}$, $\overline{w'q'}$, $\overline{w'c'}$) were excluded from the fitting exercise due to the larger scatter in this regime and significant undue influence on the curve fitting process. Cutoff values of $5\ W m^{-2}$, $2\ W m^{-2}$, and $0.2\ mgC m^{-2}$ were selected for sensible heat flux $H$, latent heat flux $LE$ and carbon dioxide flux $F_{C}$ respectively. With this quality step conducted, data from all sites are compiled into one aggregated dataset. We use a combination of curves from the literature on traditional MOST scaling as a baseline \cite{ramana_surface_2004,sfyri_scalar-flux_2018}, as well as newly fit curves for equation \eqref{eq:phi_eq}, modifying the universal constants to dynamic functions of $y_b$. The equations are then fit simultaneously to $\zeta$ and $y_b$ using an arctan loss function which is robust to outliers. The equations used for each fit are shown in table \ref{tab:equations}, and the best fit parameter values can be found in table S1. 

\section{Results} \label{sec:results}
\subsection{Traditional Surface Layer Scaling} \label{sec:trad_scaling}
\begin{figure}
\centering
\includegraphics[width=5.5in]{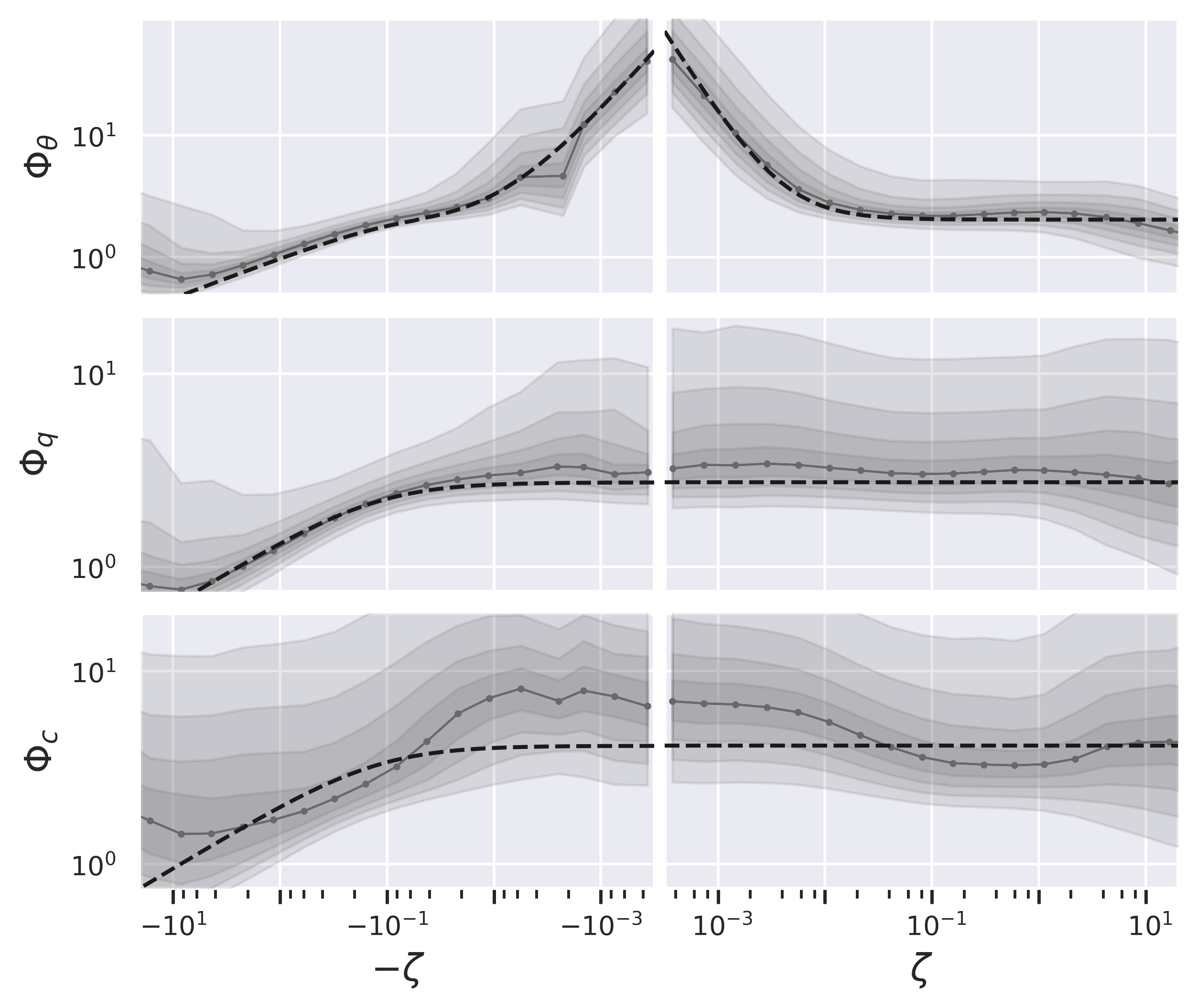}
\caption{Observed scaling for the square root of the variance of, from top to bottom, non-dimensionalized potential temperature $\Phi_{\theta}$, water vapor $\Phi_q$ and carbon dioxide $\Phi_c$. The full lines show the median values of $\Phi$ across logarithmically spaced bins of $\zeta$. The shaded areas show binned deciles; the outer area shades between the 10th and 90th percentile, then 20th to 80th, and so on.}
\label{fig:basic_scaling}
\end{figure}

Initial examination of the surface layer scalings for temperature, moisture and carbon in figure \ref{fig:basic_scaling} shows that the NEON data largely follows expectations from MOST, with some significant scatter and notable differences from the standard curves. This is particularly the case for carbon dioxide scaling, a observation commonly seen in existing literature \cite{zahn_relaxed_2023,chor_flux-variance_2017,detto_scaling_2010}. The carbon dioxide scaling under stable conditions, and to a lesser extent the water vapor scalings, show some $\zeta$ dependence; most traditional studies suggest that the scaling in this regieme is z-less, although there are some that observe a degree of $\zeta$ dependence \cite{ramana_surface_2004}. Under unstable conditions, for all three scalars, the scatter is lowest in a moderately convective regime ($-10^0<\zeta<-10^{-2}$) and higher for more convective and more neutral scaling. Across nearly all stability regimes, for all three scalars, $\Phi$ appears elevated above typical values expended from MOST, which suggests some significant deviations from traditional theory and assumptions may be impacting the results.

\begin{figure}
\centering
\includegraphics[width=4.5in]{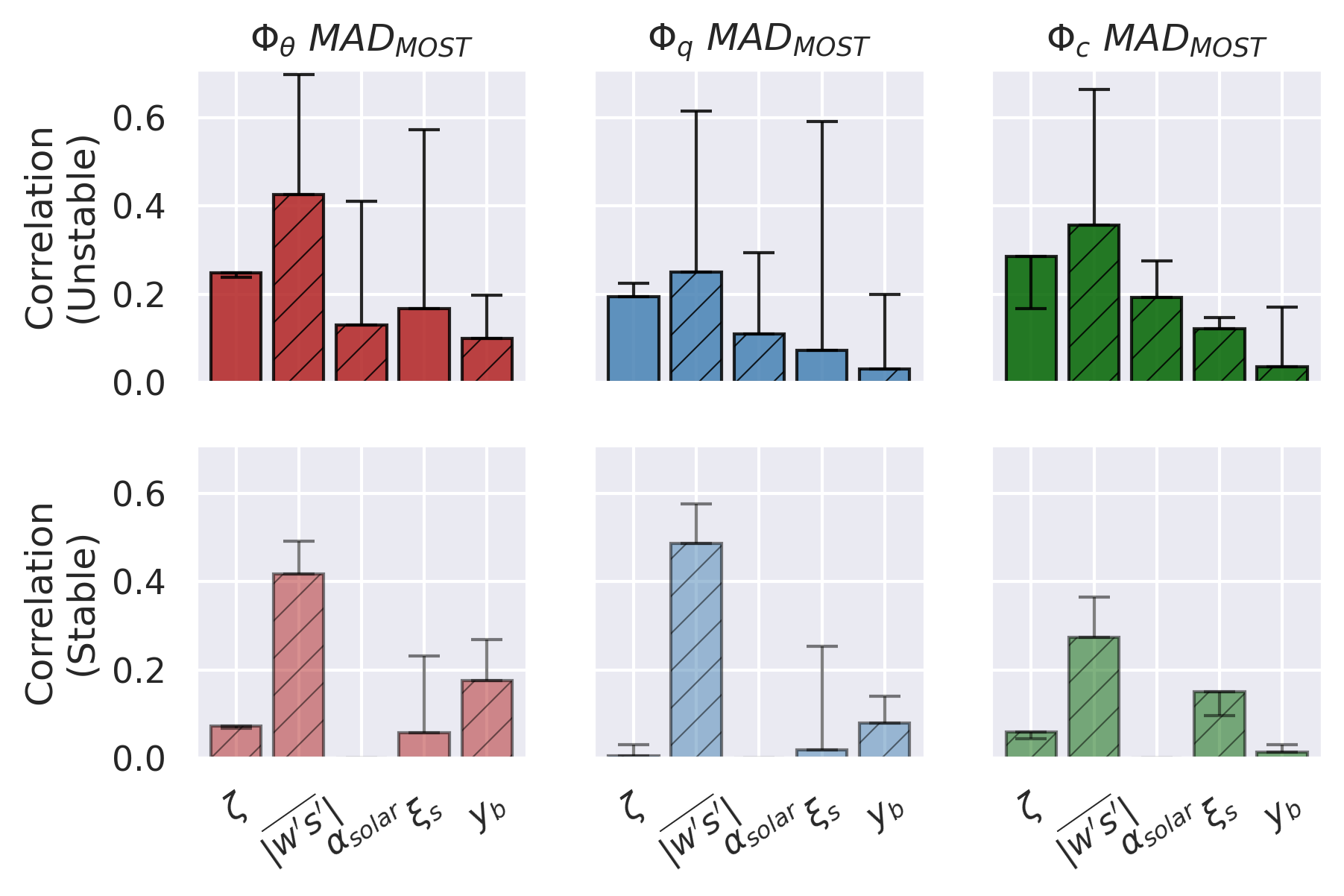}
\caption{Correlations between five variables ($\zeta$, the magnitude of the surface flux $\left|\overline{w's'}\right|$ ,$\alpha_{solar}$, stationarity metric $\xi_s$ and $y_b$) and the Median Absolute Deviation (MAD) of the data to traditional MOST for the three scalars $s$. $\Phi_{\theta}$ (left), $\Phi_q$ (middle) and $\Phi_c$ (right). The first row shows the magnitude of the spearman correlation coefficient for unstable conditions, and the second row for stable conditions. Slanted lines inside the color are used to indicate inverse (negative) correlations.}
\label{fig:correlations}
\end{figure}

As a first analysis into potential drivers of error and bias of the scaling curves to the data, we examine the correlation between various variables and the deviation from the traditional MOST relations in figure \ref{fig:correlations} using the median absolute deviation ($MAD$) as used previously for surface layer scaling in \citeA{stiperski_scaling_2019}. $MAD$ is calculated as the median deviation from the scaling curve for each data point, or
\begin{equation}
	MAD = \text{median}(|x_{scaling}-x|).
\end{equation}
due to its robustness to outliers. For unstable stratification, it is unsurprising to find that there is a correlation between $\zeta$ and $MAD$, as the scatter grows significantly for all three scalars as we approach near neutral conditions in figure \ref{fig:basic_scaling}. We also find a consistent inverse correlation between the magnitude of the flux and deviation from theory; this is well established in the literature, with a number of studies reporting the failure of traditional theory for scalars when flux is low due to increasing likelihood of a breakdown between the assumed equilibrium of variance production and dissipation as other terms in the variance budget become relevant \cite{weaver_temperature_1990,cancelli_dimensionless_2012,zahn_direct_2022,de_bruin_verification_1993}. The inverse correlation between flux magnitude and $MAD$ persists, and is actually stronger, under stable conditions. 
\par In figure \ref{fig:correlations} the correlation between $MAD$ (scaling error) and $\alpha_{solar}$ (related to spatial source heterogeneity), $\xi_s$ (the stationarity metric), and $y_b$ (anisotropy) are also examined. For $\alpha_{solar}$, there seems to be a small inverse correlation in the data for all three scalars, which may be indicative of higher error when heterogeneity is likely higher at those low solar altitudes. The temporal heterogeneity, or stationarity metric $\xi_s$, has only a weak correlation with errors using the default pruned dataset for this study. The relationship is also inverse for carbon scaling under stable and unstable conditions, defying expectations by suggesting that non-stationary data follows traditional scaling better than stationary data. This apparent inverse relationship for carbon is somewhat deceptive and is explored in more detail in sections \ref{sec:stationarity} and \ref{sec:sites}. Regardless, when we no longer confine the data to stationary periods (black bars in figure \ref{fig:correlations}), the magnitude of the stationarity metric becomes well correlated with error for carbon and water vapor under stable and unstable conditions. The correlations of the flux, solar altitude and anisotropy with $MAD$ also all become more significant. The anisotropy of turbulence, $y_b$, does not initially seem particularly correlated with error in any case. This may be deceptive, however, as previous work does suggests the relationship between $\Phi$ and $y_b$ is not always monotonically increasing \cite{stiperski_generalizing_2023,waterman_evaluating_2025}, which could result in a poor correlation coefficient despite a significant relationship between $y_b$ and the scalar scaling. All of these correlations indicate a degree of potential for relationships between various forms of heterogeneity and the surface layer scaling of heat, moisture, and carbon.

\subsection{Spatial Heterogeneity} \label{sec:spatial_het}
While a number of previous studies have investigated the impact of spatial heterogeneity on surface layer scaling \cite{detto_surface_2008,finnigan_boundary-layer_2020,bou-zeid_persistent_2020,moraes_nocturnal_2004,de_franceschi_analysis_2009,nadeau_similarity_2013,grachev_airsealand_2018, martins_turbulence_2009, zahn_scalar_2016}, they face a number of limitations, including few sites and few direct quantitative comparisons between spatial patterns of the land surface and the characteristics of the ASL. In this study, we leverage $1m$ resolution airborne remote sensing from NEON to determine lengthscales of heterogeneity, $\ell_{het}$ relevant for the scalar variances: $\ell_{het}$ of Albedo for potential temperature, $\ell_{het}$ of NDWI for water vapor, and $\ell_{het}$ of LAI for carbon dioxide. Details on the computation of the lengthscale of heterogeneity for each variable are presented in section \ref{sec:airborne}. The lengthscale provides an approximation of a patch size of Albedo, NDWI or LAI that may be considered effectively homogeneous relative to the spatial patterns in the field. 
\par To properly examine the impact that these lengthscales may have on surface layer scaling, we follow the relationships presented in \citeA{detto_surface_2008} and \citeA{belcher_turbulent_1998}. \citeA{belcher_turbulent_1998} suggested that for a boundary layer disturbed from a slow moving quasi-steady state, considered stationary and with an absence of subsidence as in MOST, there are two critical time scales to evaluate the impact of spatial patterns. First, an advective time scale $T_{adv}=\ell_{het}/\overline{u}$ and the de-correlation, or relaxation, time scale of the large, energy containing eddies $\tau=kz/u_*$. Both of these can also be considered directly as lengthscales: $\ell_{het}$, as discussed previously, and $L_t=\overline{u}\ \tau=kz(\overline{u}/u_*)$. The relationship between these two timescales can indicate when spatial heterogeneity should be important; if $T_{adv}\gg\tau$ (or $\ell_{het}\gg L_t$), then the turbulence will come into local equilibrium with the surrounding mean flow and scalar gradients before being advected, meaning that, for MOST purposes, the flow is effectively horizontally homogeneous fulfilling that key MOST assumption. If, however, $T_{adv}\ll\tau$ (or $\ell_{het}\ll L_t$) then the flow is advected without coming into local equilibrium and the homogeneity assumption of MOST is violated \cite{detto_surface_2008,belcher_turbulent_1998}. 
\par
\begin{figure}
\centering
\includegraphics[width=4in]{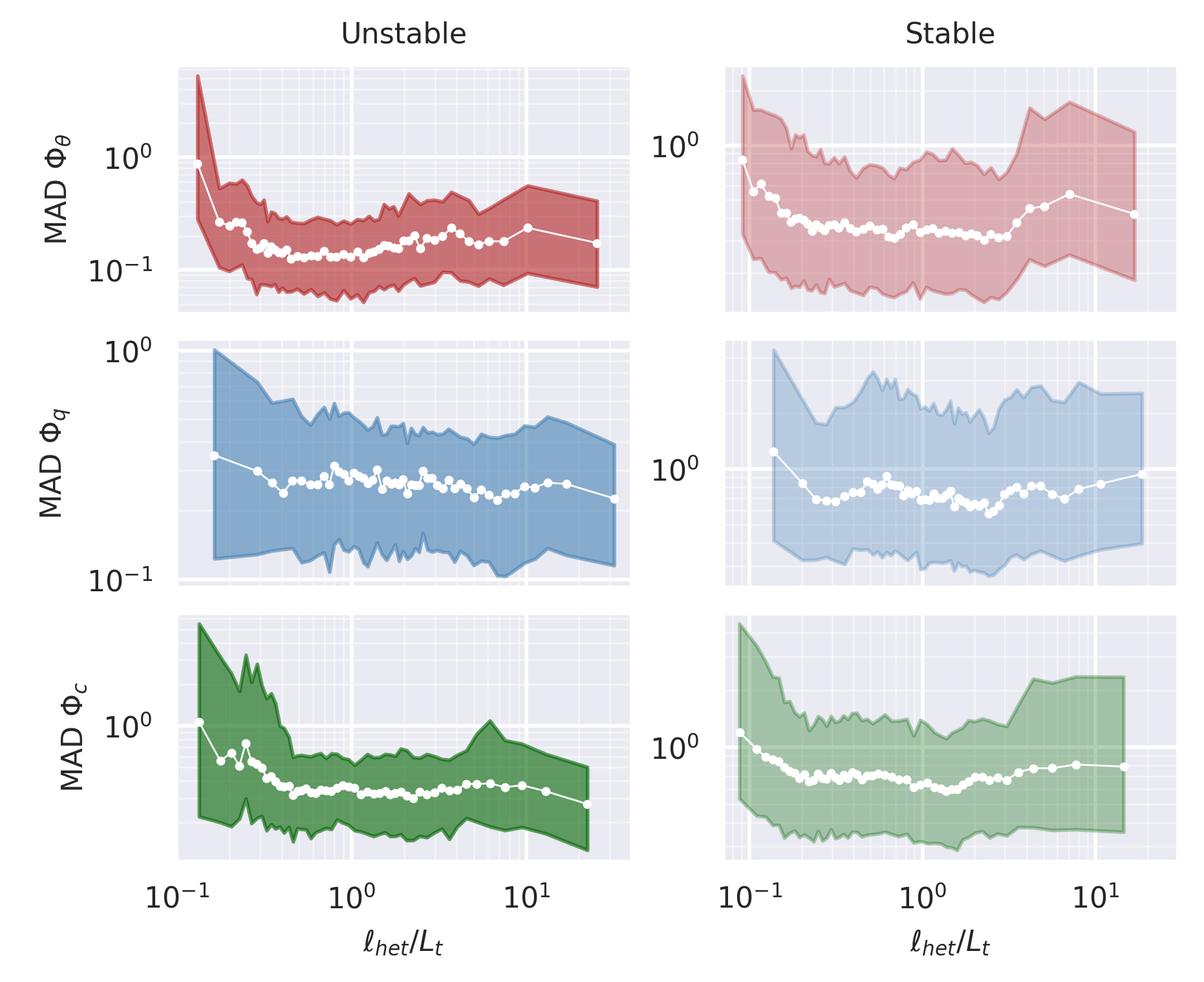}
\caption{Ratio of the lengthscale of heterogeneity to the turbulent lengthscale ($\ell_{het}/L_t$) compared to the MAD (log scaled) for $\Phi_{\theta}$ (top), $\Phi_q$ (middle) and $\Phi_c$ (bottom) under unstable (left) and stable (right) stratification. $\ell_{het}$ is determined from spatial patterns of albedo for $\theta$, NDWI for $q$ and LAI for $c$. The ratio is plotted in log scale, with 100 binned medians plotted in white, and the filled area representing the interquartile range.}
\label{fig:spatial_het}
\end{figure}

 In figure \ref{fig:spatial_het}, we examine the ratio between two lengthscales $\ell_{het}/L_t$ and the deviation from traditional MOST scaling. The lengthscales of heterogeneity are computed as integral lengthscales as described in section \ref{sec:airborne}, with the lengthscale for heat computed from albedo, the lengthscale for moisture computed from NDWI, and the lengthscale for carbon computed from LAI. There is going to be some relationship between these three variables; LAI, for example, should be closely related to spatial patterns of water vapor and temperature as well. The relationship between the lengthscale ratio and model error is still present, but weaker, when the lengthscale from LAI is used for all three scalars (not shown). Only data for the month long period surrounding the flight times at each site are used in this analysis. Sites/time periods with very low spatial variability (spatial variance in the 10th percentile or less) are also excluded, as the magnitude of heterogeneity is unlikely to be sufficient to affect MOST for these sites even if the ratio between $\ell_{het}$ and $L_t$ is very low. Under unstable conditions we see that error begins to rise significantly for all three scalars when $\ell_{het}/L_t<1$, although the exact ratio varies. The bias of the model (not shown) becomes more negative (i.e. $\Phi$ is elevated significantly above MOST predictions). The scatter in the data also tends to grow significantly. Under stable conditions, the relationship between the ratio of the lengthscales and error in the scaling is less clear; there is an obvious, albeit weak, relationship between the ratio and error for heat, a slightly weaker one for carbon, and an unclear relationship for moisture. This is not necessarily surprising; the sources and sinks of the scalars, particularly carbon dioxide and water vapor, would be much larger and more correlated with moisture and vegetation characteristics during the day when photosynthesis is active than at night. This may indicate that larger magnitudes of heterogeneity (higher spatial variance) are needed to see a clear impact. Overall, figure \ref{fig:spatial_het} show a strong and direct connection between spatial heterogeneity and the scaling of heat, carbon and moisture across a large ensemble of sites.

\subsection{Anisotropy: Heterogeneity in the Reynolds Stress} \label{sec:anisotropy}
\begin{figure}
\centering
\includegraphics[width=5.5in]{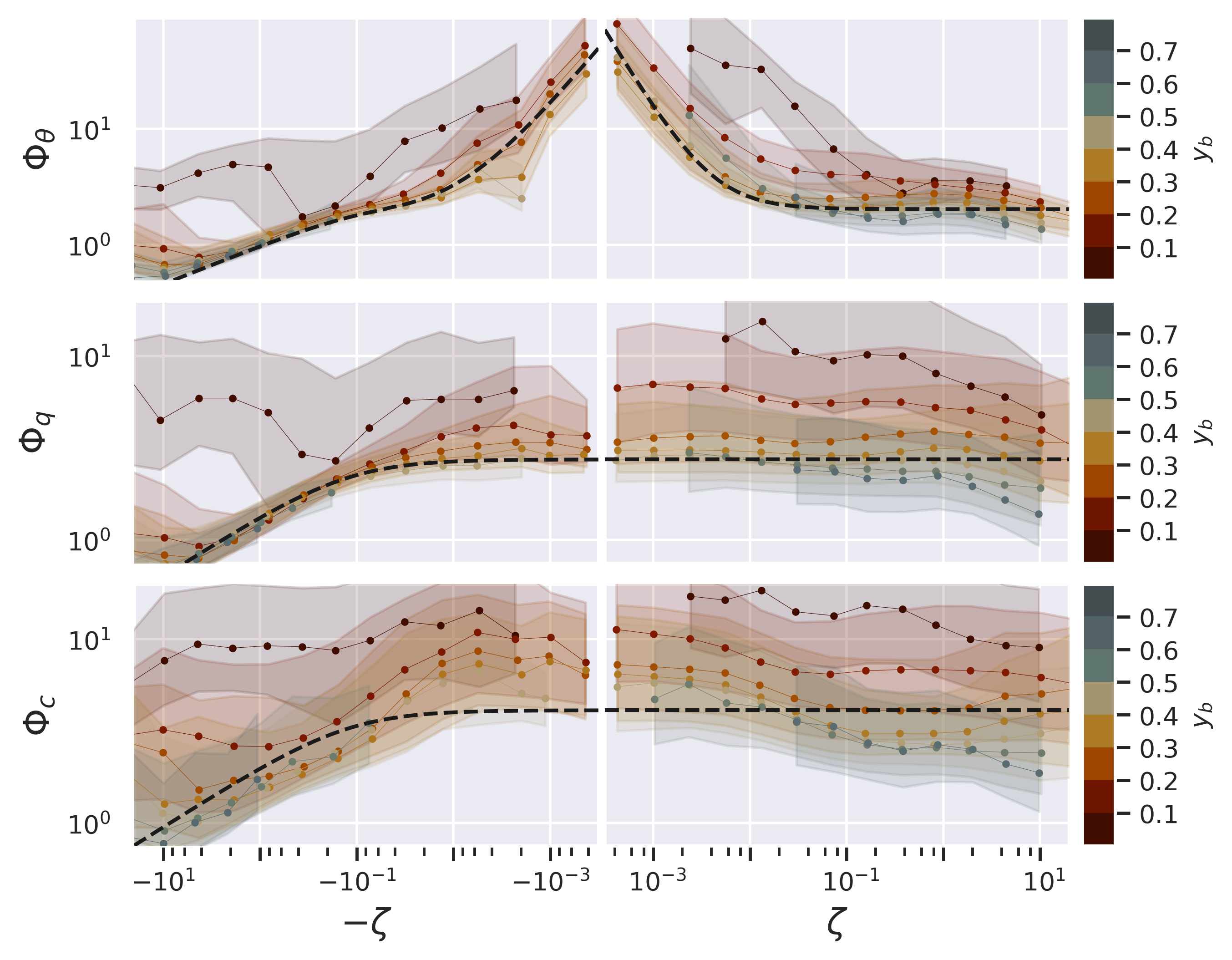}
\caption{Observed scaling for the square root of the variance of, from top to bottom, non-dimensionalized potential temperature $\Phi_{\theta}$, water vapor $\Phi_q$ and carbon dioxide $\Phi_w$. The full lines, colored according to $y_b$ (with red being anisotropic and blue being isotropic), show the median values of $\Phi$ across linearly spaced bins of anisotropy and logarithmically spaced bins of $\zeta$. The area between the 1st and 3rd quartile is shaded. Black dashed lines show a traditional MOST scaling for these non-dimensionalized quantities.}
\label{fig:aniso_scaling}
\end{figure}

The correlations in figure \ref{fig:correlations} suggest that the degree of turbulence anisotropy, $y_b$, may not be well correlated to error. Because a correlation cannot account for non-monotonically increasing relations, it remains worth examining this dependence more closely. Figure \ref{fig:aniso_scaling} shows the surface layer scaling for scalars when binned according to $y_b$. Many of the scaling relations in figure \ref{fig:aniso_scaling} show a significant degree of separation by anisotropy. Under unstable stratification, $\Phi$ follows traditional MOST scaling fairly closely for isotropic turbulence, but under very anisotropic turbulence ($y_b<0.1$) $\Phi$ is particularly elevated. For very anisotropic turbulence one can see a consistently different scaling behavior of $\Phi$ with $\zeta$  when compared to more typical turbulent anisotropy levels for all variables. Previous work on the impact of turbulence anisotropy on the flux-variance relations for velocity and potential temperature excluded very anisotropy turbulence from much of the analysis precisely due to the strongly divergent behavior \cite{stiperski_generalizing_2023}. In this study, we retain very anisotropic turbulence to highlight this divergence.  The scaling with $\zeta$ depends on anisotropy under more convective conditions, as well as near neutral conditions. In both cases, $\Phi$ increases with anisotropy, however this behavior is not immediately consistent for previous findings on the scaling of $\Phi_\theta$ which show an inverse relationship between $y_b$ and $\Phi_s$ under near neutral conditions \cite{stiperski_generalizing_2023}. For moderately convective conditions (with $\zeta$ between approximately $-10^{-1}$ and $-10^{0}$) when both shear and buoyancy are important scaling hews closely to traditional MOST outside of very anisotropic turbulence. Carbon shows significant scatter in the scaling for both unstable and stable conditions, especially under near-neutral stratification, which may be driven by site differences, a broad based scatter in $\Phi$, or both. While site differences have not yet been investigated for carbon (see section \ref{sec:sites}), previous work has shown that site differences in traditional MOST scaling are significant for the temperature variance \cite{waterman_examining_2022} and a significant driver of scatter for $\Phi_\theta$ when examined in an ensemble.
\par There is a clear difference in behavior between the scalings of the standard deviations of $\theta$, $q$ and $c$ under unstable conditions, and this gap only appears to widen when examining stable stratification. Under stable stratification, we observe more frequent scatter across all variables. For the scaling of $\Phi_\theta$, the anisotropy dependence is clear and flips as we move across $\zeta$. For stable near-neutral stratification, $\Phi_\theta$ is higher under isotropic turbulence and lower under more anisotropic turbulence, consistent with previously published findings \cite{stiperski_generalizing_2023}. It also follows traditional MOST scaling fairly well, with mid-level values of $y_b$ lining up well with traditional relations. For the bioactive scalars ($c$ and $q$), traditional MOST scaling is z-less \cite{chor_flux-variance_2017,sfyri_scalar-flux_2018}. We find this not to be the case for our dataset, with a weak $\zeta$ scaling for moisture and a relatively strong $\zeta$ scaling for carbon. As in unstable conditions, more isotropic turbulence follows traditional MOST more closely, with less $\zeta$ dependence than for very anisotropic turbulence where the relations are more obviously not z-less. Similarly to unstable stratification, under stable stratification very anisotropic turbulence diverges significantly from any scaling relation.

\begin{figure}
\centering
\includegraphics[width=5.5in]{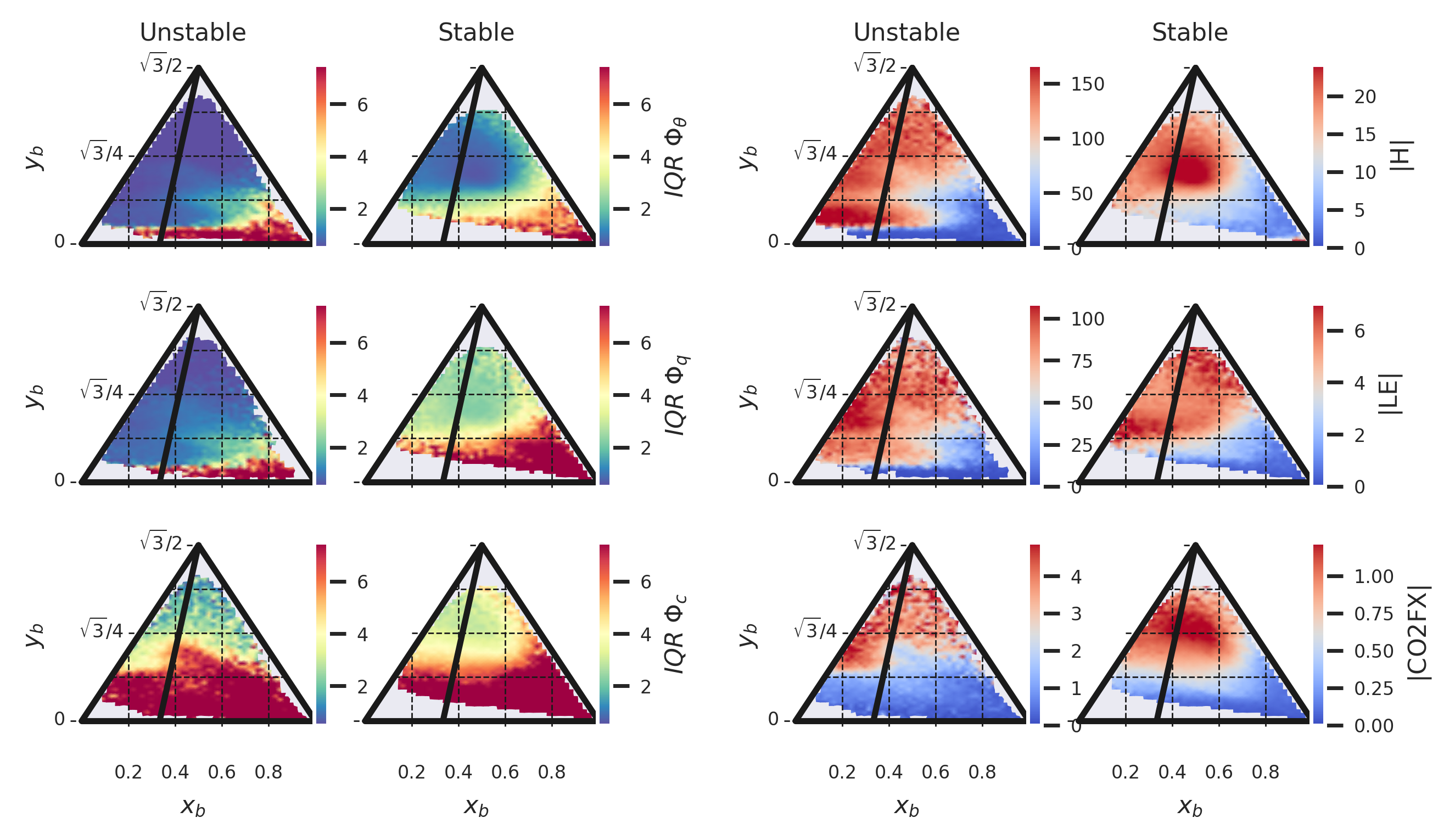}
\caption{Size of the Inter-quartile range (IQR,left) and the magnitude of the surface fluxes (right) mapped onto the barycentric map of the Lumley triangle. IQR shown for unstable (left) and stable (right) conditions for $\Phi_{\theta}$ (top), $\Phi_{q}$ (middle) and $\Phi_{c}$ (bottom) respectively. Median magnitude of the fluxes shown are sensible heat flux ($H\ (W\ m^{-2})$, top) latent heat flux ($LE\ (W\ m^{-2})$, middle) and carbon dioxide flux ($CO2FX\ (ppm\ CO_2\ m^{2})$, bottom). The plane-strain line is plotted in black for reference.}
\label{fig:lumley}
\end{figure}

For all variables, aside from $\Phi_\theta$, significant scatter is observed. To further understand the behavior of this scatter, we can look at the size of the inter-quartile range (IQR) across the anisotropy invariants $y_b$ and $x_b$. Figure \ref{fig:lumley} shows that the scatter in MOST is highest for very anisotropic, one-component turbulence (low $y_b$ and high $x_b$) under both stable and unstable conditions. Especially under unstable conditions, the scatter appears to vary primarily with $y_b$ as was found previously for the velocity variances \cite{waterman_evaluating_2025}. The Lumley triangle pattern for IQR is remarkably similar to the observed pattern for the magnitude of the scalar fluxes, with low fluxes and high IQR associated with one-component, anisotropic turbulence. This suggests that low fluxes could be the primary driver of errors in this region. Figure \ref{fig:correlations} also indicates that low fluxes could be associated with high error, and the mathematical expression for $\Phi$, equation \eqref{eq:phi_def}, indicates that for small fluxes, small differences in flux magnitude or $\sigma_s$ could affect $\Phi$ dramatically. Previous studies associate small fluxes, and large deviations from MOST, with more significant turbulent transport terms (Final term in equation \eqref{eq:var_budget}) \cite{cancelli_dimensionless_2012}. The close relationships here between anisotropy, flux magnitude, and deviation from surface layer scaling provide further evidence indicating there may be a connection between anisotropy and turbulent scalar variance transport, although additional, direct analysis of the transport terms falls outside the scope of this manuscript. 

\begin{figure}
\centering
\includegraphics[width=5.5in]{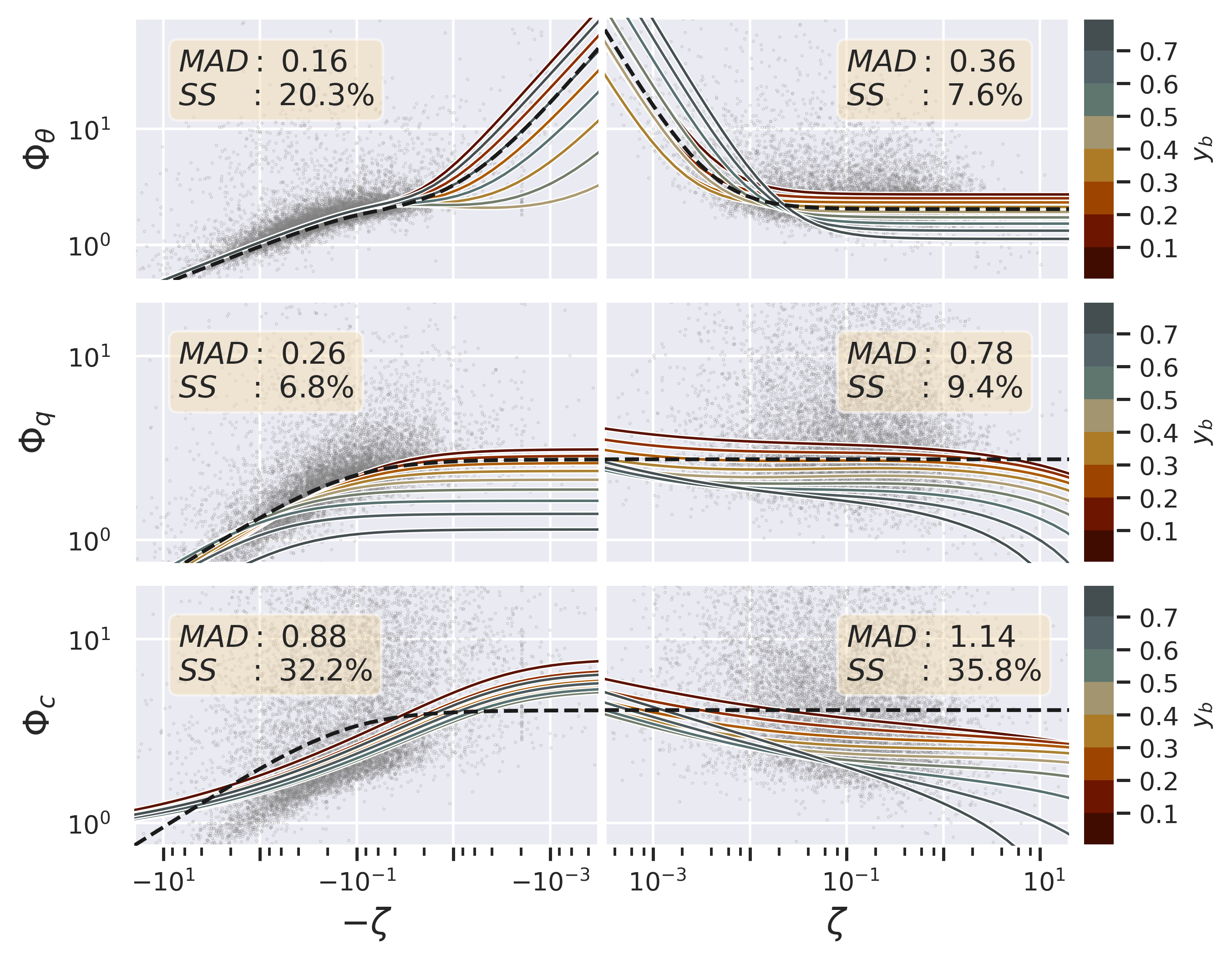}
\caption{Observed scaling for $\Phi_\theta$ (top),  $\Phi_q$ (middle) and $\Phi_c$ (bottom). Scatter shows 100,000 randomly selected points across all sites. Lines show the scaling relations by anisotropy from table \ref{tab:equations} for 10, evenly spaced $y_b$ levels between 0.1 and 0.7. Black dashed line shows a traditional MOST scaling for these non-dimensionalized quantities. A box in the corner of each plot shows the MAD of these new relations, as well as the Skill Score as a percentage improvement when evaluated against traditional MOST relations.}
\label{fig:tsw_relations}
\end{figure}

As in \citeA{stiperski_generalizing_2023} and \citeA{waterman_evaluating_2025}, novel scaling relationships can be developed as a function of anisotropy (equation \eqref{eq:phi_eqyb}). These novel scaling relationships are outlined in table \ref{tab:equations} and shown in figure \ref{fig:tsw_relations}. Fit curves are generally similar to what we see in figure \ref{fig:aniso_scaling}. There are some differences, particularly in the near neutral regime, however the amount of fit data in near neutral conditions is somewhat limited, especially for more isotropic turbulence. Model error in general remains high for carbon under stable and unstable conditions, and for moisture under stable conditions, reflecting the reality that there is significant variability in the stable scaling for both. This variability may be, at least partially, caused by scaling differences across the sites which are discussed in greater detail in section \ref{sec:sites}. To evaluate the degree of improvement the new relations provide, we show the Skill Score (SS) in figure \ref{fig:tsw_relations}, defined as 
\begin{equation}
	SS = 1-\frac{MAD_{new}}{MAD_{MOST}},
\end{equation}
In all cases, the skill score remains significant indicating at least a moderate degree of improvement by leveraging turbulent anisotropy in the scalings. Skill scores for the velocity variance scalings, which have a more direct relationship with turbulence anisotropy, were approximately twice as large \cite{waterman_evaluating_2025} on average. Skill scores are also uneven across different ecosystems, as discussed in detail in section \ref{sec:sites}. Fit parameter values for these curves are shown in table S1.

\subsection{Stationarity: Temporal Heterogeneity} \label{sec:stationarity}

\begin{figure}
\centering
\includegraphics[width=5.5in]{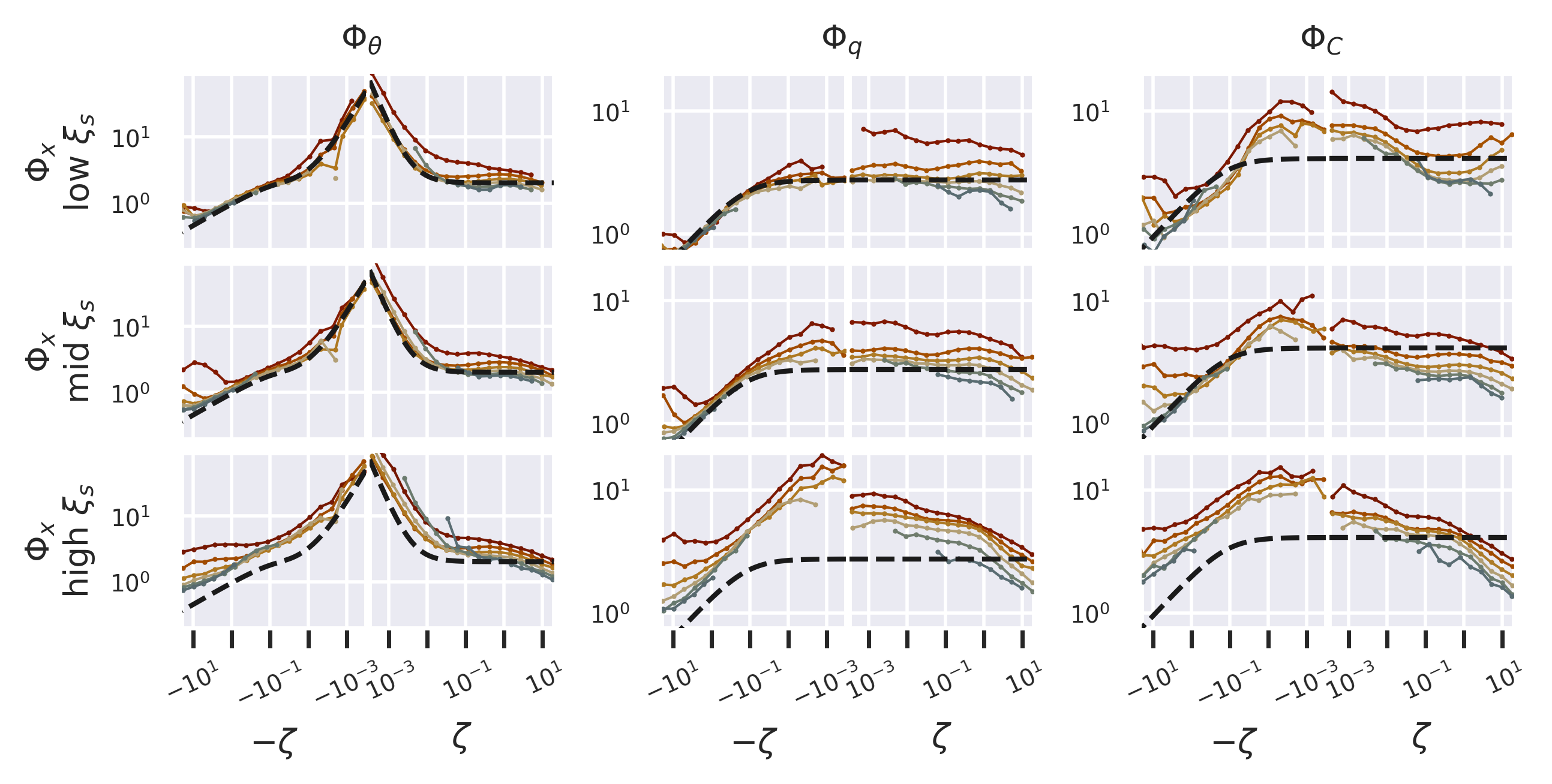}
\caption{Observed scaling for the square root of the variance of, from left to right, non-dimensionalized potential temperature $\Phi_{\theta}$, water vapor $\Phi_q$ and carbon dioxide $\Phi_w$. The full lines, colored according to $y_b$ (with red being anisotropic and blue being isotropic), show the median values of $\Phi$ across linearly spaced bins of anisotropy and logarithmically spaced bins of $\zeta$. The first row shows these scalings for stationary data (low $\xi_s$ $<$ $20\%$) , second row for mid level values of $\xi_s$ $(20$ $-$ $50\%)$, and the bottom row for more non stationary data with high levels of $\xi_s$ $>$ $50\%)$. Black dashed lines show a traditional MOST scaling for these non-dimensionalized quantities.}
\label{fig:stationarity}
\end{figure}

To explore the impact of stationarity on the scalings more fully, in this section analysis is conducted using the full NEON dataset that does not exclude data based on stationarity metrics as discussed in section \ref{sec:dataproc}. Figure \ref{fig:stationarity} shows the scalings, binned into separate curves by $y_b$, based on the stationarity metric $\xi_s$ with low values defined as $\xi_s<20$, middle values defined as $20<\xi_s<50$, and large values defined as $50<\xi_s$. For less stationary turbulence (high $\xi_s$), we see elevated $\Phi$ values in nearly all cases, as well as a stronger, and in the case of heat more log-linear, $\zeta$ dependence for the scaling. For stable stratification the impact of non-stationarity appears minimal for large $\zeta$, and more impactful for small $\zeta$ (near-neutral conditions); this could be due to relatively small fluxes expected in neutral conditions, especially if scalar similarity holds (if heat, moisture and carbon are all transported similarly). For those relatively small fluxes, the variance produced by turbulence theory would be small and the measured variance would be more susceptible to increased variance in the time series from changes over time or high intermittency. In a variance budget framework, the smaller production term in near neutral conditions would become closer to the order of magnitude of the storage term under non-stationary conditions for temperature, and often for water vapor and carbon dioxide since small fluxes of the bioactive scalars would be common, although not guaranteed, in neutral conditions. Another interesting observation from figure \ref{fig:stationarity} is that averaging periods with non-stationary statistics also seem to vary more strongly with anisotropy $y_b$. This may, again, be related to weaker fluxes for very anisotropic turbulence shown in figure \ref{fig:lumley}. The same argument from the previous point would hold; weaker fluxes and smaller production terms enhance the impact of increased variance from non-stationarity and larger storage terms on $\Phi$. 
\par An exception to nearly all the aforementioned behavior is the apparent scaling for carbon. Under both stable and unstable conditions, $\Phi$ and $\zeta$ and $y_b$ dependence are all enhanced for both high and low values of $\xi_s$ relative to moderate level non-stationarity. This apparent dissimilarity between the scaling for carbon and the scaling for moisture appears to be related to very different behavior seen at very arid (low carbon cycle activity) sites. These sites tend to be very stationary due to low carbon cycle activity, and, as a result, are overrepresented in the very stationary group. They also tend to have very low fluxes which is accordingly associated with higher errors and elevated $\Phi$ values. If these sites are excluded, behavior is similar to what would be expected based on the stationarity patterns observed for $\Phi_q$. In the following section, scaling is broken down in detail based on site characteristics with further evidence for the aforementioned phenomena.

\subsection{Scaling Differences Across Ecosystems } \label{sec:sites}

\subsubsection{Scaling Under Unstable Stratification} \label{sec:unstable_scaling}
\begin{figure}
\centering
\includegraphics[width=5.5in]{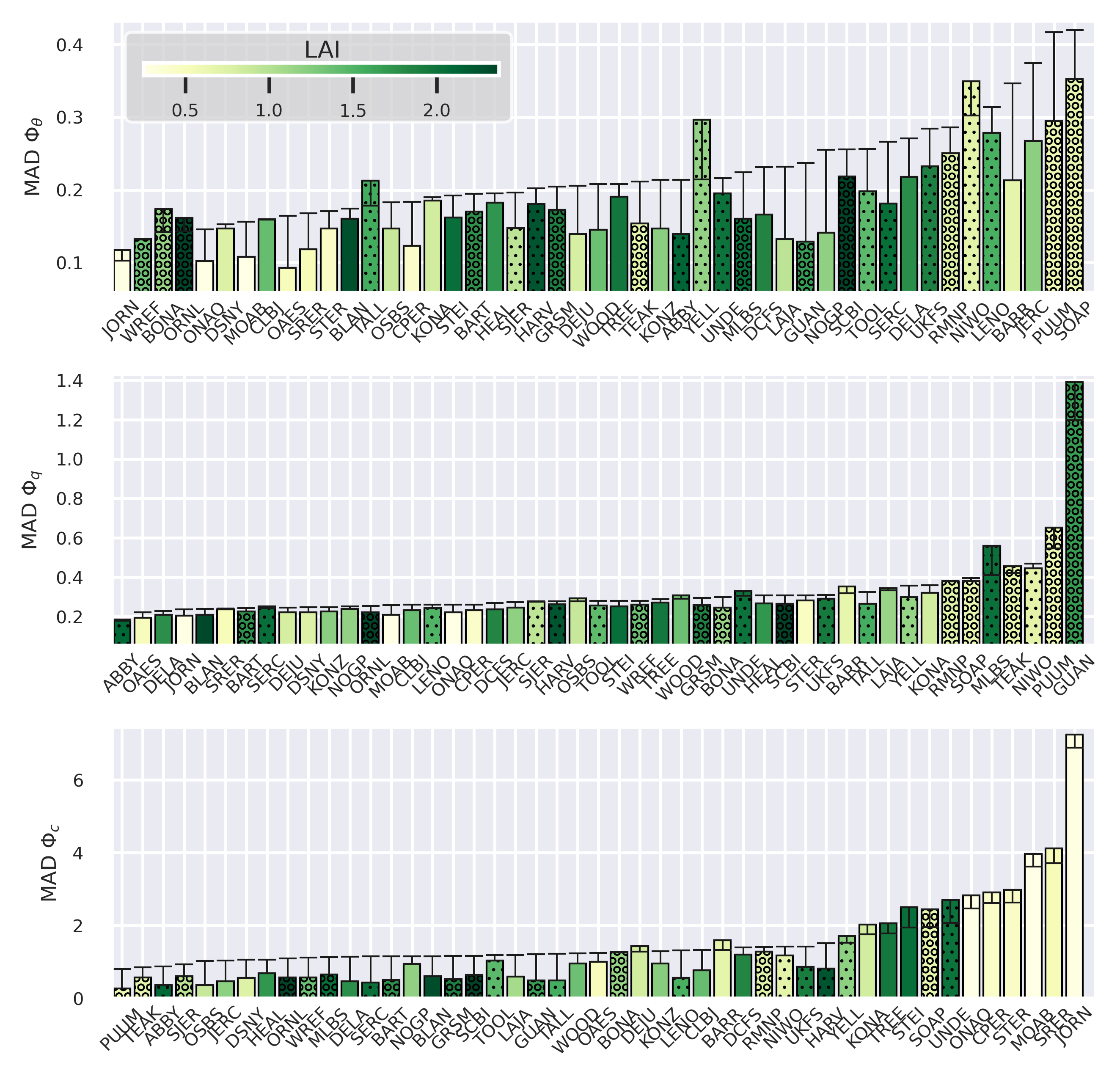}
\caption{MAD for $\Phi_{\theta}$ (top), $\Phi_q$ (middle) and $\Phi_c$ under unstable conditions for the traditional MOST relations (black line) and for the anisotropy generalized relations (colored bar) in table \ref{tab:equations} for each site in NEON. Sites are colored according to median LAI during the growing period. Sites with high complexity (median standard deviation of the digital surface model (DSM) in the flux tower footprint$>10$) are marked with black dots, and very high complexity (median standard deviation of DSM in the flux tower footprint$>20$) are marked with black circles.}
\label{fig:unstable_sitebar}
\end{figure}

\begin{figure}
\centering
\includegraphics[width=5.5in]{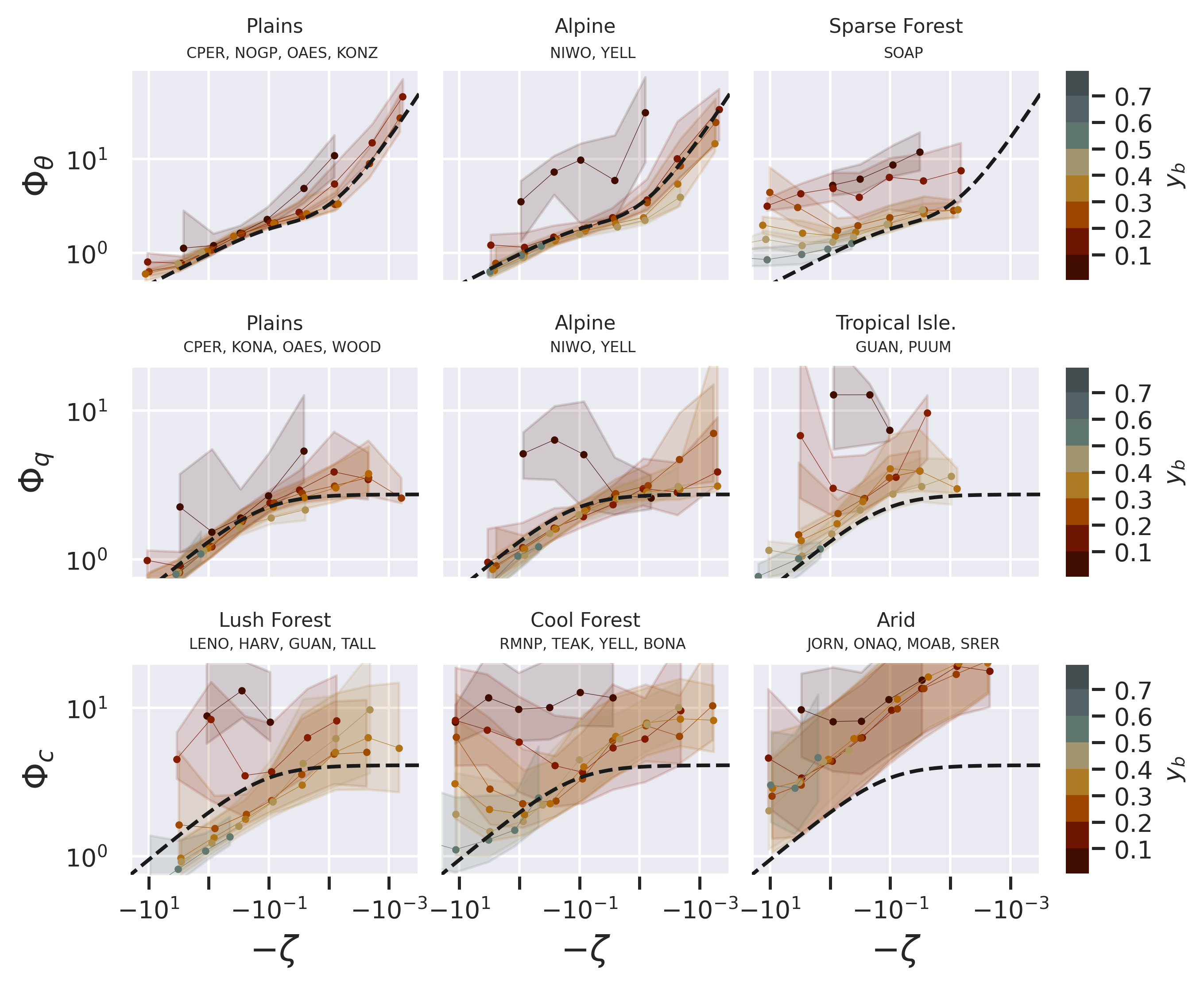}
\caption{Scaling for $\Phi_{\theta}$ (top), $\Phi_q$ (middle) and $\Phi_c$ (bottom) as in figure \ref{fig:basic_scaling} for a selection of sites with similar biases under unstable conditions. Subplot title provides a qualitative description of the group of sites, with specific sites listed beneath.}
\label{fig:sites_u}
\end{figure}

We examine the performance and nature of the scaling under unstable stratification for each of the three scalars in figures \ref{fig:unstable_sitebar} and \ref{fig:sites_u}, with each row representing one of the three scalars (top: temperature, middle: water vapor, bottom: carbon dioxide). Figure \ref{fig:unstable_sitebar} shows the performance of traditional most scaling and the modified scaling across all sites, and figure \ref{fig:sites_u} shows the scaling for a group of well performing sites (first column) and two groups of poorly performing sites (right two columns). 

\par \textbf{Temperature: } The observed scaling for $\Phi_\theta$ matches MOST fairly well under unstable conditions for the majority of the sites as is clear in the top row of figure \ref{fig:unstable_sitebar}, with the exception of a small tendency of the traditional scaling to underpredict $\Phi_{\theta}$ (see figure \ref{fig:aniso_scaling}) which is a significant contributor to the $MAD$ values. While there is no immediately obvious groupings based on the LAI in performance, comparing groups of sites with the details in figure \ref{fig:site_summary} reveal some trends. In particular, most of the best performing sites are relatively flat and homogeneous, which is expected as traditional MOST was developed over these types of terrain. What may be unexpected is that these sites also appear to see the greatest benefit from leveraging anisotropy, as is clear from the large reductions in $MAD$. Almost all sites see improvements from the modified relations, with the exception of YELL and NIWO, both sites in the Rocky Mountains with complex large scale topography.
\indent \par In figure \ref{fig:sites_u} we further explore differences in the scaling of $\Phi_\theta$. For our well performing flat plains, we see relatively little separation by $y_b$, with even more anisotropic turbulence showing similar scaling. $\Phi_\theta$, while slightly higher than traditional MOST values, follows the curve closely. The cooler alpine sites similarly follow the curve, although with lower $\Phi_{\theta}$ values; these are the only sites where the traditional MOST relations overpredict $\Phi_\theta$ on average. A close examination of SOAP, a sparse forest in the Sierras in California (figure \ref{fig:sites_u}, third column) reveals dramatically different scaling from the other sites. The relationship is nearly log-linear, and appears to be less dependent on $\zeta$ within the bounds of the available data. There is also a very strong, and consistent separation based on $y_b$. The curves appear rather similar to the non-stationary curves in figure \ref{fig:stationarity}; however this is using data corrected for non-stationarity so this is an unlikely explanation for the observed phenomena. SOAP has the highest average $x_b$ value of the NEON sites by a significant margin, and the second highest average $y_b$ value, indicating that one-component anisotropic turbulence is very common at the site. One-component and very anisotropic turbulence is associated with a number of phenomena unrelated to surface layer scaling, including large-scale attached eddies, waves and other large scale motions generally \cite{ghannam_scaling_2018,vercauteren_scale_2019,stiperski_universal_2021,gucci_sources_2023}. Attached eddies formed between canopy elements, secondary circulations and flow separation between the below and above canopy turbulence in particular would be expected for a sparse forest, and may be contributing to the observed scaling issues. SOAP is also a spatially heterogeneous environment, not only in scalar sources but also topography, which may further elevate the variances above those predicted by traditional scaling. 

\par \textbf{Water Vapor: } Traditional scaling performance is generally consistent across the sites, with especially strong performance over humid forest sites and plains where the data largely follows the traditional MOST curve with a separation by $y_b$ under more convective conditions. There are two exceptions to this pattern, with PUUM, a rainforest site in Hawai‘i and GUAN, a dry forest site in sub-tropical Puerto Rico showing significant site $MAD$ compared to other sites. Both sites are complex, and coastal. For both sites, error is significantly higher for relatively high velocity flows originating from the coastline. Fog and cloud interception (where clouds in the atmosphere flow into the land, tree canopies or sensors), a potential problem for the PUUM site in particular where fog or cloud interception are frequent \cite{delay_history_2010}, should not be included in the analysis due to the removal of saturated conditions ($RH>95\%$), but could still be a significant factor in condensation on the infrared gas analyzer and corresponding error affecting periods after these events. Since the infrared gas analyzer does not have a heater, while the relative humidity of a given time period may be less than 95\% cutoff, previous time steps may allow for accumulation of condensation which will likely persist for long time periods given the overall wet conditions. Fog and clouds, however, are unlikely to explain the observed behavior for the drier sites in Puerto Rico. Mesoscale or sub-mesoscale motions, sea breeze in particular, may be driving some of the coastal errors, however advective flows and their impacts on flux-variance relations are difficult to quantify and evaluate without multiple towers. Another possible explanation is that this divergence from scaling relations occurs due to source inhomogeneity; when the wind speed is high and the prevailing wind comes from the ocean, the tower will pick up a mix of turbulence in flows that originated over the land and the ocean, which would presumably be dry and wet respectively. This would result in an increase in the variance that is not derived from the local flux. When the atmosphere is highly unstable, and we expect fluxes to be larger, this variance may not significantly impact the flux-variance relations when compared to near-neutral conditions where the flux is low. This may explain the observation that the elevated $\Phi_q$ values at coastal sites in figure \ref{fig:sites_u} appear to be more significant under near neutral rather than more convective conditions.

\par \textbf{Carbon Dioxide: } The carbon dioxide variance scaling, and its deviation from theory, appears dependent on the strength of the local carbon cycle. This is consistent with the results in figure \ref{fig:correlations}, which show strong correlation between the magnitude of the carbon flux and model error. The scaling consistently fails with elevated $\Phi_c$ values across all $\zeta$ at drier sites with less active vegetation and performs better at sites with very significant vegetation activity. $\Phi_C$ values at lush forest sites are very similar to those that we see from traditional MOST scaling. Over more arid sites $\Phi_C$ is elevated significantly. These elevated $\Phi$ values could be caused by lower fluxes combined with inhomogeneity in the carbon sources, as in the case for the water vapor variances described above. Fluxes will be smaller, in general, at sites with low carbon cycle activity. Since carbon sources and sinks (vegetation) tend to be more sparse and inhomogeneous in these environments, variance could also be slightly elevated \cite{weaver_temperature_1990}. The smaller fluxes will make equation \eqref{eq:phi_def} more sensitive to any changes in variance due to phenomena like source spatial heterogeneity from section \ref{sec:spatial_het} which are not captured in either the traditional or modified MOST relations. The elevated $\Phi$ values at arid sites can also explain the odd behavior in figure \ref{fig:stationarity} for the stationary scaling of $\Phi_c$. Arid sites will have more stationary statistics for carbon due to low bioactivity, and also have elevated $\Phi$ values due to small fluxes, causing the divergent behavior. 

\subsubsection{Scaling Under Stable Stratification} \label{sec:stable_scaling}
\begin{figure}
\centering
\includegraphics[width=5.5in]{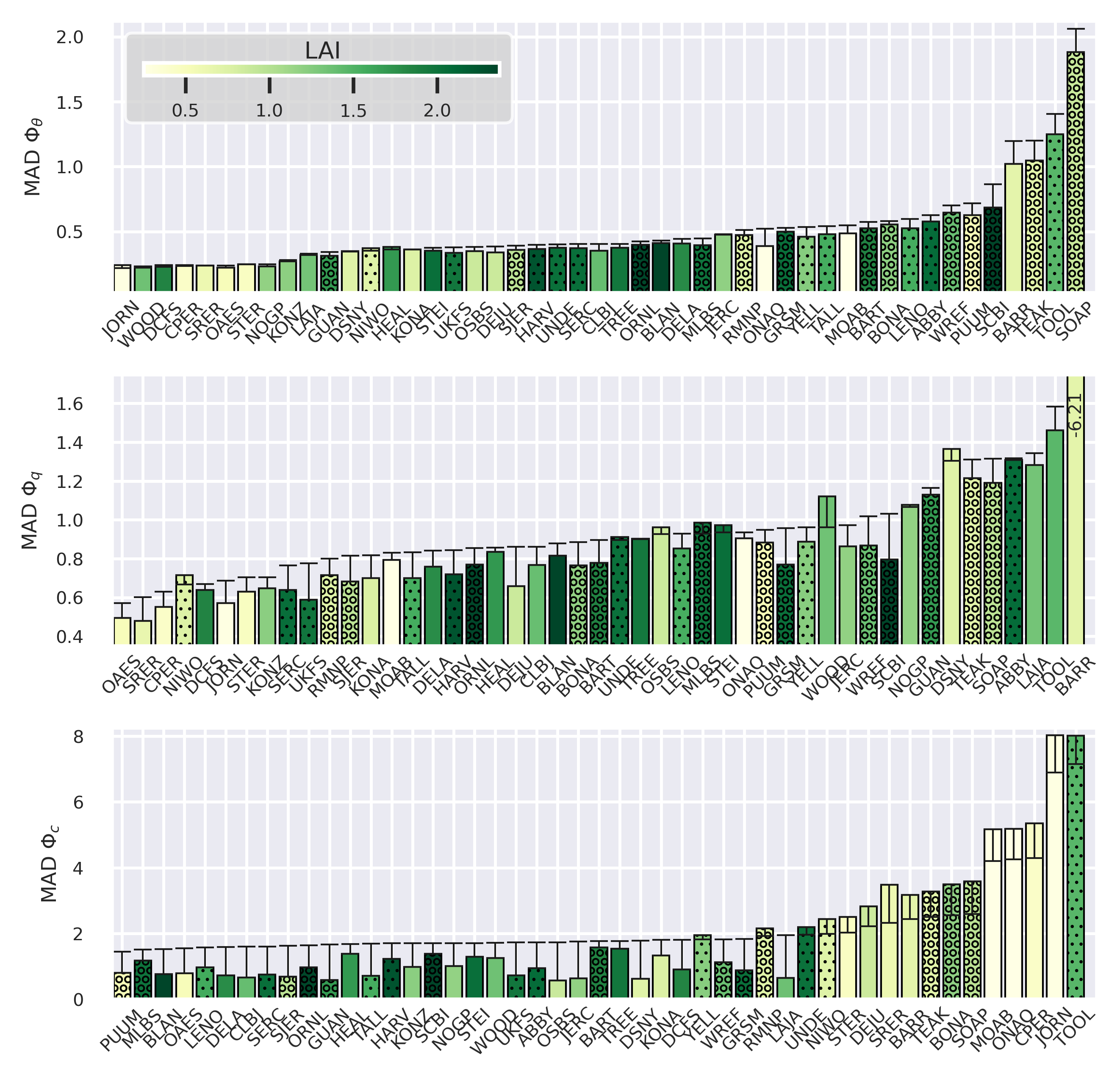}
\caption{As in figure \ref{fig:unstable_sitebar}, except for stable conditions. The MAD for BAAR is cut off to make the others more clear, and the MAD is printed directly on the bar.}
\label{fig:stable_sitebar}
\end{figure}

\begin{figure}
\centering
\includegraphics[width=5.5in]{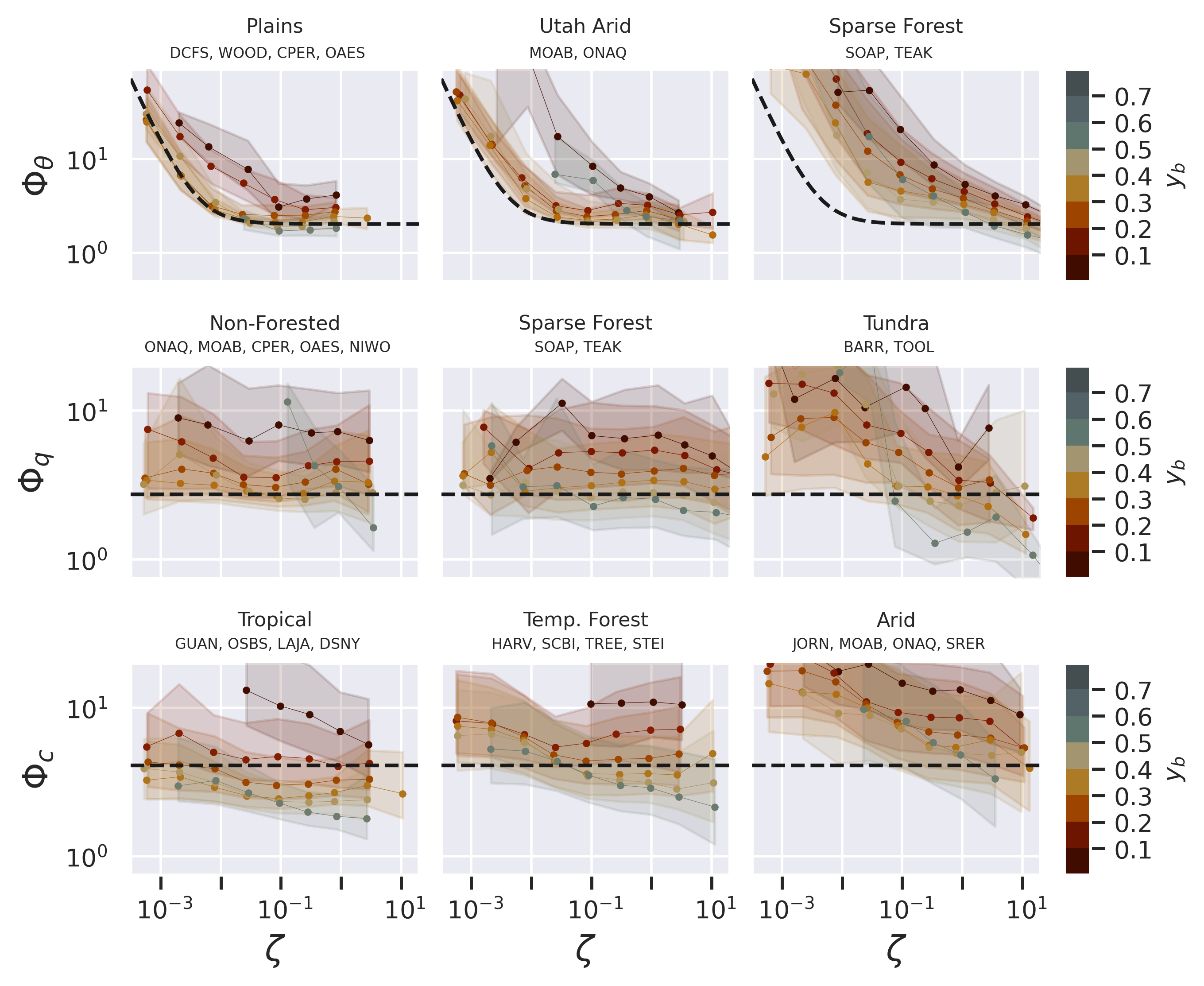}
\caption{Scaling for $\Phi_{\theta}$ (top), $\Phi_q$ (middle) and $\Phi_c$ (bottom) as in figure \ref{fig:basic_scaling} for a selection of sites with similar biases under stable conditions. Subplot title provides a qualitative description of the group of sites, with specific sites listed beneath.}
\label{fig:sites_s}
\end{figure}

We examine the performance and nature of the scaling under stable stratification for each of the three scalars in figures \ref{fig:stable_sitebar} and \ref{fig:sites_s}, with each row representing one of the three scalars (top: temperature, middle: water vapor, bottom: carbon dioxide). Figure \ref{fig:stable_sitebar} shows the performance of traditional MOST scaling and the modified scaling across all sites, and figure \ref{fig:sites_s} shows the scaling for a group of well performing sites (first column) and two groups of poorly performing sites (right two columns).

\par \textbf{Temperature: } Traditional scaling has a very clear relationship with site characteristics under stable conditions, with traditional flat, homogeneous sites with low vegetation (notably including sites with high and low mean temperatures) seem to match traditional scaling well, while also displaying a clear relationship between the scaling and anisotropy that mimics the results from the combined dataset. Notably, the sites with the poorest performance generally appear to be sites with relatively complex terrain/canopy configurations. For flat, homogeneous, non-vegetated sites the scaling appears approximately z-less between $\zeta =10^{-1.5}$ and $10^{1}$. By contrast, a small slope appears in forests, particularly tall ones, that becomes even more significant for the sparse forests at Soaproot Saddle (SOAP) and the lower Teakettle (TEAK) in the Sierras; at these sites the form of the scaling appears divorced from traditional relations with a nearly log-linear relationship. As discussed for SOAP in section \ref{sec:unstable_scaling}, SOAP is one of the most anisotropic, one-component sites in the network, with TEAK having the second highest mean $x_b$ and the lowest $y_b$ values as well. Under stable conditions, one-component turbulence is associated with internal gravity waves and continuously trending velocity variances and covariances during the averaging periods \cite{gucci_sources_2023}. 

\par \textbf{Water Vapor: } The scaling under stable conditions for water vapor shows exceptionally poor performance at BARR (Utqiaġvik). Similar to the poor performing sites for water vapor in figure \ref{fig:unstable_sitebar}, BARR is a coastal site where source inhomogeneity, advective flows, and sea breezes may be a critical factor affecting the surface layer scaling. In addition, the site is known for high prevalence of nighttime coastal fog \cite{dengel_influence_2021}, which may be influencing the measurements through condensation on the infrared gas analyzer as discussed in \ref{sec:unstable_scaling}. While scaling at BARR (and another tundra site, TOOL) is clearly not z-less, scaling at the dense forest (not shown), sparse forest, and non-forested sites appears to maintain the z-less scaling commonly found in other studies. The sparse forests, however, have an increased scaling dependency on the anisotropy as was found for temperature in both stable and unstable conditions. 

\par \textbf{Carbon Dioxide: } Much like under unstable conditions, the scaling of $\Phi_c$ appears related to vegetation activity and the strength of the carbon cycle. Lush environments with high bioactivity show more z-less scaling, less $y_b$ dependence, and lower $\Phi_c$ values than  arid sites which would be expected to have a weaker carbon cycle. The non-arid sites with poor performance tend to be cooler sites in alpine or tundra locations where bioactivity would also be low. In these locations, particularly arid sites, the scaling becomes more clearly dependent on $\zeta$ in figure \ref{fig:sites_s}. The dependence on anisotropy is also increased at these sites, as is the case for $\Phi_q$ and even $\Phi_{\theta}$; broadly, we see that anisotropy is less relevant over vegetated canopies than grasslands and similar terrain, which is not observed over unstable conditions.

\section{Discussion} \label{sec:discussion}
\subsection{Interactions Between Spatial Heterogeneity, Anisotropy and Stationarity}
The results in the previous sections highlight three major sources of deviation from traditional surface layer scaling. While most of the analysis examines spatial heterogeneity, anisotropy and stationarity independently of each other, it is important to consider the interactions between them. In figure \ref{fig:stationarity}, for example we already saw a relationship between anisotropy and stationarity, indicating that for less stationary timeseries, the anisotropy of turbulence becomes more valuable. Under stable conditions, intermittent gravity waves tend to be very anisotropic and, due to their intermittency, make the flow non-stationary \cite{gucci_sources_2023}. Large scale flows (sub-mesoscale wind variability) tend to be more anisotropic \cite{vercauteren_scale_2019,stiperski_universal_2021}, so the existence of eddies of scales closer to the averaging period may make the timeseries both non-stationary and anisotropic. A more focused study examining intermittency directly would be needed to fully evaluate this. Due to the diverging nature of very isotropic and anisotropic flows, and the non-local nature of some very anisotropic flows, it is also reasonable to conclude that the anisotropy of turbulence will have some relationship with spatial heterogeneity and how it impacts the scaling, and previous work has found some connections between vegetation patterns (heterogeneity) and $y_b$ \cite{waterman_evaluating_2025}. Others, however, over non-vegetated regions have found poor relationship between $y_b$ and surface spatial characteristics \cite{mosso_revealing_2025}. As such, additional examination to the direct link between spatial patterns and anisotropy is needed. 

\begin{figure}
\centering
\includegraphics[width=5.5in]{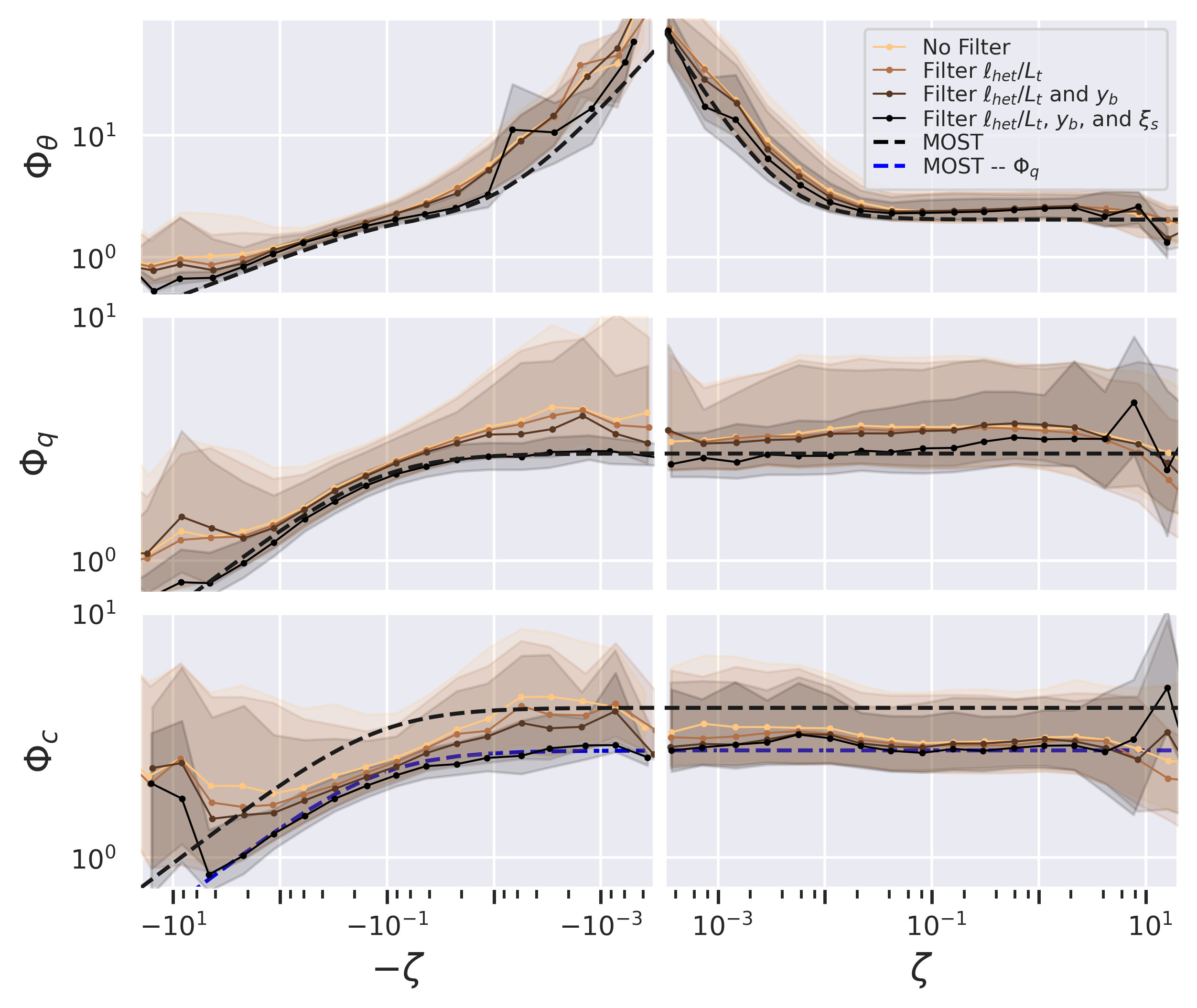}
\caption{Scaling for $\Phi_{\theta}$ (top), $\Phi_q$ (middle) and $\Phi_c$ (bottom) binned by $\zeta$ for different degrees of filtering: no filtering, removing spatially heterogeneous periods (only using $\ell_{het}/L_t>c_s$ with $c_s$ between $.5$ and $1$ depending on variable), removing heterogeneous and very isotropic/anisotropic periods (only using $0.225<y_b<0.375$), and finally removing heterogeneous, very isotropic/anisotropic, and non-stationary periods (only using $\xi_s<30$). Filled area shows the 25th to 75th percentile for the bin. Traditional MOST scaling is also shown, and for $\Phi_c$ the traditional MOST scaling for $\Phi_q$ is also shown in blue.}
\label{fig:filter_scaling}
\end{figure}

We can also examine the implications of collectively filtering for all these deviations from traditional theory. In figure \ref{fig:filter_scaling}, we show how filtering out heterogeneous, very isotropic, very anisotropic, and non-stationary data changes the scaling. Overall, we see a collapse towards the traditional MOST relations as we apply increasingly strict filters, particularly for $\Phi_q$ and $\Phi_c$. The $\Phi_c$ and $\Phi_q$ curves both lose much of their $\zeta$ dependence under stable conditions, and what remains for $\Phi_c$ under near neutral relations is likely a product of the bias from arid sites discussed in section \ref{sec:sites}. The $\Phi_c$ curves under unstable conditions also appear to be converging towards the traditional MOST lines for $\Phi_q$, suggesting strong scalar similarity. Notably, the $\Phi_\theta$ curves do not approach any constant value, illustrating the scalar dissimilarity between the bioactive scalars and potential temperature in near neutral conditions. Previous work has suggested that weaker fluxes, large divergence terms, and large dissimilarity in the spatial patterns of the scalar is related to broader scalar dissimilarity in the surface layer \cite{chor_flux-variance_2017,cancelli_dimensionless_2012}; the filters address many of these conditions. Although this implies strong scalar similarity, a more comprehensive analysis on scalar similarity would need to be completed to fully assess this claim. The temperature scaling does not see the same overall improvement from the filtering, although there are some consistent improvements under very convective conditions towards the traditional regime. This implies that, to a degree, temperature scaling has a relatively low sensitivity to non-stationarity, heterogeneity and anisotropy, although there is also the distinct possibility that a more appropriate metric for spatial heterogeneity than albedo may improve results. Albedo will be correlated with temperature, however albedo does not account for heat capacity of the surface: for example asphalt and lakes would have very similar albedo values despite drastically different anticipated sensible heat flux and surface temperatures. While filtering out heterogeneous, non-stationary, anisotropic and very isotropic turbulence does improve theory performance, it eliminates large amounts of data, reducing the usable data by over 80\%. Global and regional atmospheric modeling need to be able to represent all conditions and removing non-canonical flows can be harmful for that application. Finding ways to incorporate the processes that this misses, such as the anisotropy curves presented in figure \ref{fig:aniso_scaling}, are important for broader public impact. 

\section{Conclusion}
In this study, we examine the ASL, MOST-based flux-variance relations for heat, water vapor, and carbon dioxide across the diverse range of ecosystems present in the NEON network. In particular, this work emphasizes three forms of heterogeneity related to significant deviations from traditional MOST theory: spatial heterogeneity of the scalar sources, anisotropy of the Reynolds stress tensor, and non-stationarity of the timeseries. All three metrics have a unique relationship with the flux-variance relations, and accounting for them yields significant improvements in the performance of traditional and anisotropy modified scalings. The work also illuminates paths for further research. Results in figure \ref{fig:spatial_het} show strong promise for other more direct, quantitative analysis from observations between spatial heterogeneity and surface layer scaling; more work analyzing how well this holds through time, under what conditions it might matter, and where heterogeneity has the greatest impact are needed. In particular, analysis that considers the changing shape and directionality of the flux footprint has the potential to yield an even stronger relationship between spatial heterogeneity and the flux-variance relations. Similar analysis applied to the flux-gradient relations, used to parameterize surface fluxes in large scale ESMs and for NWP, also hold significant potential.
\par The work also benefits greatly from the diversity within NEON, showing that these analyses are not heavily biased to individual sites or types of terrain. In addition, we are able to see strong inter site variability in some cases, further illuminating situations in which MOST-based schemes fail due to particulars of local conditions. Notable examples include breakdown of carbon dioxide scaling at less bioactive sites and for the water vapor variance scaling at coastal sites with prevailing wind from the coast. New anisotropy based flux variance relations, including the new anisotropy generalized relations for $\Phi_c$ and $\Phi_q$, have the potential for significant impact if applied into surface layer parameterizations for boundary layer schemes. We illustrate the conditions where the flux variance relations can apply, and where they fail, which can guide future work to account for flows and conditions that do not meet MOST assumptions.

\section*{Open Research Section}
The raw data used for this analysis totals more than 40 TB and is publicly available from the National Ecological Observation Network data portal, for remotely sensed water indicies \cite{national_ecological_observatory_network_neon_canopy_2025,national_ecological_observatory_network_neon_canopy_2025-1}, albedo \cite{national_ecological_observatory_network_neon_albedo_2025}, and Leaf Area Index (LAI) \cite{national_ecological_observatory_network_neon_lai_2025,national_ecological_observatory_network_neon_lai_2025-1}, and turbulence information \cite{national_ecological_observatory_network_neon_bundled_2025} . Scripts to process this data and generate the results in this publication include \citeA{waterman_tswaterneon_aniso_2025} and \citeA{waterman_tswaterscalar_neon_2025}.

\acknowledgments
The authors thank the various agencies that have supported this work, including the National Science Foundation Division of Atmospheric and Geospace Sciences (NSF-AGS) Postdoctoral Fellowship (Award 2412560; PI Dr. Tyler Waterman), NSF-AGS Physical and Dynamic Meteorology Program (PDM) (Award 2414424; PI Dr. Marc Calaf) and the European Research Council (ERC) (Award 101001691; PI Dr. Ivana Stiperski).

%
%

\bibliography{Scalars}

\begin{thebibliography}{}

\bibitem [\protect \citeauthoryear {%
Babić%
, Večenaj%
\BCBL {}\ \BBA {} De~Wekker%
}{%
Babić%
\ \protect \BOthers {.}}{%
{\protect \APACyear {2016}}%
}]{%
babic_fluxvariance_2016}
\APACinsertmetastar {%
babic_fluxvariance_2016}%
\begin{APACrefauthors}%
Babić, N.%
, Večenaj, Z.%
\BCBL {}\ \BBA {} De~Wekker, S\BPBI F\BPBI J.%
\end{APACrefauthors}%
\unskip\
\newblock
\APACrefYearMonthDay{2016}{{\APACmonth{04}}}{}.
\newblock
{\BBOQ}\APACrefatitle {Flux–{Variance} {Similarity} in {Complex} {Terrain} and {Its} {Sensitivity} to {Different} {Methods} of {Treating} {Non}-stationarity} {Flux–{Variance} {Similarity} in {Complex} {Terrain} and {Its} {Sensitivity} to {Different} {Methods} of {Treating} {Non}-stationarity}.{\BBCQ}
\newblock
\APACjournalVolNumPages{Boundary-Layer Meteorology}{159}{1}{123--145}.
\newblock
\begin{APACrefURL} [{2025-02-04}]\url{http://link.springer.com/10.1007/s10546-015-0110-0} \end{APACrefURL}
\newblock
\begin{APACrefDOI} \doi{10.1007/s10546-015-0110-0} \end{APACrefDOI}
\PrintBackRefs{\CurrentBib}

\bibitem [\protect \citeauthoryear {%
Banerjee%
, Krahl%
, Durst%
\BCBL {}\ \BBA {} Zenger%
}{%
Banerjee%
\ \protect \BOthers {.}}{%
{\protect \APACyear {2007}}%
}]{%
banerjee_presentation_2007}
\APACinsertmetastar {%
banerjee_presentation_2007}%
\begin{APACrefauthors}%
Banerjee, S.%
, Krahl, R.%
, Durst, F.%
\BCBL {}\ \BBA {} Zenger, C.%
\end{APACrefauthors}%
\unskip\
\newblock
\APACrefYearMonthDay{2007}{{\APACmonth{01}}}{}.
\newblock
{\BBOQ}\APACrefatitle {Presentation of anisotropy properties of turbulence, invariants versus eigenvalue approaches} {Presentation of anisotropy properties of turbulence, invariants versus eigenvalue approaches}.{\BBCQ}
\newblock
\APACjournalVolNumPages{Journal of Turbulence}{8}{}{N32}.
\newblock
\begin{APACrefURL} [{2023-11-15}]\url{https://www.tandfonline.com/doi/full/10.1080/14685240701506896} \end{APACrefURL}
\newblock
\begin{APACrefDOI} \doi{10.1080/14685240701506896} \end{APACrefDOI}
\PrintBackRefs{\CurrentBib}

\bibitem [\protect \citeauthoryear {%
Belcher%
\ \BBA {} Hunt%
}{%
Belcher%
\ \BBA {} Hunt%
}{%
{\protect \APACyear {1998}}%
}]{%
belcher_turbulent_1998}
\APACinsertmetastar {%
belcher_turbulent_1998}%
\begin{APACrefauthors}%
Belcher, S\BPBI E.%
\BCBT {}\ \BBA {} Hunt, J\BPBI C\BPBI R.%
\end{APACrefauthors}%
\unskip\
\newblock
\APACrefYearMonthDay{1998}{{\APACmonth{01}}}{}.
\newblock
{\BBOQ}\APACrefatitle {Turbulent {Flow} {Over} {Hills} and {Waves}} {Turbulent {Flow} {Over} {Hills} and {Waves}}.{\BBCQ}
\newblock
\APACjournalVolNumPages{Annual Review of Fluid Mechanics}{30}{1}{507--538}.
\newblock
\begin{APACrefURL} [{2025-09-08}]\url{https://www.annualreviews.org/doi/10.1146/annurev.fluid.30.1.507} \end{APACrefURL}
\newblock
\begin{APACrefDOI} \doi{10.1146/annurev.fluid.30.1.507} \end{APACrefDOI}
\PrintBackRefs{\CurrentBib}

\bibitem [\protect \citeauthoryear {%
Bou-Zeid%
, Anderson%
, Katul%
\BCBL {}\ \BBA {} Mahrt%
}{%
Bou-Zeid%
\ \protect \BOthers {.}}{%
{\protect \APACyear {2020}}%
}]{%
bou-zeid_persistent_2020}
\APACinsertmetastar {%
bou-zeid_persistent_2020}%
\begin{APACrefauthors}%
Bou-Zeid, E.%
, Anderson, W.%
, Katul, G\BPBI G.%
\BCBL {}\ \BBA {} Mahrt, L.%
\end{APACrefauthors}%
\unskip\
\newblock
\APACrefYearMonthDay{2020}{{\APACmonth{12}}}{}.
\newblock
{\BBOQ}\APACrefatitle {The {Persistent} {Challenge} of {Surface} {Heterogeneity} in {Boundary}-{Layer} {Meteorology}: {A} {Review}} {The {Persistent} {Challenge} of {Surface} {Heterogeneity} in {Boundary}-{Layer} {Meteorology}: {A} {Review}}.{\BBCQ}
\newblock
\APACjournalVolNumPages{Boundary-Layer Meteorology}{177}{2-3}{227--245}.
\newblock
\begin{APACrefURL} [{2023-10-31}]\url{https://link.springer.com/10.1007/s10546-020-00551-8} \end{APACrefURL}
\newblock
\begin{APACrefDOI} \doi{10.1007/s10546-020-00551-8} \end{APACrefDOI}
\PrintBackRefs{\CurrentBib}

\bibitem [\protect \citeauthoryear {%
Buttar%
\ \protect \BOthers {.}}{%
Buttar%
\ \protect \BOthers {.}}{%
{\protect \APACyear {2022}}%
}]{%
buttar_estimation_2022}
\APACinsertmetastar {%
buttar_estimation_2022}%
\begin{APACrefauthors}%
Buttar, N\BPBI A.%
, Hu, Y.%
, Tanny, J.%
, Raza, A.%
, Niaz, Y.%
, Khan, M\BPBI I.%
\BDBL {}Bilal~Idrees, M.%
\end{APACrefauthors}%
\unskip\
\newblock
\APACrefYearMonthDay{2022}{{\APACmonth{09}}}{}.
\newblock
{\BBOQ}\APACrefatitle {Estimation of {Sensible} and {Latent} {Heat} {Fluxes} {Using} {Flux} {Variance} {Method} under {Unstable} {Conditions}: {A} {Case} {Study} of {Tea} {Plants}} {Estimation of {Sensible} and {Latent} {Heat} {Fluxes} {Using} {Flux} {Variance} {Method} under {Unstable} {Conditions}: {A} {Case} {Study} of {Tea} {Plants}}.{\BBCQ}
\newblock
\APACjournalVolNumPages{Atmosphere}{13}{10}{1545}.
\newblock
\begin{APACrefURL} [{2025-02-04}]\url{https://www.mdpi.com/2073-4433/13/10/1545} \end{APACrefURL}
\newblock
\begin{APACrefDOI} \doi{10.3390/atmos13101545} \end{APACrefDOI}
\PrintBackRefs{\CurrentBib}

\bibitem [\protect \citeauthoryear {%
Buttar%
, Yongguang%
, Tanny%
, Akram%
\BCBL {}\ \BBA {} Shabbir%
}{%
Buttar%
\ \protect \BOthers {.}}{%
{\protect \APACyear {2019}}%
}]{%
buttar_fetch_2019}
\APACinsertmetastar {%
buttar_fetch_2019}%
\begin{APACrefauthors}%
Buttar, N\BPBI A.%
, Yongguang, H.%
, Tanny, J.%
, Akram, M\BPBI W.%
\BCBL {}\ \BBA {} Shabbir, A.%
\end{APACrefauthors}%
\unskip\
\newblock
\APACrefYearMonthDay{2019}{{\APACmonth{06}}}{}.
\newblock
{\BBOQ}\APACrefatitle {Fetch {Effect} on {Flux}-{Variance} {Estimations} of {Sensible} and {Latent} {Heat} {Fluxes} of {Camellia} {Sinensis}} {Fetch {Effect} on {Flux}-{Variance} {Estimations} of {Sensible} and {Latent} {Heat} {Fluxes} of {Camellia} {Sinensis}}.{\BBCQ}
\newblock
\APACjournalVolNumPages{Atmosphere}{10}{6}{299}.
\newblock
\begin{APACrefURL} [{2025-09-11}]\url{https://www.mdpi.com/2073-4433/10/6/299} \end{APACrefURL}
\newblock
\begin{APACrefDOI} \doi{10.3390/atmos10060299} \end{APACrefDOI}
\PrintBackRefs{\CurrentBib}

\bibitem [\protect \citeauthoryear {%
Cancelli%
, Dias%
\BCBL {}\ \BBA {} Chamecki%
}{%
Cancelli%
\ \protect \BOthers {.}}{%
{\protect \APACyear {2012}}%
}]{%
cancelli_dimensionless_2012}
\APACinsertmetastar {%
cancelli_dimensionless_2012}%
\begin{APACrefauthors}%
Cancelli, D\BPBI M.%
, Dias, N\BPBI L.%
\BCBL {}\ \BBA {} Chamecki, M.%
\end{APACrefauthors}%
\unskip\
\newblock
\APACrefYearMonthDay{2012}{{\APACmonth{10}}}{}.
\newblock
{\BBOQ}\APACrefatitle {Dimensionless criteria for the production‐dissipation equilibrium of scalar fluctuations and their implications for scalar similarity} {Dimensionless criteria for the production‐dissipation equilibrium of scalar fluctuations and their implications for scalar similarity}.{\BBCQ}
\newblock
\APACjournalVolNumPages{Water Resources Research}{48}{10}{2012WR012127}.
\newblock
\begin{APACrefURL} [{2025-06-04}]\url{https://agupubs.onlinelibrary.wiley.com/doi/10.1029/2012WR012127} \end{APACrefURL}
\newblock
\begin{APACrefDOI} \doi{10.1029/2012WR012127} \end{APACrefDOI}
\PrintBackRefs{\CurrentBib}

\bibitem [\protect \citeauthoryear {%
Charrondière%
\ \BBA {} Stiperski%
}{%
Charrondière%
\ \BBA {} Stiperski%
}{%
{\protect \APACyear {2024}}%
}]{%
charrondiere_spectral_2024}
\APACinsertmetastar {%
charrondiere_spectral_2024}%
\begin{APACrefauthors}%
Charrondière, C.%
\BCBT {}\ \BBA {} Stiperski, I.%
\end{APACrefauthors}%
\unskip\
\newblock
\APACrefYearMonthDay{2024}{{\APACmonth{10}}}{}.
\newblock
{\BBOQ}\APACrefatitle {Spectral scaling of unstably stratified atmospheric flows: {Turbulence} anisotropy and the low‐frequency spread} {Spectral scaling of unstably stratified atmospheric flows: {Turbulence} anisotropy and the low‐frequency spread}.{\BBCQ}
\newblock
\APACjournalVolNumPages{Quarterly Journal of the Royal Meteorological Society}{150}{764}{4196--4216}.
\newblock
\begin{APACrefURL} [{2025-05-05}]\url{https://rmets.onlinelibrary.wiley.com/doi/10.1002/qj.4811} \end{APACrefURL}
\newblock
\begin{APACrefDOI} \doi{10.1002/qj.4811} \end{APACrefDOI}
\PrintBackRefs{\CurrentBib}

\bibitem [\protect \citeauthoryear {%
Cheng%
\ \BBA {} Xu%
}{%
Cheng%
\ \BBA {} Xu%
}{%
{\protect \APACyear {2015}}%
}]{%
cheng_improved_2015}
\APACinsertmetastar {%
cheng_improved_2015}%
\begin{APACrefauthors}%
Cheng, A.%
\BCBT {}\ \BBA {} Xu, K\BHBI M.%
\end{APACrefauthors}%
\unskip\
\newblock
\APACrefYearMonthDay{2015}{{\APACmonth{07}}}{}.
\newblock
{\BBOQ}\APACrefatitle {Improved {Low}-{Cloud} {Simulation} from the {Community} {Atmosphere} {Model} with an {Advanced} {Third}-{Order} {Turbulence} {Closure}} {Improved {Low}-{Cloud} {Simulation} from the {Community} {Atmosphere} {Model} with an {Advanced} {Third}-{Order} {Turbulence} {Closure}}.{\BBCQ}
\newblock
\APACjournalVolNumPages{Journal of Climate}{28}{14}{5737--5762}.
\newblock
\begin{APACrefURL} [{2025-09-09}]\url{http://journals.ametsoc.org/doi/10.1175/JCLI-D-14-00776.1} \end{APACrefURL}
\newblock
\begin{APACrefDOI} \doi{10.1175/JCLI-D-14-00776.1} \end{APACrefDOI}
\PrintBackRefs{\CurrentBib}

\bibitem [\protect \citeauthoryear {%
Chor%
\ \protect \BOthers {.}}{%
Chor%
\ \protect \BOthers {.}}{%
{\protect \APACyear {2017}}%
}]{%
chor_flux-variance_2017}
\APACinsertmetastar {%
chor_flux-variance_2017}%
\begin{APACrefauthors}%
Chor, T\BPBI L.%
, Dias, N\BPBI L.%
, Araújo, A.%
, Wolff, S.%
, Zahn, E.%
, Manzi, A.%
\BDBL {}Sörgel, M.%
\end{APACrefauthors}%
\unskip\
\newblock
\APACrefYearMonthDay{2017}{{\APACmonth{05}}}{}.
\newblock
{\BBOQ}\APACrefatitle {Flux-variance and flux-gradient relationships in the roughness sublayer over the {Amazon} forest} {Flux-variance and flux-gradient relationships in the roughness sublayer over the {Amazon} forest}.{\BBCQ}
\newblock
\APACjournalVolNumPages{Agricultural and Forest Meteorology}{239}{}{213--222}.
\newblock
\begin{APACrefURL} [{2024-03-17}]\url{https://linkinghub.elsevier.com/retrieve/pii/S0168192317301144} \end{APACrefURL}
\newblock
\begin{APACrefDOI} \doi{10.1016/j.agrformet.2017.03.009} \end{APACrefDOI}
\PrintBackRefs{\CurrentBib}

\bibitem [\protect \citeauthoryear {%
Cohen%
, Cavallo%
, Coniglio%
\BCBL {}\ \BBA {} Brooks%
}{%
Cohen%
\ \protect \BOthers {.}}{%
{\protect \APACyear {2015}}%
}]{%
cohen_review_2015}
\APACinsertmetastar {%
cohen_review_2015}%
\begin{APACrefauthors}%
Cohen, A\BPBI E.%
, Cavallo, S\BPBI M.%
, Coniglio, M\BPBI C.%
\BCBL {}\ \BBA {} Brooks, H\BPBI E.%
\end{APACrefauthors}%
\unskip\
\newblock
\APACrefYearMonthDay{2015}{{\APACmonth{06}}}{}.
\newblock
{\BBOQ}\APACrefatitle {A {Review} of {Planetary} {Boundary} {Layer} {Parameterization} {Schemes} and {Their} {Sensitivity} in {Simulating} {Southeastern} {U}.{S}. {Cold} {Season} {Severe} {Weather} {Environments}} {A {Review} of {Planetary} {Boundary} {Layer} {Parameterization} {Schemes} and {Their} {Sensitivity} in {Simulating} {Southeastern} {U}.{S}. {Cold} {Season} {Severe} {Weather} {Environments}}.{\BBCQ}
\newblock
\APACjournalVolNumPages{Weather and Forecasting}{30}{3}{591--612}.
\newblock
\begin{APACrefURL} [{2023-11-13}]\url{https://journals.ametsoc.org/doi/10.1175/WAF-D-14-00105.1} \end{APACrefURL}
\newblock
\begin{APACrefDOI} \doi{10.1175/WAF-D-14-00105.1} \end{APACrefDOI}
\PrintBackRefs{\CurrentBib}

\bibitem [\protect \citeauthoryear {%
De~Bruin%
, Kohsiek%
\BCBL {}\ \BBA {} Van Den~Hurk%
}{%
De~Bruin%
\ \protect \BOthers {.}}{%
{\protect \APACyear {1993}}%
}]{%
de_bruin_verification_1993}
\APACinsertmetastar {%
de_bruin_verification_1993}%
\begin{APACrefauthors}%
De~Bruin, H\BPBI A\BPBI R.%
, Kohsiek, W.%
\BCBL {}\ \BBA {} Van Den~Hurk, B\BPBI J\BPBI J\BPBI M.%
\end{APACrefauthors}%
\unskip\
\newblock
\APACrefYearMonthDay{1993}{{\APACmonth{03}}}{}.
\newblock
{\BBOQ}\APACrefatitle {A verification of some methods to determine the fluxes of momentum, sensible heat, and water vapour using standard deviation and structure parameter of scalar meteorological quantities} {A verification of some methods to determine the fluxes of momentum, sensible heat, and water vapour using standard deviation and structure parameter of scalar meteorological quantities}.{\BBCQ}
\newblock
\APACjournalVolNumPages{Boundary-Layer Meteorology}{63}{3}{231--257}.
\newblock
\begin{APACrefURL} [{2023-11-13}]\url{http://link.springer.com/10.1007/BF00710461} \end{APACrefURL}
\newblock
\begin{APACrefDOI} \doi{10.1007/BF00710461} \end{APACrefDOI}
\PrintBackRefs{\CurrentBib}

\bibitem [\protect \citeauthoryear {%
De~Franceschi%
, Zardi%
, Tagliazucca%
\BCBL {}\ \BBA {} Tampieri%
}{%
De~Franceschi%
\ \protect \BOthers {.}}{%
{\protect \APACyear {2009}}%
}]{%
de_franceschi_analysis_2009}
\APACinsertmetastar {%
de_franceschi_analysis_2009}%
\begin{APACrefauthors}%
De~Franceschi, M.%
, Zardi, D.%
, Tagliazucca, M.%
\BCBL {}\ \BBA {} Tampieri, F.%
\end{APACrefauthors}%
\unskip\
\newblock
\APACrefYearMonthDay{2009}{{\APACmonth{10}}}{}.
\newblock
{\BBOQ}\APACrefatitle {Analysis of second‐order moments in surface layer turbulence in an {Alpine} valley} {Analysis of second‐order moments in surface layer turbulence in an {Alpine} valley}.{\BBCQ}
\newblock
\APACjournalVolNumPages{Quarterly Journal of the Royal Meteorological Society}{135}{644}{1750--1765}.
\newblock
\begin{APACrefURL} [{2023-11-15}]\url{https://rmets.onlinelibrary.wiley.com/doi/10.1002/qj.506} \end{APACrefURL}
\newblock
\begin{APACrefDOI} \doi{10.1002/qj.506} \end{APACrefDOI}
\PrintBackRefs{\CurrentBib}

\bibitem [\protect \citeauthoryear {%
DeLay%
\ \BBA {} Giambelluca%
}{%
DeLay%
\ \BBA {} Giambelluca%
}{%
{\protect \APACyear {2010}}%
}]{%
delay_history_2010}
\APACinsertmetastar {%
delay_history_2010}%
\begin{APACrefauthors}%
DeLay, J.%
\BCBT {}\ \BBA {} Giambelluca, T.%
\end{APACrefauthors}%
\unskip\
\newblock
\APACrefYearMonthDay{2010}{}{}.
\newblock
{\BBOQ}\APACrefatitle {History of cloud water interception research in {Hawai}'i} {History of cloud water interception research in {Hawai}'i}.{\BBCQ}
\newblock
\BIn{} \APACrefbtitle {Tropical {Montane} {Cloud} {Forests}: {Science} for {Conservation} and {Management}} {Tropical {Montane} {Cloud} {Forests}: {Science} for {Conservation} and {Management}}\ (\BPGS\ 332--341).
\newblock
\APACaddressPublisher{Cambridge, UK}{Cambridge University Press}.
\PrintBackRefs{\CurrentBib}

\bibitem [\protect \citeauthoryear {%
Dengel%
, Billesbach%
\BCBL {}\ \BBA {} Torn%
}{%
Dengel%
\ \protect \BOthers {.}}{%
{\protect \APACyear {2021}}%
}]{%
dengel_influence_2021}
\APACinsertmetastar {%
dengel_influence_2021}%
\begin{APACrefauthors}%
Dengel, S.%
, Billesbach, D.%
\BCBL {}\ \BBA {} Torn, M\BPBI S.%
\end{APACrefauthors}%
\unskip\
\newblock
\APACrefYearMonthDay{2021}{{\APACmonth{12}}}{}.
\newblock
{\BBOQ}\APACrefatitle {Influence of {Tundra} {Polygon} {Type} and {Climate} {Variability} on {CO}$_{\textrm{2}}$ and {CH}$_{\textrm{4}}$ {Fluxes} {Near} {Utqiagvik}, {Alaska}} {Influence of {Tundra} {Polygon} {Type} and {Climate} {Variability} on {CO}$_{\textrm{2}}$ and {CH}$_{\textrm{4}}$ {Fluxes} {Near} {Utqiagvik}, {Alaska}}.{\BBCQ}
\newblock
\APACjournalVolNumPages{Journal of Geophysical Research: Biogeosciences}{126}{12}{e2021JG006262}.
\newblock
\begin{APACrefURL} [{2025-05-27}]\url{https://agupubs.onlinelibrary.wiley.com/doi/10.1029/2021JG006262} \end{APACrefURL}
\newblock
\begin{APACrefDOI} \doi{10.1029/2021JG006262} \end{APACrefDOI}
\PrintBackRefs{\CurrentBib}

\bibitem [\protect \citeauthoryear {%
Detto%
, Baldocchi%
\BCBL {}\ \BBA {} Katul%
}{%
Detto%
\ \protect \BOthers {.}}{%
{\protect \APACyear {2010}}%
}]{%
detto_scaling_2010}
\APACinsertmetastar {%
detto_scaling_2010}%
\begin{APACrefauthors}%
Detto, M.%
, Baldocchi, D.%
\BCBL {}\ \BBA {} Katul, G\BPBI G.%
\end{APACrefauthors}%
\unskip\
\newblock
\APACrefYearMonthDay{2010}{{\APACmonth{09}}}{}.
\newblock
{\BBOQ}\APACrefatitle {Scaling {Properties} of {Biologically} {Active} {Scalar} {Concentration} {Fluctuations} in the {Atmospheric} {Surface} {Layer} over a {Managed} {Peatland}} {Scaling {Properties} of {Biologically} {Active} {Scalar} {Concentration} {Fluctuations} in the {Atmospheric} {Surface} {Layer} over a {Managed} {Peatland}}.{\BBCQ}
\newblock
\APACjournalVolNumPages{Boundary-Layer Meteorology}{136}{3}{407--430}.
\newblock
\begin{APACrefURL} [{2024-03-17}]\url{http://link.springer.com/10.1007/s10546-010-9514-z} \end{APACrefURL}
\newblock
\begin{APACrefDOI} \doi{10.1007/s10546-010-9514-z} \end{APACrefDOI}
\PrintBackRefs{\CurrentBib}

\bibitem [\protect \citeauthoryear {%
Detto%
, Katul%
, Mancini%
, Montaldo%
\BCBL {}\ \BBA {} Albertson%
}{%
Detto%
\ \protect \BOthers {.}}{%
{\protect \APACyear {2008}}%
}]{%
detto_surface_2008}
\APACinsertmetastar {%
detto_surface_2008}%
\begin{APACrefauthors}%
Detto, M.%
, Katul, G.%
, Mancini, M.%
, Montaldo, N.%
\BCBL {}\ \BBA {} Albertson, J.%
\end{APACrefauthors}%
\unskip\
\newblock
\APACrefYearMonthDay{2008}{{\APACmonth{06}}}{}.
\newblock
{\BBOQ}\APACrefatitle {Surface heterogeneity and its signature in higher-order scalar similarity relationships} {Surface heterogeneity and its signature in higher-order scalar similarity relationships}.{\BBCQ}
\newblock
\APACjournalVolNumPages{Agricultural and Forest Meteorology}{148}{6-7}{902--916}.
\newblock
\begin{APACrefURL} [{2023-11-13}]\url{https://linkinghub.elsevier.com/retrieve/pii/S0168192307003188} \end{APACrefURL}
\newblock
\begin{APACrefDOI} \doi{10.1016/j.agrformet.2007.12.008} \end{APACrefDOI}
\PrintBackRefs{\CurrentBib}

\bibitem [\protect \citeauthoryear {%
Falocchi%
, Giovannini%
, Franceschi%
\BCBL {}\ \BBA {} Zardi%
}{%
Falocchi%
\ \protect \BOthers {.}}{%
{\protect \APACyear {2018}}%
}]{%
falocchi_refinement_2018}
\APACinsertmetastar {%
falocchi_refinement_2018}%
\begin{APACrefauthors}%
Falocchi, M.%
, Giovannini, L.%
, Franceschi, M\BPBI D.%
\BCBL {}\ \BBA {} Zardi, D.%
\end{APACrefauthors}%
\unskip\
\newblock
\APACrefYearMonthDay{2018}{{\APACmonth{09}}}{}.
\newblock
{\BBOQ}\APACrefatitle {A {Refinement} of the {McMillen} (1988) {Recursive} {Digital} {Filter} for the {Analysis} of {Atmospheric} {Turbulence}} {A {Refinement} of the {McMillen} (1988) {Recursive} {Digital} {Filter} for the {Analysis} of {Atmospheric} {Turbulence}}.{\BBCQ}
\newblock
\APACjournalVolNumPages{Boundary-Layer Meteorology}{168}{3}{517--523}.
\newblock
\begin{APACrefURL} [{2023-11-15}]\url{http://link.springer.com/10.1007/s10546-018-0355-5} \end{APACrefURL}
\newblock
\begin{APACrefDOI} \doi{10.1007/s10546-018-0355-5} \end{APACrefDOI}
\PrintBackRefs{\CurrentBib}

\bibitem [\protect \citeauthoryear {%
Finnigan%
\ \protect \BOthers {.}}{%
Finnigan%
\ \protect \BOthers {.}}{%
{\protect \APACyear {2020}}%
}]{%
finnigan_boundary-layer_2020}
\APACinsertmetastar {%
finnigan_boundary-layer_2020}%
\begin{APACrefauthors}%
Finnigan, J.%
, Ayotte, K.%
, Harman, I.%
, Katul, G.%
, Oldroyd, H.%
, Patton, E.%
\BDBL {}Taylor, P.%
\end{APACrefauthors}%
\unskip\
\newblock
\APACrefYearMonthDay{2020}{{\APACmonth{12}}}{}.
\newblock
{\BBOQ}\APACrefatitle {Boundary-{Layer} {Flow} {Over} {Complex} {Topography}} {Boundary-{Layer} {Flow} {Over} {Complex} {Topography}}.{\BBCQ}
\newblock
\APACjournalVolNumPages{Boundary-Layer Meteorology}{177}{2-3}{247--313}.
\newblock
\begin{APACrefURL} [{2023-11-15}]\url{https://link.springer.com/10.1007/s10546-020-00564-3} \end{APACrefURL}
\newblock
\begin{APACrefDOI} \doi{10.1007/s10546-020-00564-3} \end{APACrefDOI}
\PrintBackRefs{\CurrentBib}

\bibitem [\protect \citeauthoryear {%
Fischer%
\ \protect \BOthers {.}}{%
Fischer%
\ \protect \BOthers {.}}{%
{\protect \APACyear {2023}}%
}]{%
fischer_merging_2023}
\APACinsertmetastar {%
fischer_merging_2023}%
\begin{APACrefauthors}%
Fischer, M.%
, Katul, G.%
, Noormets, A.%
, Pozníková, G.%
, Domec, J\BHBI C.%
, Orság, M.%
\BDBL {}King, J\BPBI S.%
\end{APACrefauthors}%
\unskip\
\newblock
\APACrefYearMonthDay{2023}{{\APACmonth{11}}}{}.
\newblock
{\BBOQ}\APACrefatitle {Merging flux-variance with surface renewal methods in the roughness sublayer and the atmospheric surface layer} {Merging flux-variance with surface renewal methods in the roughness sublayer and the atmospheric surface layer}.{\BBCQ}
\newblock
\APACjournalVolNumPages{Agricultural and Forest Meteorology}{342}{}{109692}.
\newblock
\begin{APACrefURL} [{2025-02-04}]\url{https://linkinghub.elsevier.com/retrieve/pii/S0168192323003829} \end{APACrefURL}
\newblock
\begin{APACrefDOI} \doi{10.1016/j.agrformet.2023.109692} \end{APACrefDOI}
\PrintBackRefs{\CurrentBib}

\bibitem [\protect \citeauthoryear {%
Foken%
}{%
Foken%
}{%
{\protect \APACyear {2006}}%
}]{%
foken_50_2006}
\APACinsertmetastar {%
foken_50_2006}%
\begin{APACrefauthors}%
Foken, T.%
\end{APACrefauthors}%
\unskip\
\newblock
\APACrefYearMonthDay{2006}{{\APACmonth{06}}}{}.
\newblock
{\BBOQ}\APACrefatitle {50 {Years} of the {Monin}–{Obukhov} {Similarity} {Theory}} {50 {Years} of the {Monin}–{Obukhov} {Similarity} {Theory}}.{\BBCQ}
\newblock
\APACjournalVolNumPages{Boundary-Layer Meteorology}{119}{3}{431--447}.
\newblock
\begin{APACrefURL} [{2023-09-14}]\url{http://link.springer.com/10.1007/s10546-006-9048-6} \end{APACrefURL}
\newblock
\begin{APACrefDOI} \doi{10.1007/s10546-006-9048-6} \end{APACrefDOI}
\PrintBackRefs{\CurrentBib}

\bibitem [\protect \citeauthoryear {%
Foken%
\ \BBA {} Wichura%
}{%
Foken%
\ \BBA {} Wichura%
}{%
{\protect \APACyear {1996}}%
}]{%
foken_tools_1996}
\APACinsertmetastar {%
foken_tools_1996}%
\begin{APACrefauthors}%
Foken, T.%
\BCBT {}\ \BBA {} Wichura, B.%
\end{APACrefauthors}%
\unskip\
\newblock
\APACrefYearMonthDay{1996}{{\APACmonth{01}}}{}.
\newblock
{\BBOQ}\APACrefatitle {Tools for quality assessment of surface-based flux measurements} {Tools for quality assessment of surface-based flux measurements}.{\BBCQ}
\newblock
\APACjournalVolNumPages{Agricultural and Forest Meteorology}{78}{1-2}{83--105}.
\newblock
\begin{APACrefURL} [{2024-12-30}]\url{https://linkinghub.elsevier.com/retrieve/pii/0168192395022481} \end{APACrefURL}
\newblock
\begin{APACrefDOI} \doi{10.1016/0168-1923(95)02248-1} \end{APACrefDOI}
\PrintBackRefs{\CurrentBib}

\bibitem [\protect \citeauthoryear {%
Gao%
}{%
Gao%
}{%
{\protect \APACyear {1996}}%
}]{%
gao_ndwinormalized_1996}
\APACinsertmetastar {%
gao_ndwinormalized_1996}%
\begin{APACrefauthors}%
Gao, B\BHBI c.%
\end{APACrefauthors}%
\unskip\
\newblock
\APACrefYearMonthDay{1996}{{\APACmonth{12}}}{}.
\newblock
{\BBOQ}\APACrefatitle {{NDWI}—{A} normalized difference water index for remote sensing of vegetation liquid water from space} {{NDWI}—{A} normalized difference water index for remote sensing of vegetation liquid water from space}.{\BBCQ}
\newblock
\APACjournalVolNumPages{Remote Sensing of Environment}{58}{3}{257--266}.
\newblock
\begin{APACrefURL} [{2025-09-04}]\url{https://linkinghub.elsevier.com/retrieve/pii/S0034425796000673} \end{APACrefURL}
\newblock
\begin{APACrefDOI} \doi{10.1016/S0034-4257(96)00067-3} \end{APACrefDOI}
\PrintBackRefs{\CurrentBib}

\bibitem [\protect \citeauthoryear {%
Ghannam%
, Katul%
, Bou-Zeid%
, Gerken%
\BCBL {}\ \BBA {} Chamecki%
}{%
Ghannam%
\ \protect \BOthers {.}}{%
{\protect \APACyear {2018}}%
}]{%
ghannam_scaling_2018}
\APACinsertmetastar {%
ghannam_scaling_2018}%
\begin{APACrefauthors}%
Ghannam, K.%
, Katul, G\BPBI G.%
, Bou-Zeid, E.%
, Gerken, T.%
\BCBL {}\ \BBA {} Chamecki, M.%
\end{APACrefauthors}%
\unskip\
\newblock
\APACrefYearMonthDay{2018}{{\APACmonth{03}}}{}.
\newblock
{\BBOQ}\APACrefatitle {Scaling and {Similarity} of the {Anisotropic} {Coherent} {Eddies} in {Near}-{Surface} {Atmospheric} {Turbulence}} {Scaling and {Similarity} of the {Anisotropic} {Coherent} {Eddies} in {Near}-{Surface} {Atmospheric} {Turbulence}}.{\BBCQ}
\newblock
\APACjournalVolNumPages{Journal of the Atmospheric Sciences}{75}{3}{943--964}.
\newblock
\begin{APACrefURL} [{2025-09-03}]\url{https://journals.ametsoc.org/doi/10.1175/JAS-D-17-0246.1} \end{APACrefURL}
\newblock
\begin{APACrefDOI} \doi{10.1175/JAS-D-17-0246.1} \end{APACrefDOI}
\PrintBackRefs{\CurrentBib}

\bibitem [\protect \citeauthoryear {%
Grachev%
\ \protect \BOthers {.}}{%
Grachev%
\ \protect \BOthers {.}}{%
{\protect \APACyear {2018}}%
}]{%
grachev_airsealand_2018}
\APACinsertmetastar {%
grachev_airsealand_2018}%
\begin{APACrefauthors}%
Grachev, A\BPBI A.%
, Leo, L\BPBI S.%
, Fernando, H\BPBI J\BPBI S.%
, Fairall, C\BPBI W.%
, Creegan, E.%
, Blomquist, B\BPBI W.%
\BDBL {}Hocut, C\BPBI M.%
\end{APACrefauthors}%
\unskip\
\newblock
\APACrefYearMonthDay{2018}{{\APACmonth{05}}}{}.
\newblock
{\BBOQ}\APACrefatitle {Air–{Sea}/{Land} {Interaction} in the {Coastal} {Zone}} {Air–{Sea}/{Land} {Interaction} in the {Coastal} {Zone}}.{\BBCQ}
\newblock
\APACjournalVolNumPages{Boundary-Layer Meteorology}{167}{2}{181--210}.
\newblock
\begin{APACrefURL} [{2025-09-15}]\url{https://doi.org/10.1007/s10546-017-0326-2} \end{APACrefURL}
\newblock
\begin{APACrefDOI} \doi{10.1007/s10546-017-0326-2} \end{APACrefDOI}
\PrintBackRefs{\CurrentBib}

\bibitem [\protect \citeauthoryear {%
Gucci%
, Giovannini%
, Stiperski%
, Zardi%
\BCBL {}\ \BBA {} Vercauteren%
}{%
Gucci%
\ \protect \BOthers {.}}{%
{\protect \APACyear {2023}}%
}]{%
gucci_sources_2023}
\APACinsertmetastar {%
gucci_sources_2023}%
\begin{APACrefauthors}%
Gucci, F.%
, Giovannini, L.%
, Stiperski, I.%
, Zardi, D.%
\BCBL {}\ \BBA {} Vercauteren, N.%
\end{APACrefauthors}%
\unskip\
\newblock
\APACrefYearMonthDay{2023}{}{}.
\newblock
{\BBOQ}\APACrefatitle {Sources of anisotropy in the {Reynolds} stress tensor in the stable boundary layer} {Sources of anisotropy in the {Reynolds} stress tensor in the stable boundary layer}.{\BBCQ}
\newblock
\APACjournalVolNumPages{Quarterly Journal of the Royal Meteorological Society}{149}{750}{277--299}.
\newblock
\begin{APACrefURL} [{2025-09-15}]\url{https://onlinelibrary.wiley.com/doi/abs/10.1002/qj.4407} \end{APACrefURL}
\newblock
\APACrefnote{\_eprint: https://rmets.onlinelibrary.wiley.com/doi/pdf/10.1002/qj.4407}
\newblock
\begin{APACrefDOI} \doi{10.1002/qj.4407} \end{APACrefDOI}
\PrintBackRefs{\CurrentBib}

\bibitem [\protect \citeauthoryear {%
Jin%
\ \protect \BOthers {.}}{%
Jin%
\ \protect \BOthers {.}}{%
{\protect \APACyear {2019}}%
}]{%
jin_overall_2019}
\APACinsertmetastar {%
jin_overall_2019}%
\begin{APACrefauthors}%
Jin, S.%
, Homer, C.%
, Yang, L.%
, Danielson, P.%
, Dewitz, J.%
, Li, C.%
\BDBL {}Howard, D.%
\end{APACrefauthors}%
\unskip\
\newblock
\APACrefYearMonthDay{2019}{{\APACmonth{12}}}{}.
\newblock
{\BBOQ}\APACrefatitle {Overall {Methodology} {Design} for the {United} {States} {National} {Land} {Cover} {Database} 2016 {Products}} {Overall {Methodology} {Design} for the {United} {States} {National} {Land} {Cover} {Database} 2016 {Products}}.{\BBCQ}
\newblock
\APACjournalVolNumPages{Remote Sensing}{11}{24}{2971}.
\newblock
\begin{APACrefURL} [{2023-11-15}]\url{https://www.mdpi.com/2072-4292/11/24/2971} \end{APACrefURL}
\newblock
\begin{APACrefDOI} \doi{10.3390/rs11242971} \end{APACrefDOI}
\PrintBackRefs{\CurrentBib}

\bibitem [\protect \citeauthoryear {%
Kroon%
\ \BBA {} De~Bruin%
}{%
Kroon%
\ \BBA {} De~Bruin%
}{%
{\protect \APACyear {1995}}%
}]{%
kroon_crau_1995}
\APACinsertmetastar {%
kroon_crau_1995}%
\begin{APACrefauthors}%
Kroon, L.%
\BCBT {}\ \BBA {} De~Bruin, H.%
\end{APACrefauthors}%
\unskip\
\newblock
\APACrefYearMonthDay{1995}{{\APACmonth{04}}}{}.
\newblock
{\BBOQ}\APACrefatitle {The {Crau} field experiment: turbulent exchange in the surface layer under conditions of strong local advection} {The {Crau} field experiment: turbulent exchange in the surface layer under conditions of strong local advection}.{\BBCQ}
\newblock
\APACjournalVolNumPages{Journal of Hydrology}{166}{3-4}{327--351}.
\newblock
\begin{APACrefURL} [{2023-11-13}]\url{https://linkinghub.elsevier.com/retrieve/pii/002216949405092C} \end{APACrefURL}
\newblock
\begin{APACrefDOI} \doi{10.1016/0022-1694(94)05092-C} \end{APACrefDOI}
\PrintBackRefs{\CurrentBib}

\bibitem [\protect \citeauthoryear {%
Larson%
}{%
Larson%
}{%
{\protect \APACyear {2022}}%
}]{%
larson_clubb-silhs_2022}
\APACinsertmetastar {%
larson_clubb-silhs_2022}%
\begin{APACrefauthors}%
Larson, V\BPBI E.%
\end{APACrefauthors}%
\unskip\
\newblock
\APACrefYearMonthDay{2022}{{\APACmonth{03}}}{}.
\newblock
\APACrefbtitle {{CLUBB}-{SILHS}: {A} parameterization of subgrid variability in the atmosphere.} {{CLUBB}-{SILHS}: {A} parameterization of subgrid variability in the atmosphere.}
\newblock
\APACaddressPublisher{}{arXiv}.
\newblock
\begin{APACrefURL} [{2023-04-23}]\url{http://arxiv.org/abs/1711.03675} \end{APACrefURL}
\newblock
\APACrefnote{arXiv:1711.03675 [physics]}
\PrintBackRefs{\CurrentBib}

\bibitem [\protect \citeauthoryear {%
Lee%
}{%
Lee%
}{%
{\protect \APACyear {2009}}%
}]{%
lee_influence_2009}
\APACinsertmetastar {%
lee_influence_2009}%
\begin{APACrefauthors}%
Lee, Y\BHBI H.%
\end{APACrefauthors}%
\unskip\
\newblock
\APACrefYearMonthDay{2009}{{\APACmonth{09}}}{}.
\newblock
{\BBOQ}\APACrefatitle {The {Influence} of {Local} {Stability} on {Heat} and {Momentum} {Transfer} within {Open} {Canopies}} {The {Influence} of {Local} {Stability} on {Heat} and {Momentum} {Transfer} within {Open} {Canopies}}.{\BBCQ}
\newblock
\APACjournalVolNumPages{Boundary-Layer Meteorology}{132}{3}{383--399}.
\newblock
\begin{APACrefURL} [{2023-11-13}]\url{http://link.springer.com/10.1007/s10546-009-9405-3} \end{APACrefURL}
\newblock
\begin{APACrefDOI} \doi{10.1007/s10546-009-9405-3} \end{APACrefDOI}
\PrintBackRefs{\CurrentBib}

\bibitem [\protect \citeauthoryear {%
Linden%
\ \protect \BOthers {.}}{%
Linden%
\ \protect \BOthers {.}}{%
{\protect \APACyear {2020}}%
}]{%
linden_businger_2020}
\APACinsertmetastar {%
linden_businger_2020}%
\begin{APACrefauthors}%
Linden, S\BPBI J\BPBI A\BPBI v\BPBI d.%
, Wiel, B\BPBI J\BPBI H\BPBI v\BPBI d.%
, Petenko, I.%
, Heerwaarden, C\BPBI C\BPBI v.%
, Baas, P.%
\BCBL {}\ \BBA {} Jonker, H\BPBI J\BPBI J.%
\end{APACrefauthors}%
\unskip\
\newblock
\APACrefYearMonthDay{2020}{{\APACmonth{10}}}{}.
\newblock
{\BBOQ}\APACrefatitle {A {Businger} {Mechanism} for {Intermittent} {Bursting} in the {Stable} {Boundary} {Layer}} {A {Businger} {Mechanism} for {Intermittent} {Bursting} in the {Stable} {Boundary} {Layer}}.{\BBCQ}
\newblock
\APACjournalVolNumPages{Journal of the Atmospheric Sciences}{77}{10}{3343--3360}.
\newblock
\begin{APACrefURL} [{2025-10-31}]\url{https://journals.ametsoc.org/view/journals/atsc/77/10/jasD190309.xml} \end{APACrefURL}
\newblock
\APACrefnote{Publisher: American Meteorological Society Section: Journal of the Atmospheric Sciences}
\newblock
\begin{APACrefDOI} \doi{10.1175/JAS-D-19-0309.1} \end{APACrefDOI}
\PrintBackRefs{\CurrentBib}

\bibitem [\protect \citeauthoryear {%
Lock%
, Brown%
, Bush%
, Martin%
\BCBL {}\ \BBA {} Smith%
}{%
Lock%
\ \protect \BOthers {.}}{%
{\protect \APACyear {2000}}%
}]{%
lock_new_2000}
\APACinsertmetastar {%
lock_new_2000}%
\begin{APACrefauthors}%
Lock, A\BPBI P.%
, Brown, A\BPBI R.%
, Bush, M\BPBI R.%
, Martin, G\BPBI M.%
\BCBL {}\ \BBA {} Smith, R\BPBI N\BPBI B.%
\end{APACrefauthors}%
\unskip\
\newblock
\APACrefYearMonthDay{2000}{{\APACmonth{09}}}{}.
\newblock
{\BBOQ}\APACrefatitle {A {New} {Boundary} {Layer} {Mixing} {Scheme}. {Part} {I}: {Scheme} {Description} and {Single}-{Column} {Model} {Tests}} {A {New} {Boundary} {Layer} {Mixing} {Scheme}. {Part} {I}: {Scheme} {Description} and {Single}-{Column} {Model} {Tests}}.{\BBCQ}
\newblock
\APACjournalVolNumPages{Monthly Weather Review}{128}{9}{3187--3199}.
\newblock
\begin{APACrefURL} [{2025-09-09}]\url{http://journals.ametsoc.org/doi/10.1175/1520-0493(2000)128<3187:ANBLMS>2.0.CO;2} \end{APACrefURL}
\newblock
\begin{APACrefDOI} \doi{10.1175/1520-0493(2000)128<3187:ANBLMS>2.0.CO;2} \end{APACrefDOI}
\PrintBackRefs{\CurrentBib}

\bibitem [\protect \citeauthoryear {%
Lumley%
\ \BBA {} Newman%
}{%
Lumley%
\ \BBA {} Newman%
}{%
{\protect \APACyear {1977}}%
}]{%
lumley_return_1977}
\APACinsertmetastar {%
lumley_return_1977}%
\begin{APACrefauthors}%
Lumley, J\BPBI L.%
\BCBT {}\ \BBA {} Newman, G\BPBI R.%
\end{APACrefauthors}%
\unskip\
\newblock
\APACrefYearMonthDay{1977}{}{}.
\newblock
{\BBOQ}\APACrefatitle {The return to isotropy of homogeneous turbulence} {The return to isotropy of homogeneous turbulence}.{\BBCQ}
\newblock
\APACjournalVolNumPages{Journal of Fluid Mechanics}{82}{1}{161--178}.
\newblock
\begin{APACrefDOI} \doi{10.1017/S0022112077000585} \end{APACrefDOI}
\PrintBackRefs{\CurrentBib}

\bibitem [\protect \citeauthoryear {%
Mahrt%
}{%
Mahrt%
}{%
{\protect \APACyear {1999}}%
}]{%
mahrt_stratified_1999}
\APACinsertmetastar {%
mahrt_stratified_1999}%
\begin{APACrefauthors}%
Mahrt, L.%
\end{APACrefauthors}%
\unskip\
\newblock
\APACrefYearMonthDay{1999}{{\APACmonth{03}}}{}.
\newblock
{\BBOQ}\APACrefatitle {Stratified {Atmospheric} {Boundary} {Layers}} {Stratified {Atmospheric} {Boundary} {Layers}}.{\BBCQ}
\newblock
\APACjournalVolNumPages{Boundary-Layer Meteorology}{90}{3}{375--396}.
\newblock
\begin{APACrefURL} [{2023-11-15}]\url{http://link.springer.com/10.1023/A:1001765727956} \end{APACrefURL}
\newblock
\begin{APACrefDOI} \doi{10.1023/A:1001765727956} \end{APACrefDOI}
\PrintBackRefs{\CurrentBib}

\bibitem [\protect \citeauthoryear {%
Mahrt%
\ \BBA {} Bou-Zeid%
}{%
Mahrt%
\ \BBA {} Bou-Zeid%
}{%
{\protect \APACyear {2020}}%
}]{%
mahrt_non-stationary_2020}
\APACinsertmetastar {%
mahrt_non-stationary_2020}%
\begin{APACrefauthors}%
Mahrt, L.%
\BCBT {}\ \BBA {} Bou-Zeid, E.%
\end{APACrefauthors}%
\unskip\
\newblock
\APACrefYearMonthDay{2020}{{\APACmonth{12}}}{}.
\newblock
{\BBOQ}\APACrefatitle {Non-stationary {Boundary} {Layers}} {Non-stationary {Boundary} {Layers}}.{\BBCQ}
\newblock
\APACjournalVolNumPages{Boundary-Layer Meteorology}{177}{2-3}{189--204}.
\newblock
\begin{APACrefURL} [{2025-09-03}]\url{https://link.springer.com/10.1007/s10546-020-00533-w} \end{APACrefURL}
\newblock
\begin{APACrefDOI} \doi{10.1007/s10546-020-00533-w} \end{APACrefDOI}
\PrintBackRefs{\CurrentBib}

\bibitem [\protect \citeauthoryear {%
Martins%
, Moraes%
, Acevedo%
\BCBL {}\ \BBA {} Degrazia%
}{%
Martins%
\ \protect \BOthers {.}}{%
{\protect \APACyear {2009}}%
}]{%
martins_turbulence_2009}
\APACinsertmetastar {%
martins_turbulence_2009}%
\begin{APACrefauthors}%
Martins, C\BPBI A.%
, Moraes, O\BPBI L\BPBI L.%
, Acevedo, O\BPBI C.%
\BCBL {}\ \BBA {} Degrazia, G\BPBI A.%
\end{APACrefauthors}%
\unskip\
\newblock
\APACrefYearMonthDay{2009}{{\APACmonth{10}}}{}.
\newblock
{\BBOQ}\APACrefatitle {Turbulence {Intensity} {Parameters} over a {Very} {Complex} {Terrain}} {Turbulence {Intensity} {Parameters} over a {Very} {Complex} {Terrain}}.{\BBCQ}
\newblock
\APACjournalVolNumPages{Boundary-Layer Meteorology}{133}{1}{35--45}.
\newblock
\begin{APACrefURL} [{2023-11-15}]\url{http://link.springer.com/10.1007/s10546-009-9413-3} \end{APACrefURL}
\newblock
\begin{APACrefDOI} \doi{10.1007/s10546-009-9413-3} \end{APACrefDOI}
\PrintBackRefs{\CurrentBib}

\bibitem [\protect \citeauthoryear {%
Metzger%
\ \protect \BOthers {.}}{%
Metzger%
\ \protect \BOthers {.}}{%
{\protect \APACyear {2019}}%
}]{%
metzger_neon_2019}
\APACinsertmetastar {%
metzger_neon_2019}%
\begin{APACrefauthors}%
Metzger, S.%
, Ayres, E.%
, Durden, D.%
, Florian, C.%
, Lee, R.%
, Lunch, C.%
\BDBL {}Zulueta, R\BPBI C.%
\end{APACrefauthors}%
\unskip\
\newblock
\APACrefYearMonthDay{2019}{{\APACmonth{11}}}{}.
\newblock
{\BBOQ}\APACrefatitle {From {NEON} {Field} {Sites} to {Data} {Portal}: {A} {Community} {Resource} for {Surface}–{Atmosphere} {Research} {Comes} {Online}} {From {NEON} {Field} {Sites} to {Data} {Portal}: {A} {Community} {Resource} for {Surface}–{Atmosphere} {Research} {Comes} {Online}}.{\BBCQ}
\newblock
\APACjournalVolNumPages{Bulletin of the American Meteorological Society}{100}{11}{2305--2325}.
\newblock
\begin{APACrefURL} [{2025-03-26}]\url{https://journals.ametsoc.org/view/journals/bams/100/11/bams-d-17-0307.1.xml} \end{APACrefURL}
\newblock
\begin{APACrefDOI} \doi{10.1175/BAMS-D-17-0307.1} \end{APACrefDOI}
\PrintBackRefs{\CurrentBib}

\bibitem [\protect \citeauthoryear {%
Metzger%
\ \protect \BOthers {.}}{%
Metzger%
\ \protect \BOthers {.}}{%
{\protect \APACyear {2022}}%
}]{%
metzger_neon_2022}
\APACinsertmetastar {%
metzger_neon_2022}%
\begin{APACrefauthors}%
Metzger, S.%
, Durden, D.%
, Xu, K.%
, Pingintha-Durden, N.%
, Luo, H.%
, Florian, C.%
\BCBL {}\ \BBA {} Thibault, K.%
\end{APACrefauthors}%
\unskip\
\newblock
\APACrefYearMonthDay{2022}{}{}.
\newblock
{\BBOQ}\APACrefatitle {{NEON} {ALGORITHMTHEORETICAL} {BASIS} {DOCUMENT} ({ATBD}): {EDDY}-{COVARIANCE} {DATA} {PRODUCTS} {BUNDLE}} {{NEON} {ALGORITHMTHEORETICAL} {BASIS} {DOCUMENT} ({ATBD}): {EDDY}-{COVARIANCE} {DATA} {PRODUCTS} {BUNDLE}}.{\BBCQ}
\newblock

\PrintBackRefs{\CurrentBib}

\bibitem [\protect \citeauthoryear {%
Monin%
\ \BBA {} Obukhov%
}{%
Monin%
\ \BBA {} Obukhov%
}{%
{\protect \APACyear {1954}}%
}]{%
monin_basic_1954}
\APACinsertmetastar {%
monin_basic_1954}%
\begin{APACrefauthors}%
Monin, A\BPBI S.%
\BCBT {}\ \BBA {} Obukhov, A\BPBI M.%
\end{APACrefauthors}%
\unskip\
\newblock
\APACrefYearMonthDay{1954}{}{}.
\newblock
{\BBOQ}\APACrefatitle {Basic laws of turbulent mixing in the surface layer of the atmosphere} {Basic laws of turbulent mixing in the surface layer of the atmosphere}.{\BBCQ}
\newblock
\APACjournalVolNumPages{Tr. Akad. Nauk SSSR Geophiz.}{}{}{}.
\PrintBackRefs{\CurrentBib}

\bibitem [\protect \citeauthoryear {%
Moraes%
, Acevedo%
, Da~Silva%
, Magnago%
\BCBL {}\ \BBA {} Siqueira%
}{%
Moraes%
\ \protect \BOthers {.}}{%
{\protect \APACyear {2004}}%
}]{%
moraes_nocturnal_2004}
\APACinsertmetastar {%
moraes_nocturnal_2004}%
\begin{APACrefauthors}%
Moraes, O\BPBI L\BPBI L.%
, Acevedo, O\BPBI C.%
, Da~Silva, R.%
, Magnago, R.%
\BCBL {}\ \BBA {} Siqueira, A\BPBI C.%
\end{APACrefauthors}%
\unskip\
\newblock
\APACrefYearMonthDay{2004}{{\APACmonth{07}}}{}.
\newblock
{\BBOQ}\APACrefatitle {Nocturnal {Surface}-{Layer} {Characteristics} at the {Bottom} of a {Valley}} {Nocturnal {Surface}-{Layer} {Characteristics} at the {Bottom} of a {Valley}}.{\BBCQ}
\newblock
\APACjournalVolNumPages{Boundary-Layer Meteorology}{112}{1}{159--177}.
\newblock
\begin{APACrefURL} [{2023-11-15}]\url{http://link.springer.com/10.1023/B:BOUN.0000020163.36907.f9} \end{APACrefURL}
\newblock
\begin{APACrefDOI} \doi{10.1023/B:BOUN.0000020163.36907.f9} \end{APACrefDOI}
\PrintBackRefs{\CurrentBib}

\bibitem [\protect \citeauthoryear {%
Mosso%
, Calaf%
\BCBL {}\ \BBA {} Stiperski%
}{%
Mosso%
\ \protect \BOthers {.}}{%
{\protect \APACyear {2024}}%
}]{%
mosso_fluxgradient_2024}
\APACinsertmetastar {%
mosso_fluxgradient_2024}%
\begin{APACrefauthors}%
Mosso, S.%
, Calaf, M.%
\BCBL {}\ \BBA {} Stiperski, I.%
\end{APACrefauthors}%
\unskip\
\newblock
\APACrefYearMonthDay{2024}{{\APACmonth{07}}}{}.
\newblock
{\BBOQ}\APACrefatitle {Flux‐gradient relations and their dependence on turbulence anisotropy} {Flux‐gradient relations and their dependence on turbulence anisotropy}.{\BBCQ}
\newblock
\APACjournalVolNumPages{Quarterly Journal of the Royal Meteorological Society}{150}{763}{3346--3367}.
\newblock
\begin{APACrefURL} [{2025-01-07}]\url{https://rmets.onlinelibrary.wiley.com/doi/10.1002/qj.4762} \end{APACrefURL}
\newblock
\begin{APACrefDOI} \doi{10.1002/qj.4762} \end{APACrefDOI}
\PrintBackRefs{\CurrentBib}

\bibitem [\protect \citeauthoryear {%
Mosso%
, Lapo%
\BCBL {}\ \BBA {} Stiperski%
}{%
Mosso%
\ \protect \BOthers {.}}{%
{\protect \APACyear {2025}}%
}]{%
mosso_revealing_2025}
\APACinsertmetastar {%
mosso_revealing_2025}%
\begin{APACrefauthors}%
Mosso, S.%
, Lapo, K.%
\BCBL {}\ \BBA {} Stiperski, I.%
\end{APACrefauthors}%
\unskip\
\newblock
\APACrefYearMonthDay{2025}{{\APACmonth{03}}}{}.
\newblock
\APACrefbtitle {Revealing the drivers of turbulence anisotropy over flat and complex terrain: an interpretable machine learning approach.} {Revealing the drivers of turbulence anisotropy over flat and complex terrain: an interpretable machine learning approach.}
\newblock
\APACaddressPublisher{}{arXiv}.
\newblock
\begin{APACrefURL} [{2025-05-06}]\url{http://arxiv.org/abs/2503.15276} \end{APACrefURL}
\newblock
\APACrefnote{arXiv:2503.15276 [physics]}
\newblock
\begin{APACrefDOI} \doi{10.48550/arXiv.2503.15276} \end{APACrefDOI}
\PrintBackRefs{\CurrentBib}

\bibitem [\protect \citeauthoryear {%
Nadeau%
, Pardyjak%
, Higgins%
\BCBL {}\ \BBA {} Parlange%
}{%
Nadeau%
\ \protect \BOthers {.}}{%
{\protect \APACyear {2013}}%
}]{%
nadeau_similarity_2013}
\APACinsertmetastar {%
nadeau_similarity_2013}%
\begin{APACrefauthors}%
Nadeau, D\BPBI F.%
, Pardyjak, E\BPBI R.%
, Higgins, C\BPBI W.%
\BCBL {}\ \BBA {} Parlange, M\BPBI B.%
\end{APACrefauthors}%
\unskip\
\newblock
\APACrefYearMonthDay{2013}{{\APACmonth{06}}}{}.
\newblock
{\BBOQ}\APACrefatitle {Similarity {Scaling} {Over} a {Steep} {Alpine} {Slope}} {Similarity {Scaling} {Over} a {Steep} {Alpine} {Slope}}.{\BBCQ}
\newblock
\APACjournalVolNumPages{Boundary-Layer Meteorology}{147}{3}{401--419}.
\newblock
\begin{APACrefURL} [{2023-11-15}]\url{http://link.springer.com/10.1007/s10546-012-9787-5} \end{APACrefURL}
\newblock
\begin{APACrefDOI} \doi{10.1007/s10546-012-9787-5} \end{APACrefDOI}
\PrintBackRefs{\CurrentBib}

\bibitem [\protect \citeauthoryear {%
Nakanishi%
\ \BBA {} Niino%
}{%
Nakanishi%
\ \BBA {} Niino%
}{%
{\protect \APACyear {2009}}%
}]{%
nakanishi_development_2009}
\APACinsertmetastar {%
nakanishi_development_2009}%
\begin{APACrefauthors}%
Nakanishi, M.%
\BCBT {}\ \BBA {} Niino, H.%
\end{APACrefauthors}%
\unskip\
\newblock
\APACrefYearMonthDay{2009}{}{}.
\newblock
{\BBOQ}\APACrefatitle {Development of an {Improved} {Turbulence} {Closure} {Model} for the {Atmospheric} {Boundary} {Layer}} {Development of an {Improved} {Turbulence} {Closure} {Model} for the {Atmospheric} {Boundary} {Layer}}.{\BBCQ}
\newblock
\APACjournalVolNumPages{Journal of the Meteorological Society of Japan. Ser. II}{87}{5}{895--912}.
\newblock
\begin{APACrefURL} [{2024-03-01}]\url{http://www.jstage.jst.go.jp/article/jmsj/87/5/87_5_895/_article} \end{APACrefURL}
\newblock
\begin{APACrefDOI} \doi{10.2151/jmsj.87.895} \end{APACrefDOI}
\PrintBackRefs{\CurrentBib}

\bibitem [\protect \citeauthoryear {%
{National Ecological Observatory Network (NEON)}%
}{%
{National Ecological Observatory Network (NEON)}%
}{%
{\protect \APACyear {2025}}%
{\protect \APACexlab {{\protect \BCnt {1}}}}}]{%
national_ecological_observatory_network_neon_albedo_2025}
\APACinsertmetastar {%
national_ecological_observatory_network_neon_albedo_2025}%
\begin{APACrefauthors}%
{National Ecological Observatory Network (NEON)}.%
\end{APACrefauthors}%
\unskip\
\newblock
\APACrefYearMonthDay{2025{\protect \BCnt {1}}}{}{}.
\newblock
\APACrefbtitle {Albedo - spectrometer - mosaic ({DP3}.30011.001).} {Albedo - spectrometer - mosaic ({DP3}.30011.001).}
\newblock
\APACaddressPublisher{}{National Ecological Observatory Network (NEON)}.
\newblock
\begin{APACrefURL} \url{https://data.neonscience.org/data-products/DP3.30011.001/RELEASE-2025} \end{APACrefURL}
\newblock
\begin{APACrefDOI} \doi{10.48443/PK80-Z448} \end{APACrefDOI}
\PrintBackRefs{\CurrentBib}

\bibitem [\protect \citeauthoryear {%
{National Ecological Observatory Network (NEON)}%
}{%
{National Ecological Observatory Network (NEON)}%
}{%
{\protect \APACyear {2025}}%
{\protect \APACexlab {{\protect \BCnt {2}}}}}]{%
national_ecological_observatory_network_neon_bundled_2025}
\APACinsertmetastar {%
national_ecological_observatory_network_neon_bundled_2025}%
\begin{APACrefauthors}%
{National Ecological Observatory Network (NEON)}.%
\end{APACrefauthors}%
\unskip\
\newblock
\APACrefYearMonthDay{2025{\protect \BCnt {2}}}{}{}.
\newblock
\APACrefbtitle {Bundled data products - eddy covariance ({DP4}.00200.001).} {Bundled data products - eddy covariance ({DP4}.00200.001).}
\newblock
\APACaddressPublisher{}{National Ecological Observatory Network (NEON)}.
\newblock
\begin{APACrefURL} \url{https://data.neonscience.org/data-products/DP4.00200.001/RELEASE-2025} \end{APACrefURL}
\newblock
\begin{APACrefDOI} \doi{10.48443/R7ZP-Y487} \end{APACrefDOI}
\PrintBackRefs{\CurrentBib}

\bibitem [\protect \citeauthoryear {%
{National Ecological Observatory Network (NEON)}%
}{%
{National Ecological Observatory Network (NEON)}%
}{%
{\protect \APACyear {2025}}%
{\protect \APACexlab {{\protect \BCnt {3}}}}}]{%
national_ecological_observatory_network_neon_canopy_2025-1}
\APACinsertmetastar {%
national_ecological_observatory_network_neon_canopy_2025-1}%
\begin{APACrefauthors}%
{National Ecological Observatory Network (NEON)}.%
\end{APACrefauthors}%
\unskip\
\newblock
\APACrefYearMonthDay{2025{\protect \BCnt {3}}}{}{}.
\newblock
\APACrefbtitle {Canopy water indices - bidirectional mosaic ({DP3}.30019.002).} {Canopy water indices - bidirectional mosaic ({DP3}.30019.002).}
\newblock
\APACaddressPublisher{}{National Ecological Observatory Network (NEON)}.
\newblock
\begin{APACrefURL} \url{https://data.neonscience.org/data-products/DP3.30019.002} \end{APACrefURL}
\PrintBackRefs{\CurrentBib}

\bibitem [\protect \citeauthoryear {%
{National Ecological Observatory Network (NEON)}%
}{%
{National Ecological Observatory Network (NEON)}%
}{%
{\protect \APACyear {2025}}%
{\protect \APACexlab {{\protect \BCnt {4}}}}}]{%
national_ecological_observatory_network_neon_canopy_2025}
\APACinsertmetastar {%
national_ecological_observatory_network_neon_canopy_2025}%
\begin{APACrefauthors}%
{National Ecological Observatory Network (NEON)}.%
\end{APACrefauthors}%
\unskip\
\newblock
\APACrefYearMonthDay{2025{\protect \BCnt {4}}}{}{}.
\newblock
\APACrefbtitle {Canopy water indices - mosaic ({DP3}.30019.001).} {Canopy water indices - mosaic ({DP3}.30019.001).}
\newblock
\APACaddressPublisher{}{National Ecological Observatory Network (NEON)}.
\newblock
\begin{APACrefURL} \url{https://data.neonscience.org/data-products/DP3.30019.001/RELEASE-2025} \end{APACrefURL}
\newblock
\begin{APACrefDOI} \doi{10.48443/C1HC-MF71} \end{APACrefDOI}
\PrintBackRefs{\CurrentBib}

\bibitem [\protect \citeauthoryear {%
{National Ecological Observatory Network (NEON)}%
}{%
{National Ecological Observatory Network (NEON)}%
}{%
{\protect \APACyear {2025}}%
{\protect \APACexlab {{\protect \BCnt {5}}}}}]{%
national_ecological_observatory_network_neon_lai_2025-1}
\APACinsertmetastar {%
national_ecological_observatory_network_neon_lai_2025-1}%
\begin{APACrefauthors}%
{National Ecological Observatory Network (NEON)}.%
\end{APACrefauthors}%
\unskip\
\newblock
\APACrefYearMonthDay{2025{\protect \BCnt {5}}}{}{}.
\newblock
\APACrefbtitle {{LAI} - spectrometer - bidirectional mosaic ({DP3}.30012.002).} {{LAI} - spectrometer - bidirectional mosaic ({DP3}.30012.002).}
\newblock
\APACaddressPublisher{}{National Ecological Observatory Network (NEON)}.
\newblock
\begin{APACrefURL} \url{https://data.neonscience.org/data-products/DP3.30012.002} \end{APACrefURL}
\PrintBackRefs{\CurrentBib}

\bibitem [\protect \citeauthoryear {%
{National Ecological Observatory Network (NEON)}%
}{%
{National Ecological Observatory Network (NEON)}%
}{%
{\protect \APACyear {2025}}%
{\protect \APACexlab {{\protect \BCnt {6}}}}}]{%
national_ecological_observatory_network_neon_lai_2025}
\APACinsertmetastar {%
national_ecological_observatory_network_neon_lai_2025}%
\begin{APACrefauthors}%
{National Ecological Observatory Network (NEON)}.%
\end{APACrefauthors}%
\unskip\
\newblock
\APACrefYearMonthDay{2025{\protect \BCnt {6}}}{}{}.
\newblock
\APACrefbtitle {{LAI} - spectrometer - mosaic ({DP3}.30012.001).} {{LAI} - spectrometer - mosaic ({DP3}.30012.001).}
\newblock
\APACaddressPublisher{}{National Ecological Observatory Network (NEON)}.
\newblock
\begin{APACrefURL} \url{https://data.neonscience.org/data-products/DP3.30012.001/RELEASE-2025} \end{APACrefURL}
\newblock
\begin{APACrefDOI} \doi{10.48443/PWFV-2J87} \end{APACrefDOI}
\PrintBackRefs{\CurrentBib}

\bibitem [\protect \citeauthoryear {%
{National Ecological Observatory Network (NEON)}%
}{%
{National Ecological Observatory Network (NEON)}%
}{%
{\protect \APACyear {2025}}%
{\protect \APACexlab {{\protect \BCnt {7}}}}}]{%
national_ecological_observatory_network_neon_phenology_2025}
\APACinsertmetastar {%
national_ecological_observatory_network_neon_phenology_2025}%
\begin{APACrefauthors}%
{National Ecological Observatory Network (NEON)}.%
\end{APACrefauthors}%
\unskip\
\newblock
\APACrefYearMonthDay{2025{\protect \BCnt {7}}}{}{}.
\newblock
\APACrefbtitle {Phenology images ({DP1}.00033.001).} {Phenology images ({DP1}.00033.001).}
\newblock
\APACaddressPublisher{}{National Ecological Observatory Network (NEON)}.
\newblock
\begin{APACrefURL} \url{https://data.neonscience.org/data-products/DP1.00033.001} \end{APACrefURL}
\PrintBackRefs{\CurrentBib}

\bibitem [\protect \citeauthoryear {%
Ohtaki%
}{%
Ohtaki%
}{%
{\protect \APACyear {1985}}%
}]{%
ohtaki_similarity_1985}
\APACinsertmetastar {%
ohtaki_similarity_1985}%
\begin{APACrefauthors}%
Ohtaki, E.%
\end{APACrefauthors}%
\unskip\
\newblock
\APACrefYearMonthDay{1985}{{\APACmonth{05}}}{}.
\newblock
{\BBOQ}\APACrefatitle {On the similarity in atmospheric fluctuations of carbon dioxide, water vapor and temperature over vegetated fields} {On the similarity in atmospheric fluctuations of carbon dioxide, water vapor and temperature over vegetated fields}.{\BBCQ}
\newblock
\APACjournalVolNumPages{Boundary-Layer Meteorology}{32}{1}{25--37}.
\newblock
\begin{APACrefURL} [{2024-03-16}]\url{http://link.springer.com/10.1007/BF00120712} \end{APACrefURL}
\newblock
\begin{APACrefDOI} \doi{10.1007/BF00120712} \end{APACrefDOI}
\PrintBackRefs{\CurrentBib}

\bibitem [\protect \citeauthoryear {%
Plotnick%
, Gardner%
, Hargrove%
, Prestegaard%
\BCBL {}\ \BBA {} Perlmutter%
}{%
Plotnick%
\ \protect \BOthers {.}}{%
{\protect \APACyear {1996}}%
}]{%
plotnick_lacunarity_1996}
\APACinsertmetastar {%
plotnick_lacunarity_1996}%
\begin{APACrefauthors}%
Plotnick, R\BPBI E.%
, Gardner, R\BPBI H.%
, Hargrove, W\BPBI W.%
, Prestegaard, K.%
\BCBL {}\ \BBA {} Perlmutter, M.%
\end{APACrefauthors}%
\unskip\
\newblock
\APACrefYearMonthDay{1996}{{\APACmonth{05}}}{}.
\newblock
{\BBOQ}\APACrefatitle {Lacunarity analysis: {A} general technique for the analysis of spatial patterns} {Lacunarity analysis: {A} general technique for the analysis of spatial patterns}.{\BBCQ}
\newblock
\APACjournalVolNumPages{Physical Review E}{53}{5}{5461--5468}.
\newblock
\begin{APACrefURL} [{2025-09-04}]\url{https://link.aps.org/doi/10.1103/PhysRevE.53.5461} \end{APACrefURL}
\newblock
\begin{APACrefDOI} \doi{10.1103/PhysRevE.53.5461} \end{APACrefDOI}
\PrintBackRefs{\CurrentBib}

\bibitem [\protect \citeauthoryear {%
Pope%
}{%
Pope%
}{%
{\protect \APACyear {2000}}%
}]{%
pope_turbulent_2000}
\APACinsertmetastar {%
pope_turbulent_2000}%
\begin{APACrefauthors}%
Pope, S\BPBI B.%
\end{APACrefauthors}%
\unskip\
\newblock
\APACrefYear{2000}.
\newblock
\APACrefbtitle {Turbulent {Flows}} {Turbulent {Flows}}.
\newblock
\APACaddressPublisher{}{Cambridge University Press}.
\PrintBackRefs{\CurrentBib}

\bibitem [\protect \citeauthoryear {%
Ramana%
, Krishnan%
\BCBL {}\ \BBA {} Kunhikrishnan%
}{%
Ramana%
\ \protect \BOthers {.}}{%
{\protect \APACyear {2004}}%
}]{%
ramana_surface_2004}
\APACinsertmetastar {%
ramana_surface_2004}%
\begin{APACrefauthors}%
Ramana, M.%
, Krishnan, P.%
\BCBL {}\ \BBA {} Kunhikrishnan, P\BPBI K.%
\end{APACrefauthors}%
\unskip\
\newblock
\APACrefYearMonthDay{2004}{{\APACmonth{04}}}{}.
\newblock
{\BBOQ}\APACrefatitle {Surface {Boundary}-{Layer} {Characteristics} {Over} a {Tropical} {Inland} {Station}: {Seasonal} {Features}} {Surface {Boundary}-{Layer} {Characteristics} {Over} a {Tropical} {Inland} {Station}: {Seasonal} {Features}}.{\BBCQ}
\newblock
\APACjournalVolNumPages{Boundary-Layer Meteorology}{111}{1}{153--157}.
\newblock
\begin{APACrefURL} [{2024-03-20}]\url{http://link.springer.com/10.1023/B:BOUN.0000010999.25921.1a} \end{APACrefURL}
\newblock
\begin{APACrefDOI} \doi{10.1023/B:BOUN.0000010999.25921.1a} \end{APACrefDOI}
\PrintBackRefs{\CurrentBib}

\bibitem [\protect \citeauthoryear {%
Rannik%
}{%
Rannik%
}{%
{\protect \APACyear {1998}}%
}]{%
rannik_surface_1998}
\APACinsertmetastar {%
rannik_surface_1998}%
\begin{APACrefauthors}%
Rannik, U.%
\end{APACrefauthors}%
\unskip\
\newblock
\APACrefYearMonthDay{1998}{{\APACmonth{04}}}{}.
\newblock
{\BBOQ}\APACrefatitle {On the surface layer similarity at a complex forest site} {On the surface layer similarity at a complex forest site}.{\BBCQ}
\newblock
\APACjournalVolNumPages{Journal of Geophysical Research: Atmospheres}{103}{D8}{8685--8697}.
\newblock
\begin{APACrefURL} [{2024-03-17}]\url{https://agupubs.onlinelibrary.wiley.com/doi/10.1029/98JD00086} \end{APACrefURL}
\newblock
\begin{APACrefDOI} \doi{10.1029/98JD00086} \end{APACrefDOI}
\PrintBackRefs{\CurrentBib}

\bibitem [\protect \citeauthoryear {%
Salmaso%
, Cal%
\BCBL {}\ \BBA {} Calaf%
}{%
Salmaso%
\ \protect \BOthers {.}}{%
{\protect \APACyear {2025}}%
}]{%
salmaso_canopy_2025}
\APACinsertmetastar {%
salmaso_canopy_2025}%
\begin{APACrefauthors}%
Salmaso, G.%
, Cal, R\BPBI B.%
\BCBL {}\ \BBA {} Calaf, M.%
\end{APACrefauthors}%
\unskip\
\newblock
\APACrefYearMonthDay{2025}{{\APACmonth{10}}}{}.
\newblock
\APACrefbtitle {Canopy {Heterogeneity} {Footprints} in the {Roughness} {Sublayer}.} {Canopy {Heterogeneity} {Footprints} in the {Roughness} {Sublayer}.}
\newblock
\APACaddressPublisher{}{Preprints}.
\newblock
\begin{APACrefURL} [{2025-10-31}]\url{https://essopenarchive.org/users/670372/articles/1304551-canopy-heterogeneity-footprints-in-the-roughness-sublayer?commit=2b9ed385f324d5856030582d1c915a830571dfd9} \end{APACrefURL}
\newblock
\begin{APACrefDOI} \doi{10.22541/essoar.175941910.09851123/v1} \end{APACrefDOI}
\PrintBackRefs{\CurrentBib}

\bibitem [\protect \citeauthoryear {%
Sfyri%
\ \protect \BOthers {.}}{%
Sfyri%
\ \protect \BOthers {.}}{%
{\protect \APACyear {2018}}%
}]{%
sfyri_scalar-flux_2018}
\APACinsertmetastar {%
sfyri_scalar-flux_2018}%
\begin{APACrefauthors}%
Sfyri, E.%
, Rotach, M\BPBI W.%
, Stiperski, I.%
, Bosveld, F\BPBI C.%
, Lehner, M.%
\BCBL {}\ \BBA {} Obleitner, F.%
\end{APACrefauthors}%
\unskip\
\newblock
\APACrefYearMonthDay{2018}{{\APACmonth{10}}}{}.
\newblock
{\BBOQ}\APACrefatitle {Scalar-{Flux} {Similarity} in the {Layer} {Near} the {Surface} {Over} {Mountainous} {Terrain}} {Scalar-{Flux} {Similarity} in the {Layer} {Near} the {Surface} {Over} {Mountainous} {Terrain}}.{\BBCQ}
\newblock
\APACjournalVolNumPages{Boundary-Layer Meteorology}{169}{1}{11--46}.
\newblock
\begin{APACrefURL} [{2024-03-14}]\url{http://link.springer.com/10.1007/s10546-018-0365-3} \end{APACrefURL}
\newblock
\begin{APACrefDOI} \doi{10.1007/s10546-018-0365-3} \end{APACrefDOI}
\PrintBackRefs{\CurrentBib}

\bibitem [\protect \citeauthoryear {%
Shih%
\ \protect \BOthers {.}}{%
Shih%
\ \protect \BOthers {.}}{%
{\protect \APACyear {2025}}%
}]{%
shih_challenges_2025}
\APACinsertmetastar {%
shih_challenges_2025}%
\begin{APACrefauthors}%
Shih, C\BHBI H.%
, Anderson, R\BPBI G.%
, Skaggs, T\BPBI H.%
, Juang, J\BHBI Y.%
, Chen, Y\BHBI Y.%
, Jang, Y\BHBI S.%
\BDBL {}Lo, M\BHBI H.%
\end{APACrefauthors}%
\unskip\
\newblock
\APACrefYearMonthDay{2025}{{\APACmonth{03}}}{}.
\newblock
{\BBOQ}\APACrefatitle {Challenges and limitations of applying the flux variance similarity ({FVS}) method to partition evapotranspiration in a montane cloud forest} {Challenges and limitations of applying the flux variance similarity ({FVS}) method to partition evapotranspiration in a montane cloud forest}.{\BBCQ}
\newblock
\APACjournalVolNumPages{Agricultural and Forest Meteorology}{362}{}{110391}.
\newblock
\begin{APACrefURL} [{2025-09-11}]\url{https://linkinghub.elsevier.com/retrieve/pii/S0168192325000115} \end{APACrefURL}
\newblock
\begin{APACrefDOI} \doi{10.1016/j.agrformet.2025.110391} \end{APACrefDOI}
\PrintBackRefs{\CurrentBib}

\bibitem [\protect \citeauthoryear {%
Siqueira%
\ \BBA {} Katul%
}{%
Siqueira%
\ \BBA {} Katul%
}{%
{\protect \APACyear {2002}}%
}]{%
siqueira_estimating_2002}
\APACinsertmetastar {%
siqueira_estimating_2002}%
\begin{APACrefauthors}%
Siqueira, M.%
\BCBT {}\ \BBA {} Katul, G.%
\end{APACrefauthors}%
\unskip\
\newblock
\APACrefYearMonthDay{2002}{{\APACmonth{04}}}{}.
\newblock
{\BBOQ}\APACrefatitle {Estimating {Heat} {Sources} {And} {Fluxes} {In} {Thermally} {Stratified} {Canopy} {Flows} {Using} {Higher}-{Order} {Closure} {Models}} {Estimating {Heat} {Sources} {And} {Fluxes} {In} {Thermally} {Stratified} {Canopy} {Flows} {Using} {Higher}-{Order} {Closure} {Models}}.{\BBCQ}
\newblock
\APACjournalVolNumPages{Boundary-Layer Meteorology}{103}{1}{125--142}.
\newblock
\begin{APACrefURL} [{2025-09-15}]\url{https://doi.org/10.1023/A:1014526305879} \end{APACrefURL}
\newblock
\begin{APACrefDOI} \doi{10.1023/A:1014526305879} \end{APACrefDOI}
\PrintBackRefs{\CurrentBib}

\bibitem [\protect \citeauthoryear {%
Stiperski%
\ \BBA {} Calaf%
}{%
Stiperski%
\ \BBA {} Calaf%
}{%
{\protect \APACyear {2018}}%
}]{%
stiperski_dependence_2018}
\APACinsertmetastar {%
stiperski_dependence_2018}%
\begin{APACrefauthors}%
Stiperski, I.%
\BCBT {}\ \BBA {} Calaf, M.%
\end{APACrefauthors}%
\unskip\
\newblock
\APACrefYearMonthDay{2018}{{\APACmonth{04}}}{}.
\newblock
{\BBOQ}\APACrefatitle {Dependence of near‐surface similarity scaling on the anisotropy of atmospheric turbulence} {Dependence of near‐surface similarity scaling on the anisotropy of atmospheric turbulence}.{\BBCQ}
\newblock
\APACjournalVolNumPages{Quarterly Journal of the Royal Meteorological Society}{144}{712}{641--657}.
\newblock
\begin{APACrefURL} [{2025-06-04}]\url{https://rmets.onlinelibrary.wiley.com/doi/10.1002/qj.3224} \end{APACrefURL}
\newblock
\begin{APACrefDOI} \doi{10.1002/qj.3224} \end{APACrefDOI}
\PrintBackRefs{\CurrentBib}

\bibitem [\protect \citeauthoryear {%
Stiperski%
\ \BBA {} Calaf%
}{%
Stiperski%
\ \BBA {} Calaf%
}{%
{\protect \APACyear {2023}}%
}]{%
stiperski_generalizing_2023}
\APACinsertmetastar {%
stiperski_generalizing_2023}%
\begin{APACrefauthors}%
Stiperski, I.%
\BCBT {}\ \BBA {} Calaf, M.%
\end{APACrefauthors}%
\unskip\
\newblock
\APACrefYearMonthDay{2023}{{\APACmonth{03}}}{}.
\newblock
{\BBOQ}\APACrefatitle {Generalizing {Monin}-{Obukhov} {Similarity} {Theory} (1954) for {Complex} {Atmospheric} {Turbulence}} {Generalizing {Monin}-{Obukhov} {Similarity} {Theory} (1954) for {Complex} {Atmospheric} {Turbulence}}.{\BBCQ}
\newblock
\APACjournalVolNumPages{Physical Review Letters}{130}{12}{124001}.
\newblock
\begin{APACrefURL} [{2024-03-11}]\url{https://link.aps.org/doi/10.1103/PhysRevLett.130.124001} \end{APACrefURL}
\newblock
\begin{APACrefDOI} \doi{10.1103/PhysRevLett.130.124001} \end{APACrefDOI}
\PrintBackRefs{\CurrentBib}

\bibitem [\protect \citeauthoryear {%
Stiperski%
, Calaf%
\BCBL {}\ \BBA {} Rotach%
}{%
Stiperski%
\ \protect \BOthers {.}}{%
{\protect \APACyear {2019}}%
}]{%
stiperski_scaling_2019}
\APACinsertmetastar {%
stiperski_scaling_2019}%
\begin{APACrefauthors}%
Stiperski, I.%
, Calaf, M.%
\BCBL {}\ \BBA {} Rotach, M\BPBI W.%
\end{APACrefauthors}%
\unskip\
\newblock
\APACrefYearMonthDay{2019}{{\APACmonth{02}}}{}.
\newblock
{\BBOQ}\APACrefatitle {Scaling, {Anisotropy}, and {Complexity} in {Near}‐{Surface} {Atmospheric} {Turbulence}} {Scaling, {Anisotropy}, and {Complexity} in {Near}‐{Surface} {Atmospheric} {Turbulence}}.{\BBCQ}
\newblock
\APACjournalVolNumPages{Journal of Geophysical Research: Atmospheres}{124}{3}{1428--1448}.
\newblock
\begin{APACrefURL} [{2025-01-01}]\url{https://agupubs.onlinelibrary.wiley.com/doi/10.1029/2018JD029383} \end{APACrefURL}
\newblock
\begin{APACrefDOI} \doi{10.1029/2018JD029383} \end{APACrefDOI}
\PrintBackRefs{\CurrentBib}

\bibitem [\protect \citeauthoryear {%
Stiperski%
, Chamecki%
\BCBL {}\ \BBA {} Calaf%
}{%
Stiperski%
, Chamecki%
\BCBL {}\ \BBA {} Calaf%
}{%
{\protect \APACyear {2021}}%
}]{%
stiperski_anisotropy_2021}
\APACinsertmetastar {%
stiperski_anisotropy_2021}%
\begin{APACrefauthors}%
Stiperski, I.%
, Chamecki, M.%
\BCBL {}\ \BBA {} Calaf, M.%
\end{APACrefauthors}%
\unskip\
\newblock
\APACrefYearMonthDay{2021}{{\APACmonth{09}}}{}.
\newblock
{\BBOQ}\APACrefatitle {Anisotropy of {Unstably} {Stratified} {Near}-{Surface} {Turbulence}} {Anisotropy of {Unstably} {Stratified} {Near}-{Surface} {Turbulence}}.{\BBCQ}
\newblock
\APACjournalVolNumPages{Boundary-Layer Meteorology}{180}{3}{363--384}.
\newblock
\begin{APACrefURL} [{2023-08-21}]\url{https://link.springer.com/10.1007/s10546-021-00634-0} \end{APACrefURL}
\newblock
\begin{APACrefDOI} \doi{10.1007/s10546-021-00634-0} \end{APACrefDOI}
\PrintBackRefs{\CurrentBib}

\bibitem [\protect \citeauthoryear {%
Stiperski%
, Katul%
\BCBL {}\ \BBA {} Calaf%
}{%
Stiperski%
, Katul%
\BCBL {}\ \BBA {} Calaf%
}{%
{\protect \APACyear {2021}}%
}]{%
stiperski_universal_2021}
\APACinsertmetastar {%
stiperski_universal_2021}%
\begin{APACrefauthors}%
Stiperski, I.%
, Katul, G\BPBI G.%
\BCBL {}\ \BBA {} Calaf, M.%
\end{APACrefauthors}%
\unskip\
\newblock
\APACrefYearMonthDay{2021}{{\APACmonth{05}}}{}.
\newblock
{\BBOQ}\APACrefatitle {Universal {Return} to {Isotropy} of {Inhomogeneous} {Atmospheric} {Boundary} {Layer} {Turbulence}} {Universal {Return} to {Isotropy} of {Inhomogeneous} {Atmospheric} {Boundary} {Layer} {Turbulence}}.{\BBCQ}
\newblock
\APACjournalVolNumPages{Physical Review Letters}{126}{19}{194501}.
\newblock
\begin{APACrefURL} [{2025-09-15}]\url{https://link.aps.org/doi/10.1103/PhysRevLett.126.194501} \end{APACrefURL}
\newblock
\APACrefnote{Publisher: American Physical Society}
\newblock
\begin{APACrefDOI} \doi{10.1103/PhysRevLett.126.194501} \end{APACrefDOI}
\PrintBackRefs{\CurrentBib}

\bibitem [\protect \citeauthoryear {%
Stull%
}{%
Stull%
}{%
{\protect \APACyear {1988}}%
}]{%
stull_mean_1988}
\APACinsertmetastar {%
stull_mean_1988}%
\begin{APACrefauthors}%
Stull, R\BPBI B.%
\end{APACrefauthors}%
\unskip\
\newblock
\APACrefYearMonthDay{1988}{}{}.
\newblock
{\BBOQ}\APACrefatitle {Mean {Boundary} {Layer} {Characteristics}} {Mean {Boundary} {Layer} {Characteristics}}.{\BBCQ}
\newblock
\BIn{} R\BPBI B.~Stull\ (\BED), \APACrefbtitle {An {Introduction} to {Boundary} {Layer} {Meteorology}} {An {Introduction} to {Boundary} {Layer} {Meteorology}}\ (\BPGS\ 1--27).
\newblock
\APACaddressPublisher{Dordrecht}{Springer Netherlands}.
\newblock
\begin{APACrefURL} [{2023-11-13}]\url{http://link.springer.com/10.1007/978-94-009-3027-8_1} \end{APACrefURL}
\newblock
\begin{APACrefDOI} \doi{10.1007/978-94-009-3027-8_1} \end{APACrefDOI}
\PrintBackRefs{\CurrentBib}

\bibitem [\protect \citeauthoryear {%
Vercauteren%
, Boyko%
, Faranda%
\BCBL {}\ \BBA {} Stiperski%
}{%
Vercauteren%
\ \protect \BOthers {.}}{%
{\protect \APACyear {2019}}%
}]{%
vercauteren_scale_2019}
\APACinsertmetastar {%
vercauteren_scale_2019}%
\begin{APACrefauthors}%
Vercauteren, N.%
, Boyko, V.%
, Faranda, D.%
\BCBL {}\ \BBA {} Stiperski, I.%
\end{APACrefauthors}%
\unskip\
\newblock
\APACrefYearMonthDay{2019}{{\APACmonth{07}}}{}.
\newblock
{\BBOQ}\APACrefatitle {Scale interactions and anisotropy in stable boundary layers} {Scale interactions and anisotropy in stable boundary layers}.{\BBCQ}
\newblock
\APACjournalVolNumPages{Quarterly Journal of the Royal Meteorological Society}{145}{722}{1799--1813}.
\newblock
\begin{APACrefURL} [{2024-12-31}]\url{http://arxiv.org/abs/1809.07031} \end{APACrefURL}
\newblock
\APACrefnote{arXiv:1809.07031 [physics]}
\newblock
\begin{APACrefDOI} \doi{10.1002/qj.3524} \end{APACrefDOI}
\PrintBackRefs{\CurrentBib}

\bibitem [\protect \citeauthoryear {%
Waterman%
}{%
Waterman%
}{%
{\protect \APACyear {2025}}%
{\protect \APACexlab {{\protect \BCnt {1}}}}}]{%
waterman_tswaterneon_aniso_2025}
\APACinsertmetastar {%
waterman_tswaterneon_aniso_2025}%
\begin{APACrefauthors}%
Waterman, T.%
\end{APACrefauthors}%
\unskip\
\newblock
\APACrefYearMonthDay{2025{\protect \BCnt {1}}}{{\APACmonth{09}}}{}.
\newblock
\APACrefbtitle {tswater/neon\_aniso: scalar\_draft\_v0.1.} {tswater/neon\_aniso: scalar\_draft\_v0.1.}
\newblock
\APACaddressPublisher{}{Zenodo}.
\newblock
\begin{APACrefURL} \url{https://doi.org/10.5281/zenodo.17179371} \end{APACrefURL}
\newblock
\begin{APACrefDOI} \doi{10.5281/zenodo.17179371} \end{APACrefDOI}
\PrintBackRefs{\CurrentBib}

\bibitem [\protect \citeauthoryear {%
Waterman%
}{%
Waterman%
}{%
{\protect \APACyear {2025}}%
{\protect \APACexlab {{\protect \BCnt {2}}}}}]{%
waterman_tswaterscalar_neon_2025}
\APACinsertmetastar {%
waterman_tswaterscalar_neon_2025}%
\begin{APACrefauthors}%
Waterman, T.%
\end{APACrefauthors}%
\unskip\
\newblock
\APACrefYearMonthDay{2025{\protect \BCnt {2}}}{{\APACmonth{09}}}{}.
\newblock
\APACrefbtitle {tswater/scalar\_neon: scalar\_draft\_v0.1.} {tswater/scalar\_neon: scalar\_draft\_v0.1.}
\newblock
\APACaddressPublisher{}{Zenodo}.
\newblock
\begin{APACrefURL} \url{https://doi.org/10.5281/zenodo.17179373} \end{APACrefURL}
\newblock
\begin{APACrefDOI} \doi{10.5281/zenodo.17179373} \end{APACrefDOI}
\PrintBackRefs{\CurrentBib}

\bibitem [\protect \citeauthoryear {%
Waterman%
, Bragg%
, Katul%
\BCBL {}\ \BBA {} Chaney%
}{%
Waterman%
\ \protect \BOthers {.}}{%
{\protect \APACyear {2022}}%
}]{%
waterman_examining_2022}
\APACinsertmetastar {%
waterman_examining_2022}%
\begin{APACrefauthors}%
Waterman, T.%
, Bragg, A\BPBI D.%
, Katul, G.%
\BCBL {}\ \BBA {} Chaney, N.%
\end{APACrefauthors}%
\unskip\
\newblock
\APACrefYearMonthDay{2022}{{\APACmonth{04}}}{}.
\newblock
{\BBOQ}\APACrefatitle {Examining {Parameterizations} of {Potential} {Temperature} {Variance} {Across} {Varied} {Landscapes} for {Use} in {Earth} {System} {Models}} {Examining {Parameterizations} of {Potential} {Temperature} {Variance} {Across} {Varied} {Landscapes} for {Use} in {Earth} {System} {Models}}.{\BBCQ}
\newblock
\APACjournalVolNumPages{Journal of Geophysical Research: Atmospheres}{127}{8}{}.
\newblock
\begin{APACrefURL} [{2022-12-19}]\url{https://onlinelibrary.wiley.com/doi/10.1029/2021JD036236} \end{APACrefURL}
\newblock
\begin{APACrefDOI} \doi{10.1029/2021JD036236} \end{APACrefDOI}
\PrintBackRefs{\CurrentBib}

\bibitem [\protect \citeauthoryear {%
Waterman%
, Stiperski%
, Chaney%
\BCBL {}\ \BBA {} Calaf%
}{%
Waterman%
\ \protect \BOthers {.}}{%
{\protect \APACyear {2025}}%
}]{%
waterman_evaluating_2025}
\APACinsertmetastar {%
waterman_evaluating_2025}%
\begin{APACrefauthors}%
Waterman, T.%
, Stiperski, I.%
, Chaney, N.%
\BCBL {}\ \BBA {} Calaf, M.%
\end{APACrefauthors}%
\unskip\
\newblock
\APACrefYearMonthDay{2025}{{\APACmonth{07}}}{}.
\newblock
\APACrefbtitle {Evaluating {Anisotropy}-based {Monin}-{Obukhov} {Similarity} {Theory} over {Canopies} and {Complex} {Terrain}.} {Evaluating {Anisotropy}-based {Monin}-{Obukhov} {Similarity} {Theory} over {Canopies} and {Complex} {Terrain}.}
\newblock
\APACaddressPublisher{}{arXiv}.
\newblock
\begin{APACrefURL} [{2025-09-11}]\url{http://arxiv.org/abs/2502.13970} \end{APACrefURL}
\newblock
\APACrefnote{arXiv:2502.13970 [physics]}
\newblock
\begin{APACrefDOI} \doi{10.48550/arXiv.2502.13970} \end{APACrefDOI}
\PrintBackRefs{\CurrentBib}

\bibitem [\protect \citeauthoryear {%
Weaver%
}{%
Weaver%
}{%
{\protect \APACyear {1990}}%
}]{%
weaver_temperature_1990}
\APACinsertmetastar {%
weaver_temperature_1990}%
\begin{APACrefauthors}%
Weaver, H\BPBI L.%
\end{APACrefauthors}%
\unskip\
\newblock
\APACrefYearMonthDay{1990}{{\APACmonth{10}}}{}.
\newblock
{\BBOQ}\APACrefatitle {Temperature and humidity flux-variance relations determined by one-dimensional eddy correlation} {Temperature and humidity flux-variance relations determined by one-dimensional eddy correlation}.{\BBCQ}
\newblock
\APACjournalVolNumPages{Boundary-Layer Meteorology}{53}{1-2}{77--91}.
\newblock
\begin{APACrefURL} [{2023-11-13}]\url{http://link.springer.com/10.1007/BF00122464} \end{APACrefURL}
\newblock
\begin{APACrefDOI} \doi{10.1007/BF00122464} \end{APACrefDOI}
\PrintBackRefs{\CurrentBib}

\bibitem [\protect \citeauthoryear {%
Wyngaard%
\ \BBA {} Coté%
}{%
Wyngaard%
\ \BBA {} Coté%
}{%
{\protect \APACyear {1971}}%
}]{%
wyngaard_budgets_1971}
\APACinsertmetastar {%
wyngaard_budgets_1971}%
\begin{APACrefauthors}%
Wyngaard, J\BPBI C.%
\BCBT {}\ \BBA {} Coté, O\BPBI R.%
\end{APACrefauthors}%
\unskip\
\newblock
\APACrefYearMonthDay{1971}{{\APACmonth{03}}}{}.
\newblock
{\BBOQ}\APACrefatitle {The {Budgets} of {Turbulent} {Kinetic} {Energy} and {Temperature} {Variance} in the {Atmospheric} {Surface} {Layer}} {The {Budgets} of {Turbulent} {Kinetic} {Energy} and {Temperature} {Variance} in the {Atmospheric} {Surface} {Layer}}.{\BBCQ}
\newblock
\APACjournalVolNumPages{Journal of the Atmospheric Sciences}{28}{2}{190--201}.
\newblock
\begin{APACrefURL} [{2025-09-15}]\url{https://journals.ametsoc.org/view/journals/atsc/28/2/1520-0469_1971_028_0190_tbotke_2_0_co_2.xml} \end{APACrefURL}
\newblock
\APACrefnote{Publisher: American Meteorological Society Section: Journal of the Atmospheric Sciences}
\newblock
\begin{APACrefDOI} \doi{10.1175/1520-0469(1971)028<0190:TBOTKE>2.0.CO;2} \end{APACrefDOI}
\PrintBackRefs{\CurrentBib}

\bibitem [\protect \citeauthoryear {%
Zahn%
\ \protect \BOthers {.}}{%
Zahn%
\ \protect \BOthers {.}}{%
{\protect \APACyear {2022}}%
}]{%
zahn_direct_2022}
\APACinsertmetastar {%
zahn_direct_2022}%
\begin{APACrefauthors}%
Zahn, E.%
, Bou-Zeid, E.%
, Good, S\BPBI P.%
, Katul, G\BPBI G.%
, Thomas, C\BPBI K.%
, Ghannam, K.%
\BDBL {}Kustas, W\BPBI P.%
\end{APACrefauthors}%
\unskip\
\newblock
\APACrefYearMonthDay{2022}{{\APACmonth{03}}}{}.
\newblock
{\BBOQ}\APACrefatitle {Direct partitioning of eddy-covariance water and carbon dioxide fluxes into ground and plant components} {Direct partitioning of eddy-covariance water and carbon dioxide fluxes into ground and plant components}.{\BBCQ}
\newblock
\APACjournalVolNumPages{Agricultural and Forest Meteorology}{315}{}{108790}.
\newblock
\begin{APACrefURL} [{2025-02-04}]\url{https://linkinghub.elsevier.com/retrieve/pii/S0168192321004767} \end{APACrefURL}
\newblock
\begin{APACrefDOI} \doi{10.1016/j.agrformet.2021.108790} \end{APACrefDOI}
\PrintBackRefs{\CurrentBib}

\bibitem [\protect \citeauthoryear {%
Zahn%
, Bou‐Zeid%
\BCBL {}\ \BBA {} Dias%
}{%
Zahn%
\ \protect \BOthers {.}}{%
{\protect \APACyear {2023}}%
}]{%
zahn_relaxed_2023}
\APACinsertmetastar {%
zahn_relaxed_2023}%
\begin{APACrefauthors}%
Zahn, E.%
, Bou‐Zeid, E.%
\BCBL {}\ \BBA {} Dias, N\BPBI L.%
\end{APACrefauthors}%
\unskip\
\newblock
\APACrefYearMonthDay{2023}{{\APACmonth{04}}}{}.
\newblock
{\BBOQ}\APACrefatitle {Relaxed {Eddy} {Accumulation} {Outperforms} {Monin}‐{Obukhov} {Flux} {Models} {Under} {Non}‐{Ideal} {Conditions}} {Relaxed {Eddy} {Accumulation} {Outperforms} {Monin}‐{Obukhov} {Flux} {Models} {Under} {Non}‐{Ideal} {Conditions}}.{\BBCQ}
\newblock
\APACjournalVolNumPages{Geophysical Research Letters}{50}{7}{e2023GL103099}.
\newblock
\begin{APACrefURL} [{2024-03-18}]\url{https://agupubs.onlinelibrary.wiley.com/doi/10.1029/2023GL103099} \end{APACrefURL}
\newblock
\begin{APACrefDOI} \doi{10.1029/2023GL103099} \end{APACrefDOI}
\PrintBackRefs{\CurrentBib}

\bibitem [\protect \citeauthoryear {%
Zahn%
\ \protect \BOthers {.}}{%
Zahn%
\ \protect \BOthers {.}}{%
{\protect \APACyear {2016}}%
}]{%
zahn_scalar_2016}
\APACinsertmetastar {%
zahn_scalar_2016}%
\begin{APACrefauthors}%
Zahn, E.%
, Dias, N\BPBI L.%
, Araújo, A.%
, Sá, L\BPBI D\BPBI A.%
, Sörgel, M.%
, Trebs, I.%
\BDBL {}Manzi, A.%
\end{APACrefauthors}%
\unskip\
\newblock
\APACrefYearMonthDay{2016}{{\APACmonth{09}}}{}.
\newblock
{\BBOQ}\APACrefatitle {Scalar turbulent behavior in the roughness sublayer of an {Amazonian} forest} {Scalar turbulent behavior in the roughness sublayer of an {Amazonian} forest}.{\BBCQ}
\newblock
\APACjournalVolNumPages{Atmospheric Chemistry and Physics}{16}{17}{11349--11366}.
\newblock
\begin{APACrefURL} [{2025-02-04}]\url{https://acp.copernicus.org/articles/16/11349/2016/} \end{APACrefURL}
\newblock
\begin{APACrefDOI} \doi{10.5194/acp-16-11349-2016} \end{APACrefDOI}
\PrintBackRefs{\CurrentBib}

\end{thebibliography}

%
%
%
%
%

\end{document}